\documentclass[journal]{IEEEtran}
\usepackage{mathrsfs}
\usepackage{amssymb}
\usepackage[mathcal]{euscript}
\usepackage{multicol}
\usepackage{cite}
\usepackage{stfloats}
\usepackage{graphicx}
\usepackage{float}
\usepackage{psfrag}
\usepackage{subfigure}
\usepackage{url}
\usepackage{stfloats}
\usepackage{amsmath}
\usepackage{float}
\usepackage{array}
\usepackage{amsthm}
\usepackage{booktabs}  
\usepackage{algorithm} 
\usepackage{algorithmic} 
\usepackage{amsmath,amsfonts,amssymb,amsbsy,bm,paralist,theorem,ifthen,color}
\ifCLASSINFOpdf
\else
\fi
\begin{document}

\title{\textcolor[rgb]{0.00,0.00,0.00}{Fog-Assisted Multi-User SWIPT Networks: Local Computing or Offloading}}

\author{Haina Zheng,~\IEEEmembership{Student~Member,~IEEE}, Ke Xiong,~\IEEEmembership{Member,~IEEE}, Pingyi Fan,~\IEEEmembership{Senior~Member,~IEEE}, \\Zhangdui Zhong,~\IEEEmembership{Senior~Member,~IEEE}, Khaled Ben Letaief,~\IEEEmembership{Fellow,~IEEE}

\thanks{This work was supported in part by the General Program of the National Natural Science Foundation of China (NSFC) under grant no. 61671051, in part by the Beijing Natural Science Foundation under grant no. 4162049, and in part by the major projects of Beijing Municipal Science and Technology Commission under grant no. Z181100003218010, and also in part by national key research and development program under grant no. 2016YFE0200900. (\textit{Corresponding author: Ke Xiong}.)}
\thanks{H. N. Zheng and K. Xiong  are with the School of Computer and Information Technology, and with the Beijing Key Lab of Traffic Data Analysis and Mining, Beijing Jiaotong University, Beijing, P. R. China, and also with the Beijing Key Laboratory of Security and Privacy in Intelligent Transportation, Beijing Jiaotong University, Beijing 100044, China. (e-mail: kxiong@bjtu.edu.cn).}

\thanks{P. Y. Fan is with the National Laboratory for Information Science and Technology and the Department of Electronic Engineering, Tsinghua University, Beijing 100084, China.}

\thanks{Z. D. Zhong is with the State Key Lab of Rail Traffic Control and Safety and is also with Beijing Engineering Research Center of High-speed Railway Broadband Mobile Communications, Beijing Jiaotong University, Beijing 100044, China.}

\thanks{K. B. Letaief is with the School of Engineering, Hong Kong University
of Science and Technology (HKUST), Clear Water Bay, Hong Kong (e-mail:
eekhaled@ece.ust.hk).}}
\maketitle

\begin{abstract}
This paper investigates a fog computing-assisted multi-user simultaneous wireless information and power transfer (SWIPT) \textcolor[rgb]{0.00,0.00,0.00}{network}, where multiple \textcolor[rgb]{0.00,0.00,0.00}{sensors} with power splitting (PS) receiver architectures receive information and harvest energy from a hybrid access point (HAP), and then process the received data by using local computing mode or fog offloading mode. For such a system, an optimization problem is formulated to minimize the \textcolor[rgb]{0.00,0.00,0.00}{sensors'} required \textcolor[rgb]{0.00,0.00,0.00}{energy} while guaranteeing their required information \textcolor[rgb]{0.00,0.00,0.00}{transmissions} and processing rates by jointly optimizing the multi-user scheduling, the time assignment, the sensors' transmit powers and the PS ratios. Since the problem is a \textcolor[rgb]{0.00,0.00,0.00}{mixed integer programming} (MIP) problem and cannot \textcolor[rgb]{0.00,0.00,0.00}{be} solved with existing solution methods, we solve it by applying problem decomposition, variable substitutions and theoretical analysis. For a scheduled sensor, the closed-form and semi-closed-form solutions to achieve its minimal required energy are derived, and then an efficient multi-user scheduling scheme is presented, which can \textcolor[rgb]{0.00,0.00,0.00}{achieve} the \textcolor[rgb]{0.00,0.00,0.00}{suboptimal user scheduling} with low \textcolor[rgb]{0.00,0.00,0.00}{computational} complexity. Numerical results demonstrate our obtained theoretical results, which show that for each sensor, when \textcolor[rgb]{0.00,0.00,0.00}{it} is located close to the HAP \textcolor[rgb]{0.00,0.00,0.00}{or} the fog server \textcolor[rgb]{0.00,0.00,0.00}{(FS)}, the fog offloading mode is the better choice; \textcolor[rgb]{0.00,0.00,0.00}{otherwise,} the local computing mode should be selected. The system \textcolor[rgb]{0.00,0.00,0.00}{performances} in a frame-by-frame manner \textcolor[rgb]{0.00,0.00,0.00}{are} also simulated, which \textcolor[rgb]{0.00,0.00,0.00}{show} that using the energy stored in the \textcolor[rgb]{0.00,0.00,0.00}{batteries} and that harvested from the signals transmitted by previous scheduled sensors can further decrease the \textcolor[rgb]{0.00,0.00,0.00}{total required energy of the sensors}.
\end{abstract}

\begin{IEEEkeywords}
Energy harvesting, simultaneous wireless information and power transfer, fog computing, mode selection, local computing, fog offloading.
\end{IEEEkeywords}

\section{Introduction} \label{intr}
\subsection{Background}
With the rapid development of Internet of Things (IoT) and wireless sensor networks (WSNs), various wireless devices are required to access Internet, motivating lots of data-driven computation-intensive and latency-sensitive mobile intelligent applications, such as \textcolor[rgb]{0.00,0.00,0.00}{augmented reality/virtual reality (AR/VR)}, interactive gaming, \textcolor[rgb]{0.00,0.00,0.00}{autonomous driving and industrial control etc} \cite{ref_overview}. These emerging applications require real-time computations and communications, bringing serious challenges to small-size wireless devices with limited computing capability \cite{ref_resarchintesert1,ref_resarchintesert2}. To effectively overcome these challenges and well support the computation-intensive and latency-sensitive applications with quality of service (QoS) requirements, fog computing (FC), a new paradigm similar to mobile edge computing (MEC) \cite{ref_white}, has been presented as a promising solution, as it is capable of offloading computing tasks at sensors (mobile users (MUs)) to their nearby fog servers \textcolor[rgb]{0.00,0.00,0.00}{(FSs)}. Once a FS finishes the assigned computing task, it will feedback the calculated results to the \textcolor[rgb]{0.00,0.00,0.00}{MUs}. Since FS has relatively strong enough \textcolor[rgb]{0.00,0.00,0.00}{computing} capability, the system performance in terms of task processing latency can be greatly improved \cite{ref_fog}.

Besides computing resources, huge computation-intensive \textcolor[rgb]{0.00,0.00,0.00}{and latency-sensitive} applications in IoT and WSNs also incur a great number of energy consumption at MUs \cite{ref_ict}. However, most \textcolor[rgb]{0.00,0.00,0.00}{sensors} are powered by batteries with limited energy capacities. Thus, how to provide sustainable energy supply to prolong the \textcolor[rgb]{0.00,0.00,0.00}{lifetimes} of the energy-constrained \textcolor[rgb]{0.00,0.00,0.00}{sensors} and reduce the management cost caused by frequent replacement of batteries \textcolor[rgb]{0.00,0.00,0.00}{become} critical
\cite{ref_eh1,ref_eh2,ref_eh21,ref_eh3,ref_eh4}. To resolve \textcolor[rgb]{0.00,0.00,0.00}{these problems}, energy harvesting (EH) has been regarded as a promising technology, since it is able to provide energy to \textcolor[rgb]{0.00,0.00,0.00}{sensors} by utilizing external natural energy sources (e.g. solar and wind \textcolor[rgb]{0.00,0.00,0.00}{etc})\cite{ref_zhong3,ref_zhong4} or \textcolor[rgb]{0.00,0.00,0.00}{harvesting} energy from radio frequency (RF) signals. Compared with traditional natural energy sources, RF signals are less affected by weather \textcolor[rgb]{0.00,0.00,0.00}{or} other external environmental conditions, and can be efficiently controlled and designed, so RF-based EH has greatly potential to provide stable energy to low-power energy-constrained networks including IoTs. Moreover, as RF signals also carry information when they deliver energy, the concept of simultaneous wireless information and power transfer (SWIPT) was proposed and studied in \cite{ref_08} and \cite{ref_10} from an information theoretical perspective. Later, in order to make SWIPT implementable, Zhang et al. \cite{ref_IER} presented two practical receiver architectures, i.e., time switching (TS) and power splitting (PS). Since then, both TS and PS have been widely studied in various wireless systems, see e.g. \cite{ref_LEH1,ref_LEH2,ref_KeX1,ref_KeX2}.

As \textcolor[rgb]{0.00,0.00,0.00}{FC} and SWIPT are two promising technologies that have great potential to be employed in future IoTs and WSNs, \textcolor[rgb]{0.00,0.00,0.00}{integrate} them into a single system and inherit their benefits \textcolor[rgb]{0.00,0.00,0.00}{become} very significant.

\subsection{Related Work}
So far, lots of works on SWIPT or FC can be found in the literature \textcolor[rgb]{0.00,0.00,0.00}{\cite{ref_f,ref_singleuser,ref_multiuser,ref_miso,ref_mimo,ref_outage,ref_ca,ref_csi,ref_nl,ref_anquan,ref_mec2,ref_mec3,ref_mec4,ref_mec5,ref_latency1,ref_consumption,ref_tradeoff,ref_mec9}}. For SWIPT, some works investigated the optimal resource allocation including time assignment, \textcolor[rgb]{0.00,0.00,0.00}{transmit power}, and energy beamforming \textcolor[rgb]{0.00,0.00,0.00}{vector in} various wireless systems \textcolor[rgb]{0.00,0.00,0.00}{\cite{ref_singleuser,ref_multiuser,ref_miso}},
and others focused on the system performance analysis in terms of outage probability \cite{ref_outage} and ergodic capacity \cite{ref_ca} in fading channels or designed the SWIPT systems with some practical limitations, e.g., imperfect channel state information (CSI)\cite{ref_csi}, nonlinear EH circuit features \cite{ref_nl} and communication secrecy requirements\cite{ref_anquan}.
For FC/MEC, different types of offloading frameworks and policies were presented, see e.g.,  \cite{ref_mec2,ref_mec3,ref_mec4,ref_mec5}, and multi-objective optimal resource allocations were studied, see e.g., \cite{ref_latency1,ref_consumption,ref_tradeoff,ref_mec9}, \textcolor[rgb]{0.00,0.00,0.00}{to improve system performance }in various scenarios.

However, existing works mentioned above just studied SWIPT and FC separately, which \textcolor[rgb]{0.00,0.00,0.00}{did not} exploit their benefits contemporaneously in a single system. \textcolor[rgb]{0.00,0.00,0.00}{ In order to inherit the benefits of both technologies, recently, some works began
to investigate FC with EH together. For example, in \cite{ref_re3}, dynamic computation offloading strategy was optimized for MEC system with EH devices, where however, only traditional natural source based EH technology was considered rather than the RF-based EH. Although in \cite{ref_wpt1,ref_wpt2,ref_wpt3,ref_wpt4}, the FC systems with RF-based EH were studied, where however, only wireless power transfer (WPT) was adopted rather than SWIPT. Most recently, a few works studied fog computing-assisted SWIPT networks see e.g., \cite{ref_39,ref_40,ref_41,ref_42,ref_43}. However, they did not jointly design the task offloading or only considered the single user scenario or only involved the TS SWIPT receiver. Specifically, in \cite{ref_39}, the authors studied the resource allocation for two-hop fog-assisted SWIPT OFDM networks, where however, the computation offloading was not involved. In \cite{ref_40}, the power minimization problem was studied in a SWIPT-aided fog computing networks with dog offloading, where however, the fog server was just used to assign tasks rather than participate in computing. In \cite{ref_41}, the authors studied the optimal resource allocation in ultra-low power fog-computing SWIPT-based networks, where however, only the TS receiver architecture was adopted and only single-user was considered. In \cite{ref_42}, the authors extended the work in \cite{ref_41} to a multi-user scenario, but it still only studied the TS receiver architecture. In \cite{ref_43}, the power minimization problem was investigated in SWIPT-aware fog computing system with PS receiver architecture, where however, only single-user was considered.}

\subsection{Motivations and Contributions}
\textcolor[rgb]{0.00,0.00,0.00}{As it was shown in \cite{ref_singleuser,ref_44} that PS receiver architecture is able to achieve the better performance in terms of the larger energy-rate region and higher end-to-end information rate than the TS one. Therefore, in this paper, we focus on a multiuser fog computing-assisted SWIPT networks with PS receiver architectures.} To the best of the authors' knowledge, \textcolor[rgb]{0.00,0.00,0.00}{this is the first work on the multi-user fog computing-assisted PS SWIPT neworks. For such a system, where} each sensor has to first harvest energy and receive information from a hybrid access point (HAP) that is with fixed power supply, and then tries to process the received information itself (namely, local computing mode) or offload the computing task to a nearby FS (namely, fog offloading mode) with the harvested energy, \textcolor[rgb]{0.00,0.00,0.00}{we desire to answer the following fundamental questions.}
\begin{enumerate}
  \item \textcolor[rgb]{0.00,0.00,0.00}{How to optimally schedule the sensors to minimize their total required energy?}
  \item For each sensor, what is its performance limit in terms of the minimal required \textcolor[rgb]{0.00,0.00,0.00}{energy} and what is the corresponding optimal resource allocation?
  \item For each sensor, is there one of the two modes (i.e., local computing or fog offloading) always superior to the other one, \textcolor[rgb]{0.00,0.00,0.00}{and} for a given system configuration, which one should be the better choice?
\end{enumerate}
The main contributions of this work are summarized as follows.

\begin{itemize}
 \item \emph{Firstly,} a multi-user scheduling framework is presented based on \textcolor[rgb]{0.00,0.00,0.00}{time division multiple access (TDMA)} manner, where for each time block, only one MU is scheduled to be served and for each time frame composed of multiple time blocks, all MUs can be served. Particulary, for each scheduled \textcolor[rgb]{0.00,0.00,0.00}{MU, either} local computing mode or fog offloading mode can be selected.
  \item \emph{Secondly,} to reduce the total energy requirement of all MUs, \textcolor[rgb]{0.00,0.00,0.00}{an energy}-minimization optimization problem is formulated by jointly optimizing the user scheduling order, the mode selections, the time assignments, the transmit \textcolor[rgb]{0.00,0.00,0.00}{powers} at MUs, and the PS ratios under the required information rates and energy harvesting constraints.
  \item \emph{Thirdly,} since the optimization problem is a \textcolor[rgb]{0.00,0.00,0.00}{mixed integer programming} (MIP) problem and cannot be directly solved by using standard convex solution methods, we first optimize the rest variables by fixing the user scheduling order, and then decompose the new problem into two sub-optimization problems with a given mode selection.
      \textcolor[rgb]{0.00,0.00,0.00}{Note that compared with existing works on WPT-assisted FC networks, the new variables associated with the SWIPT receiver architectures (i.e., the PS ratios) are jointly optimized with  time assignments and the transmit powers at the MU. The coupling of these variables  makes each sub-problem non-convex, which cannot be directly solved by using known convex problem solution method. Therefore, by using the perspective function and some mathematical tackles, we fortunately find an efficient way to solve them and obtain some closed-form and semi-closed-form solutions to the two sub-problems, which characterize the quantitative relationships between the system performance and prameters.}
      Then the optimal mode selection is determined by choosing the \textcolor[rgb]{0.00,0.00,0.00}{one} with the less \textcolor[rgb]{0.00,0.00,0.00}{energy} requirement. Finally, with the obtained optimal mode selections, time assignments, transmit \textcolor[rgb]{0.00,0.00,0.00}{powers} at MUs, and PS ratios, the optimal user scheduling \textcolor[rgb]{0.00,0.00,0.00}{scheme} is achieved by serving the sensor who requires the minimal \textcolor[rgb]{0.00,0.00,0.00}{energy} among the unscheduled ones for a give time block.
  \item \emph{Fourthly,} for better understanding the system and providing some simple deployment policy, the quantitative relationship between some system parameters, e.g., the number of logic operations per bit, the scaling factor of the task result, and the minimal required energy of both modes are analyzed, the quantitative relationship between the location of the MU for fixed HAP and FS and the minimal required energy of both modes is also studied by simulations. It is found that when the number of logic operations per bit is lower than a certain \textcolor[rgb]{0.00,0.00,0.00}{threshold} or the scaling factor of the task result is higher than a certain \textcolor[rgb]{0.00,0.00,0.00}{threshold}, the local computing mode is a better choice; \textcolor[rgb]{0.00,0.00,0.00}{otherwise}, the fog offloading mode should be selected. Besides, when the location of MU is close to \textcolor[rgb]{0.00,0.00,0.00}{HAP or FS, the} fog offloading mode is a better choice and for the rest locations, \textcolor[rgb]{0.00,0.00,0.00}{the} local computing mode should be selected.

\end{itemize}

\begin{figure*}[htb]
\centering
\includegraphics[width=0.58\textwidth]{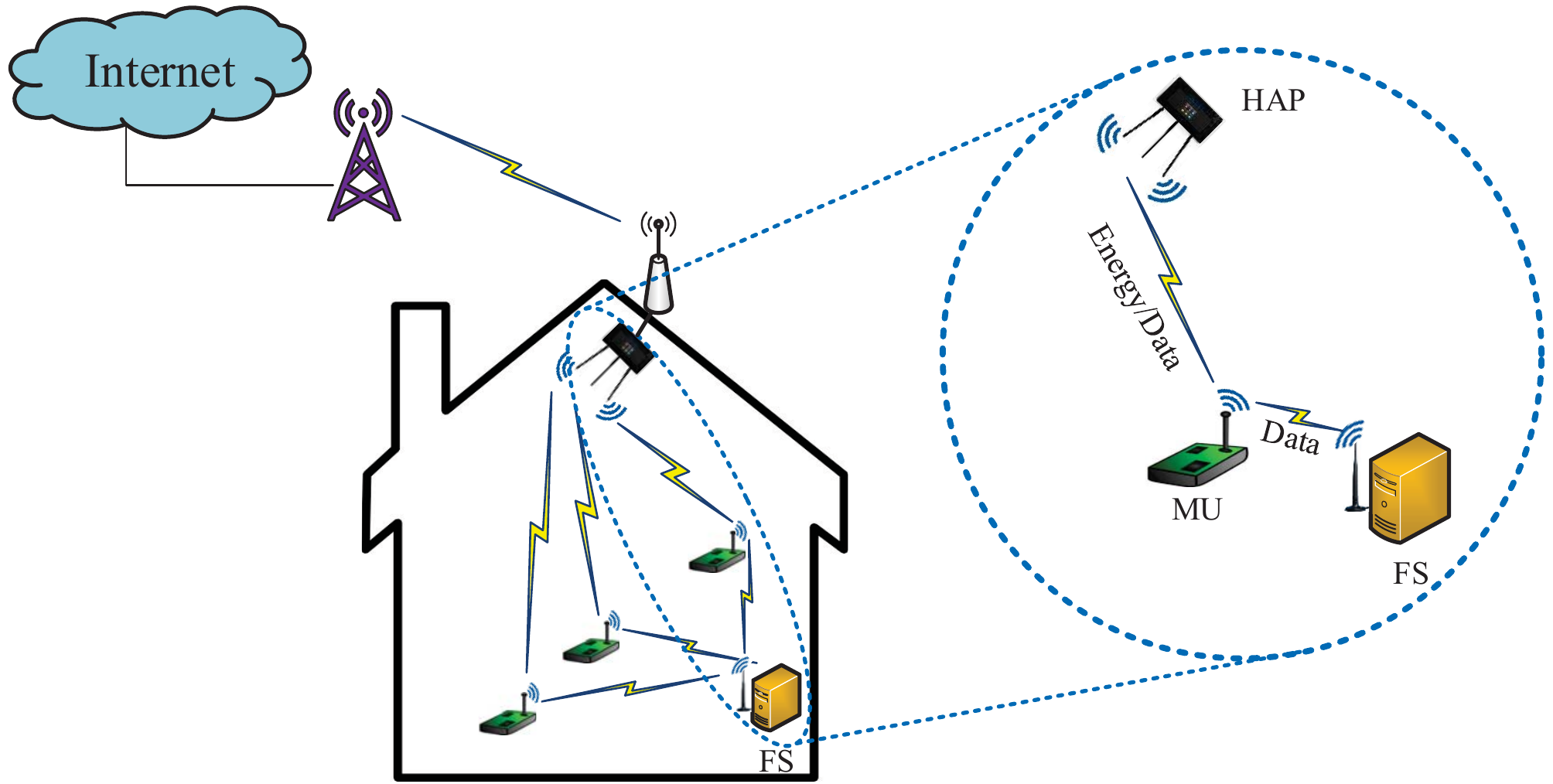}
\caption{Illustration of \textcolor[rgb]{0.00,0.00,0.00}{the multi-user fog computing-assisted PS SWIPT system} in a smart home scenario}
\label{fig_System_model}\vspace{-0.1 in}
\end{figure*}

\begin{figure*}[htb]
\centering
\includegraphics[width=0.98\textwidth]{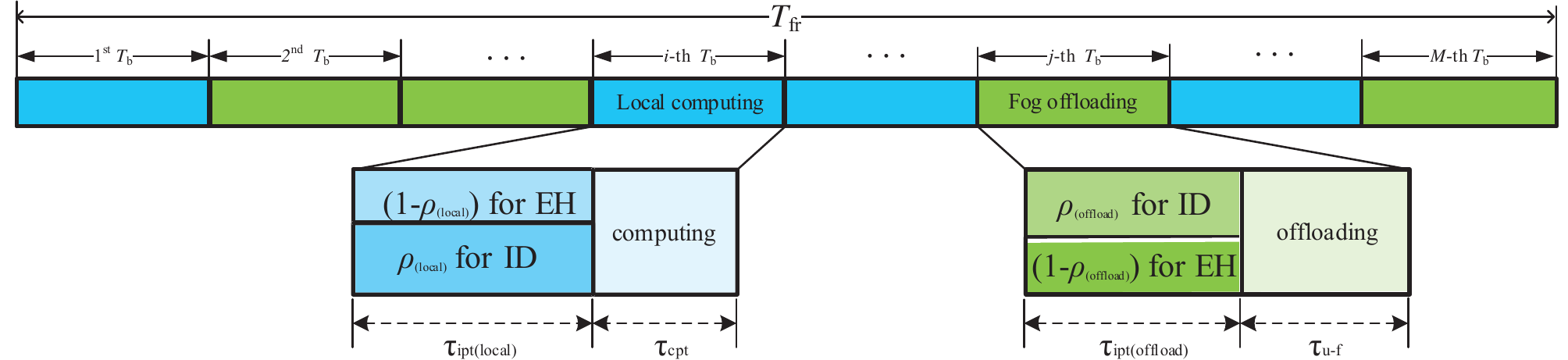}
\caption{Illustration of time frame structure}
\label{fig_time_model}\vspace{-0.1 in}
\end{figure*}

The rest of this paper is organized as follows. Section \ref{model} describes the system model. In Section \ref{solution}, the problem formulation and solution are given, including the optimal solutions for the formulation problem and our proposed user scheduling scheme. Section \ref{discuss} \textcolor[rgb]{0.00,0.00,0.00}{analyzes the system performance}. Section \ref{frame} discusses how to run the system in a frame by frame continuous scenario. Section \ref{results} provides some simulation results and finally, Section \ref{conclusion} summarizes the paper.

\section{system model}\label{model}
We consider a multi-user fog computing-assisted \textcolor[rgb]{0.00,0.00,0.00}{PS} SWIPT system consisting of a multi-antenna HAP, $M$ single-antenna MUs and a single-antenna FS, as illustrated in Figure~\ref{fig_System_model}.

The HAP desires to transmit data to MUs, and once a MU receives the data, it will process the data immediately for use. It is assumed that each MU is energy-constrained and only with very limited stored energy, and the HAP is with sufficient power supply, so that the HAP is able to charge MUs with its transmitted signals. \textcolor[rgb]{0.00,0.00,0.00}{PS} receiver architectures are employed at all MUs, so they are able to decode information and harvest energy simultaneously from the received RF signals transmitted by the HAP. A FS is deployed \textcolor[rgb]{0.00,0.00,0.00}{closed} to the MUs. As a result, the computing task of MUs can be accomplished either by MUs themselves (i.e., \emph{local computing mode}) or helped by the FS (i.e., \emph{fog offloading mode}).

Let $\mathcal{M} \triangleq \{1,...,M\}$ \textcolor[rgb]{0.00,0.00,0.00}{denotes} the set of MUs and $m \in \mathcal{M}$ represents the $m$-th MU. Denote $T_{\textrm{fr}}$ as a time frame for the multi-user system, in which all the MUs are required to be served. To do so, each time frame with the interval of $T_{\textrm{fr}}$ is divided into $M$ blocks with equal time interval $\tfrac{T_{\textrm{fr}}}{M}$. For convenience, with a little abuse of notations, we define $T_{\textrm{b}}$ $\triangleq$ $\tfrac{T_{\textrm{fr}}}{M}$. Let $T_{\textrm{req}}$ be the \textcolor[rgb]{0.00,0.00,0.00}{maximal  delay tolerance} of MUs. In order to satisfy the delay requirement, $T_{\textrm{b}}$ is chosen such that $T_{\textrm{b}}\leqslant T_{\textrm{req}}$, and the task associated with MU $m$ must be completed within $T_{\textrm{b}}$.

Block fading channel model is assumed, so in each time block, all channel coefficients are regarded as constants. To be general, both large-scale fading and small-scale fading are considered. For the large-scale fading, the line-of-sight (LoS) component associated with the channel is modelled by using the \textcolor[rgb]{0.00,0.00,0.00}{International Telecommunication Union (ITU)} indoor channel model as \cite{ref_re13}:
\begin{flalign}\label{fla_los}
L = 20\log f_{c} + n\log d - 28,
\end{flalign}
in which $L$ is the total path loss, $f_c$ is the frequency of \textcolor[rgb]{0.00,0.00,0.00}{carrier}, $d$ is the distance between the transmitter and the reciever and $n$ is the corresponding power loss coefficient. For the small-scale fading, the channel coefficients may change independently from current block to the next following Rician distribution. Without loss of generality, the time interval of each block is assumed to be equal to $T_{\textrm{b}}$.

Define $\Psi \triangleq \{\psi_{m, t}\}_{M\times M}$, representing the user scheduled matrix with $\psi_{m,t}\in \{0,1\}$\textcolor[rgb]{0.00,0.00,0.00}{, where} $m, t\in\{1,2,...,M\}$.  $\psi_{m,t}=1$
indicates that in the $t$-th time block, MU $m$ is scheduled and served; \textcolor[rgb]{0.00,0.00,0.00}{otherwise}, $\psi_{m,t}=0$ means that in the $t$-th time block, MU $m$ is not scheduled and served.

In order to avoid the inter-user interference, in each $T_{\textrm{b}}$, only one MU is allowed to be scheduled and served, and in order to make sure all the MUs be served in $T_{\textrm{fr}}$, each \textcolor[rgb]{0.00,0.00,0.00}{MU} is only scheduled once during $T_{\textrm{fr}}$. Therefore, it is satisfied that

\begin{equation} \label{fla_c8}
\left\{ \begin{aligned}
                  &\sum\nolimits_{m=1}^M\psi_{m,t}=1,  \forall t,\\
                  &\sum\nolimits_{t=1}^M\psi_{m,t}=1,  \forall m.
                          \end{aligned} \right.
                          \end{equation}

Fig.~\ref{fig_time_model} illustrates our presented transmission protocol. For each $T_{\textrm{b}}$, it is divided into several time slots to complete the local computing or fog offloading. For local computing, it is divided into two parts and for fog offloading, it is divided into four parts.

For both modes, in their first part with time interval $\tau_{\textrm{ipt}}$, \textcolor[rgb]{0.00,0.00,0.00}{each MU} decodes the received data and harvests energy from the transmitted signals by HAP. Denote the RF signal symbol transmitted by the HAP as $s$, which is originated from independent Gaussian codebooks, i.e., $s \sim \mathcal{CN} (0, 1)$.  The \textcolor[rgb]{0.00,0.00,0.00}{MU $m$'s} beamforming  vector is $\bm{w}^{(m)} \in \mathbb{C}^{N_A\times 1}$, where $N_A$ is the number of antennas deployed at the HAP. The channel vector from the HAP to MU $m$ is denoted with $\bm{h}^{(m)}_{\textrm{AP-u}} \in \mathbb{C}^{N_A\times 1}$. Assuming that perfect channel state information (CSI) is
known by MU $m$, \textcolor[rgb]{0.00,0.00,0.00}{which can be realized by channel estimation and fed back to HAP and FS. Such assumptions have been widely adopted for the optimal design and performance limit analysis of wireless communication systems, see e.g., \cite{ref_39,ref_40,ref_41,ref_42,ref_43}.}
The received signal at MU $m$ is given by
\begin{flalign}
y^{(m)} = \sqrt{P_{\textrm{AP}}}\bm{h_{\textrm{AP-u}}}^{(m)H} \bm{w}^{(m)} s + n,
\end{flalign}
where $P_\textrm{AP}$ is the available transmit power of the HAP, and $n \sim \mathcal{CN}$ (0, $\sigma^{2}_{n}$) is the noise received at \textcolor[rgb]{0.00,0.00,0.00}{each MU identically}, which obeys the circularly symmetric complex Gaussian distribution. Since the channel between the HAP and \textcolor[rgb]{0.00,0.00,0.00}{each MU $m$} is a typical multiple input single output (MISO) channel, by using the maximum rate transmission (MRT) strategy, the optimal $\bm{w}^{(m)^*}$ related to $\bm{h}^{(m)}_{\textrm{AP-u}}$ can be given by \cite{ref_ketsp}
\begin{flalign}\label{fla_2}
\bm{w}^{(m)^*} = \frac{\bm{h}^{(m)}_{\textrm{AP-u}}}{\|\bm{h}^{(m)}_{\textrm{AP-u}}\|}.
\end{flalign}
With PS SWIPT receiver architecture, a part of the received signals' power is \textcolor[rgb]{0.00,0.00,0.00}{inputted} into the EH circuit for energy harvesting and the rest part of signals' power at MU $m$ is input into the information decoding (ID) circuit for information receiving. Let \textcolor[rgb]{0.00,0.00,0.00}{$\rho^{(m)}$} $\in$ (0, 1) be the power splitting factor \textcolor[rgb]{0.00,0.00,0.00}{of MU $m$}. The harvested energy at MU $m$ in its scheduled $T_{\textrm{b}}$ can be given by
\begin{flalign}
E^{(m)}_{\textrm{e}} = \eta\left(1-\rho^{(m)}\right)P_{\textrm{AP}}\left|\bm{h}_{\textrm{AP-u}}^{(m)H} \bm{w}^{(m)}\right|^2\tau^{(m)}_{\textrm{ipt}},
\end{flalign}
where $\eta\in (0,1)$ \textcolor[rgb]{0.00,0.00,0.00}{denotes} the energy conversion efficiency of the EH circuit.
To fully utilize the broadcast feature of wireless channels, we assume that MU $m$ also accumulates the energy from the signals transmitted by the HAP to its previous MUs. Thus, the total harvested energy at MU $m$ in the \textcolor[rgb]{0.00,0.00,0.00}{previous} $m$ time blocks is
\begin{flalign}\label{meh}
E^{(m)}_{\textrm{eh}} = E^{(m)}_{\textrm{e}}+\eta P_{\textrm{AP}}\sum\nolimits_{i=1}^{m-1}\left|\bm{h}_{\textrm{AP-u}}^{(i)H} \bm{w}^{(i)}\right|^2\tau^{(i)}_{\textrm{ipt}}\textcolor[rgb]{0.00,0.00,0.00}{.}
\end{flalign}
Since when MU $m$ is scheduled, its previous $(m-1)$ MUs have been served. Therefore, the second term in (\ref{meh}), i.e., $\eta P_{\textrm{AP}}\sum\nolimits_{i=1}^{m-1}\left|\bm{h}_{\textrm{AP-u}}^{(i)H} \bm{w}^{(i)}\right|^2\tau^{(i)}_{\textrm{ipt}}$ is determined, which is \textcolor[rgb]{0.00,0.00,0.00}{a} constant to MU $m$. Hereafter, $\eta P_{\textrm{AP}}\sum\nolimits_{i=1}^{m-1}\left|\bm{h}_{\textrm{AP-u}}^{(i)H} \bm{w}^{(i)}\right|^2\tau^{(i)}_{\textrm{ipt}}$ is denoted by $\iota$ \textcolor[rgb]{0.00,0.00,0.00}{for} notational simplicity.
The average achievable information rate $R_\textrm{AP-u}$ over $T_{\textrm{b}}$  at MU $m$ can be given by

\begin{small}
\begin{flalign}
R^{(m)}_{\textrm{AP-u}} = B\frac{\tau^{(m)}_{\textrm{ipt}}}{T_{\textrm{b}}}\log\left(1+\frac{\rho^{(m)} P_{\textrm{AP}}\left|\bm{h}_{\textrm{AP-u}}^{(m)H} \bm{w}^{(m)}\right|^2}{\sigma_{n}^2}\right) ,
\end{flalign}
\end{small}
where $B$ is the system frequency bandwidth. Following \cite{ref_circuit},
we assume that the consumed energy for information decoding at MU $m$ is proportional to the received information amount. Therefore, the required energy for information decoding at MU $m$ can be given by
\begin{small}
\begin{flalign}
E^{(m)}_{\textrm{id}}\! &= \xi R^{(m)}_{\textrm{AP-u}}T_{\textrm{b}} \nonumber\\\quad
&= \xi B\log\left(\!1+\frac{\rho^{(m)} P_{\textrm{AP}}\!\left|\bm{h}_{\textrm{AP-u}}^{(m)H} \bm{w}^{(m)}\right|^2}{\sigma_{n}^2}\!\right)\tau^{(m)}_{\textrm{ipt}},
\end{flalign}
\end{small}
where $\xi$ (Joule/bit) is \textcolor[rgb]{0.00,0.00,0.00}{a} constant, which is used to characterize the energy \textcolor[rgb]{0.00,0.00,0.00}{requirement} for decoding \textcolor[rgb]{0.00,0.00,0.00}{one} bit.

\subsubsection{Local Computing Mode}
As mentioned previously, \textcolor[rgb]{0.00,0.00,0.00}{for MU $m$}, when the local computing mode is selected, the time block with interval of $T_{\textrm{b}}$ is divided into two parts, and in the second part, i.e., $\tau^{(m)}_{\textrm{cpt}}$, MU $m$ processes the received data \textcolor[rgb]{0.00,0.00,0.00}{by} itself.

To do so, some energy is required for data processing. \textcolor[rgb]{0.00,0.00,0.00}{As described in \cite{ref_re12}, the energy requirement} is larger than the Landauer limit by a factor of $M_\textrm{c}$, i.e., $M_\textrm{c}N_0\ln\!2$, where $M_\textrm{c}$ is a time-dependent immaturity factor of the technology and $N_0$ is the thermal noise spectral density. With such a computing energy requirement model, the local computing energy requirement at MU $m$ can be expressed by
\begin{flalign}\label{flaecpt}
E^{(m)}_{\textrm{cpt}} = F_{0}\alpha M_{\textrm{c}}N_{0}\ln\!2 KR^{(m)}_{\textrm{AP-u}}T_{\textrm{b}},
\end{flalign}
\textcolor[rgb]{0.00,0.00,0.00}{where $F_0$ is the fanout, i.e., the number of loading logic gates, $\alpha$ is the activity factor, respectively. $K$ is the number of logic operations per bit and a linear model w.r.t $K$ to evaluate the computational complexity of local computing. That is, more bits are required to be processed, more computation should be performed by local computing. According to \cite{ref_46} and \cite{ref_f0}, the value of $K$ depends on the specific algorithms, but at high bit rate, the computing operation with linear complexity is often expected to reduce the computational complexity and power consumption. Hence, similar to \cite{ref_41,ref_46}, we also adopt the linear model in (\ref{flaecpt}) to characterize the local computing energy consumption in this paper.}
\textcolor[rgb]{0.00,0.00,0.00}{$R^{(m)}_{\textrm{AP-u}}T_{\textrm{b}}$ represents the data size received in each $T_{\textrm{b}}$.}

\textcolor[rgb]{0.00,0.00,0.00}{As a result, the total required energy at  MU $m$ can be described by
\begin{flalign}\label{lcenergy}
E^{(m)}_{\textrm{u}}= E^{(m)}_{\textrm{id}} + E^{(m)}_{\textrm{cpt}}  - E^{(m)}_{\textrm{eh}}-E^{(m)}_{\textrm{s}},
\end{flalign}
where $E^{(m)}_{\textrm{s}}$ denotes the remaining energy stored in the battery of MU $m$ after the last transmission. When $E^{(m)}_{\textrm{u}}\geq 0$, it means that MU $m$'s harvested energy is less than the required energy, i.e., $E^{(m)}_{\textrm{id}} + E^{(m)}_{\textrm{cpt}}$. In this case, the battery has to discharge a certain amount of energy, i.e., $E^{(m)}_{\textrm{u}}$, to help accomplish the local computing.  When $E^{(m)}_{\textrm{u}}< 0$, it implies that the harvested energy is more than the required energy $E^{(m)}_{\textrm{id}} + E^{(m)}_{\textrm{cpt}}$.}
\subsubsection{Fog Offloading Mode}
When the fog offloading mode is selected, the time block with interval of $T_{\textrm{b}}$ is divided into four parts. \textcolor[rgb]{0.00,0.00,0.00}{In the first parts, the process is similar to the local computing mode,} and in the last three parts, MU $m$ offloads the decoded data to the FS, and then, waits the FS to process the data and feedback \textcolor[rgb]{0.00,0.00,0.00}{the result}.

Let $h^{(m)}_{\textrm{u-f}}$ be the complex-valued channel coefficient from MU $m$ to the FS. The average achievable information rate $R^{(m)}_\textrm{u-f}$  associated with the offloading over $T_{\textrm{b}}$ can be given by
\begin{small}
\begin{flalign}
R^{(m)}_{\textrm{u-f}} = B\frac{\tau^{(m)}_{\textrm{u-f}}}{T_{\textrm{b}}}\log\left(1 +\frac{\left|h^{(m)}_{\textrm{u-f}}\right|^2P^{(m)}_{\textrm{u-f}}}{\sigma_{\textrm{\textcolor[rgb]{0.00,0.00,0.00}{s}}}^2}\right),
\end{flalign}
\end{small}
where $P^{(m)}_\textrm{u-f}$ denotes the transmit power at MU $m$, $\sigma^{2}_{\textrm{s}}$ is the noise power at the \textcolor[rgb]{0.00,0.00,0.00}{FS}, $\tau^{(m)}_{\textrm{u-f}}$ is the time used for task offloading from MU $m$ to the FS.
Correspondingly, the energy \textcolor[rgb]{0.00,0.00,0.00}{required} for task offloading at MU $m$ is
\begin{flalign}\label{flaeu-f}
E^{(m)}_{\textrm{u-f}} = P^{(m)}_{\textrm{u-f}}\tau^{(m)}_{\textrm{u-f}},
\end{flalign}
where $P^{(m)}_{\textrm{u-f}}$ is constrained by the maximal available transmit power $P^{\textrm{(max)}}_{\textrm{u-f}}$, i.e.,
\begin{flalign}\label{fla_p}
P^{(m)}_{\textrm{u-f}}\leq P^{\textrm{(max)}}_{\textrm{u-f}}.
\end{flalign}
Once the FS receives the data, it will perform computing over the data. The required time for computing is
\begin{flalign}
\tau^{(m)}_{\textrm{fogcpt}}=\frac{KR^{(m)}_{\textrm{u-f}}T_{\textrm{b}}}{f_\textrm{fogcpt}},
\end{flalign}
where $K$ has the same meaning with that in (\ref{flaecpt}), $R^{(m)}_{\textrm{u-f}}T_{\textrm{b}}$ represents the \textcolor[rgb]{0.00,0.00,0.00}{amount} of the data transmitted from MU $m$ to the FS, and $f_\textrm{fogcpt}$ denotes the logic operations per second of the FS.

After accomplishing the computing operations, the FS transmits the computed result to MU $m$. \textcolor[rgb]{0.00,0.00,0.00}{For simplicity,} it is assumed that the data amount of the computed result is proportional to that of the input one. By defining a scaling factor $\beta^{(m)}$, the number of bits of the computed result of MU $m$ can be given by
\begin{flalign}\label{14}
N^{(m)}_{\textrm{fd}} = \beta^{(m)} R^{(m)}_{\textrm{u-f}}T_{\textrm{b}}.
\end{flalign}
When $\beta^{(m)}\leq1$, it indicates that the computing task of MU $m$ is similar to a data compression processing and when $\beta^{(m)} > 1$, the computing task of MU $m$ is similar to a data unzip processing.

Let $h^{(m)}_{\textrm{f-u}}$ be the complex-valued channel coefficient from the FS to MU $m$. The average achievable information rate $R^{(m)}_{\textrm{f-u}}$  from the FS to MU $m$ for computing result feedback is
\begin{small}
\begin{flalign}\label{15}
R^{(m)}_{\textrm{f-u}} = B\frac{\tau^{(m)}_{\textrm{f-u}}}{T_{\textrm{b}}}\log\left(1+\frac{\left|h^{(m)}_{\textrm{f-u}}\right|^2P_{\textrm{f-u}}}{\sigma_{\textrm{f}}^2}\right) ,
\end{flalign}
\end{small}
where $P_{\textrm{f-u}}$ denotes the transmit power of the FS, and $\sigma^{2}_{\textrm{f}}$ is the received noise power. $\tau^{(m)}_{\textrm{f-u}}$ is the transmission time from the FS to MU $m$. Thus, \textcolor[rgb]{0.00,0.00,0.00}{the time assignment must satisfy that}
\begin{flalign}\label{fla_time}
\tau^{(m)}_{\textrm{ipt}}+\tau^{(m)}_{\textrm{u-f}}+\tau^{(m)}_{\textrm{fogcpt}}+\tau^{(m)}_{\textrm{f-u}}\leq T_{\textrm{b}}.
\end{flalign}
Moreover, in order to guarantee all computed result be transmitted back to MU $m$ within $\tau^{(m)}_{\textrm{f-u}}$, it should satisfy that
\begin{flalign}\label{fla_beita}
R^{(m)}_{\textrm{f-u}}T_{\textrm{b}} \ge \beta^{(m)} R^{(m)}_{\textrm{u-f}}T_{\textrm{b}}.
\end{flalign}

\textcolor[rgb]{0.00,0.00,0.00}{Similar to the local computing, the total required energy at MU $m$ by using fog offloading mode is
\begin{flalign}\label{lcenergy1}
E^{(m)}_{\textrm{u}}= E^{(m)}_{\textrm{id}} +  E^{(m)}_{\textrm{u-f}} - E^{(m)}_{\textrm{eh}}-E^{(m)}_{\textrm{s}}.
\end{flalign}
When $E^{(m)}_{\textrm{u}}\geq 0$, the harvested energy is less than the total required energy $E^{(m)}_{\textrm{id}} + E^{(m)}_{\textrm{u-f}}$.  When $E^{(m)}_{\textrm{u}}< 0$, the harvested energy is more than the total required energy $E^{(m)}_{\textrm{id}} + E^{(m)}_{\textrm{u-f}}$.}

As mentioned previously, for MU $m$, it may select either local computing mode or fog offloading mode. Let $\theta^{(m)}$ be the mode selection indicator with $\theta^{(m)}\in\{0,1\}$, where $\theta^{(m)} = 1$ implies the local computing mode being selected and $\theta^{(m)} = 0$ indicates the fog offloading mode being selected. The total energy requirement of MU $m$ can be given by
\begin{flalign}
E^{(m)}_{\textrm{u}}\!=E_{\textrm{id}}^{(m)}\!+ \theta^{(m)} E_{\textrm{cpt}}^{(m)}\!+ ( 1 - \theta^{(m)} ) E_{\textrm{u-f}}^{(m)}\!- E_{\textrm{eh}}^{(m)}-E^{(m)}_{\textrm{s}},
\end{flalign}
\section{OPTIMAL PROBLEM FORMULATION AND SOLUTION}\label{solution}
\subsection{Problem formulation}
This subsection formulates an optimization problem to minimize the total energy requirement of all MUs by jointly optimizing the user scheduling order,
the mode selections, the time assignments, the transmit
\textcolor[rgb]{0.00,0.00,0.00}{powers} at MUs, and the PS ratios under the required
information rates and energy harvesting constraints,
in order to prolong their \textcolor[rgb]{0.00,0.00,0.00}{lifetimes} while guaranteeing \textcolor[rgb]{0.00,0.00,0.00}{their} minimal required information \textcolor[rgb]{0.00,0.00,0.00}{transmissions} and processing \textcolor[rgb]{0.00,0.00,0.00}{rates}.
The optimization problem can be mathematically expressed by

\begin{flalign} \label{fla_comp1}
\textbf{P}: {\kern 2pt} &\!\!\!\!\!\mathop {\min }\limits_{\substack{\Psi, \bm{\tau}^{(m)}_{\textrm{local}}, \bm{\tau}^{(m)}_{\textrm{offload}} \\ \bm{\rho}^{(m)}, P_{\textrm{u-f}}^{(m)}, \theta^{(m)}}} \,
 \!\!\sum\nolimits_{m=1}^ME^{(m)}_{\textrm{u}}, \\
 \!\!\!\!\!\!\!\!\!\!{\rm{s}}{\rm{.t}}{\rm{. }}~ &  R^{(m)}_{\textrm{AP-u}}\ge R_{\textrm{th}}, \tag{21a}\label{20a}\nonumber \\ \quad
 \ {\rm{ }}~ &   \theta^{(m)}\tau^{(m)}_{\textrm{cpt}}f_{\textrm{op}} + ( 1\!-\! \theta^{(m)} ) KR^{(m)}_{\textrm{u-f}}T_{\textrm{b}} \ge KR^{(m)}_{\textrm{AP-u}}T_{\textrm{b}}, \tag{21b}\label{20b}\nonumber \\ \quad
 \ {\rm{ }}~ &   \tau^{(m)}_{\textrm{fogcpt}}f_{\textrm{fogop}} \ge KR^{(m)}_{\textrm{u-f}}T_{\textrm{b}},  \tag{21c}\label{20c}\nonumber \\ \quad
  \ {\rm{ }}~ &   R^{(m)}_{\textrm{f-u}} \ge \beta^{(m)} R^{(m)}_{\textrm{u-f}}, \tag{21d}\label{20d} \nonumber \\ \quad
 \ {\rm{ }}~ &   \tau^{(m)}_{\textrm{ipt}}\!\!+ \theta^{(m)}\tau^{(m)}_{\textrm{cpt}}\!+ ( 1\!-\! \theta^{(m)} ) ( \tau^{(m)}_{\textrm{u-f}} \!+ \tau^{(m)}_{\textrm{fogcpt}} + \tau^{(m)}_{\textrm{f-u}}) \leq T_{\textrm{b}}, \nonumber \\ \quad
 &0 \preceq \bm{\tau}^{(m)}_{\textrm{local}}, \bm{\tau}^{(m)}_{\textrm{offload}} \preceq T_{\textrm{b}}, \tag{21e}\label{20e}\nonumber \\ \quad
 \ {\rm{ }}~ &  0 \preceq \bm{\rho}^{(m)} \preceq 1, \tag{21f}\label{20f}\nonumber \\ \quad
 \ {\rm{ }}~ &   (\ref{fla_c8}),(\ref{fla_p}),\nonumber
\end{flalign}
where $\bm{\tau_{\textrm{local}}}^{(m)}$ = [$\tau_{\textrm{ipt}}^{(m)}$, $\tau_{\textrm{cpt}}^{(m)}$] and $\bm{\tau_{\textrm{offload}}}^{(m)}$ = [$\tau_{\textrm{ipt}}^{(m)}$, $\tau_{\textrm{u-f}}^{(m)}$, $\tau_{\textrm{fogcpt}}^{(m)}$, $\tau_{\textrm{f-u}}^{(m)}$] denote the time assignment vector associated with the two modes for MU $m$, respectively. $\bm{\rho}^{(m)}=[\rho_{\textrm{(local)}}^{(m)}, \rho_{\textrm{(offload)}}^{(m)}]$ denotes the power splitting vector associated with the two modes.
Constraint (\ref{20a}) means that the information transmission rate from the HAP to the MU \textcolor[rgb]{0.00,0.00,0.00}{$m$} should be no less than a predefined threshold $R_{\textrm{th}}$. Constraint (\ref{20b}) means that no matter which mode is selected, the total number of logic operations at the MU \textcolor[rgb]{0.00,0.00,0.00}{$m$} or at the FS must not be smaller than the minimal required operations of the task. Particularly, $\tau_{\textrm{cpt}}^{(m)}f_{\textrm{op}}$ represents the operations at the MU \textcolor[rgb]{0.00,0.00,0.00}{$m$} (where $f_{\textrm{op}}$ denotes the peak operations per second at the MU \textcolor[rgb]{0.00,0.00,0.00}{$m$ identically}) and $KR_{\textrm{u-f}}^{(m)}T_{\textrm{b}}$ means the \textcolor[rgb]{0.00,0.00,0.00}{number of} offloading operations, respectively.
Constraint (\ref{20c}) is similar to Constraint (\ref{20b}), which describes the computing capability constraints at the FS. Constraint (\ref{20d}) is used to guarantee the calculated result to be completely fed back from the FS to the MU \textcolor[rgb]{0.00,0.00,0.00}{$m$}. Constraint (\ref{20e}) implies that the sum of the assigned time intervals should not \textcolor[rgb]{0.00,0.00,0.00}{be} larger than $T_{\textrm{b}}$. Constraint (\ref{20f}) indicates that the transmit power at the MU \textcolor[rgb]{0.00,0.00,0.00}{$m$} cannot exceed the maximal available transmit power at the MU \textcolor[rgb]{0.00,0.00,0.00}{$m$}.

For problem $\textbf{P}$, $\Psi$ is a matrix with discrete binary elements and $\theta^{(m)}$ is a discrete binary variable \textcolor[rgb]{0.00,0.00,0.00}{of MU $m$}, which make problem $\textbf{P}$ difficult to solve. Therefore, we deal with it by using the following method.

\textbf{(A1)}: For a given $\Psi$, we find the joint optimal $\{\bm{\tau_{\textrm{local}}}^{(m)*}, \rho^{(m)*}_{\textrm{(local)}}\}$ for the local computing mode and the joint optimal $\{\bm{\tau_{\textrm{offload}}}^{(m)*}, \rho^{(m)*}_{\textrm{(offload)}}, P^{(m)*}_{\textrm{u-f}}\}$ for the fog offloading mode. Based on the obtained results, the optimal $\theta^{(m)*}$ is determined according to the minimal required \textcolor[rgb]{0.00,0.00,0.00}{energy}. As the closed-form solutions are derived, the optimal joint mode selection, time assignment and  power allocation associated with MU $m$ is achieved with low \textcolor[rgb]{0.00,0.00,0.00}{computational} complexity.

\textbf{(A2)}: Based on the obtained optimal results $\{\bm{\tau_{\textrm{local}}}^{(m)*}, \bm{\tau_{\textrm{offload}}}^{(m)*}, \bm{\rho}^{(m)*}, P_{\textrm{u-f}}^{(m)*}, \theta^{(m)*}\}$, a user scheduling scheme is presented to find the approximate optimal $\Psi^*$, which is also with low \textcolor[rgb]{0.00,0.00,0.00}{computational} complexity.

The detailed information of our presented solving approach is described in Section III.$B$. To achieve notation simplicity, we omit the superscript ``$(m)$" of the notations in the sequel.

\subsection{Solving Approach}

\textbf{\emph{I. Optimal $\{\bm{\tau^*_{\textrm{local}}}, \bm{\tau^*_{\textrm{offload}}}, \bm{\rho^*}, P^*_{\textrm{u-f}}, \theta^*\}$ for a given $\Psi$}}

\textit{\textbf{Step 1}: Optimization of the two modes.} With a fixed $\Psi$, for each MU $m$, Problem $\textbf{P}$ is simplified to be the following Problem $\textbf{P}_{0}$.
\begin{flalign} \label{fla_comp1}
\textbf{P}_{0}: {\kern 2pt} &\!\!\!\!\!\!\mathop {\min }\limits_{\substack{\bm{\tau_{\textrm{local}}}, \bm{\tau_{\textrm{offload}}}, \bm{\rho}\\ P_{\textrm{u-f}}, \theta}} \,
\!\!\!E_{\textrm{id}} + \theta E_{\textrm{cpt}} + ( 1 - \theta) E_{\textrm{u-f}} - E_{\textrm{eh}}-E_{\textrm{s}}, \\
&\,\,\,\,{\rm{s}}{\rm{.t}}{\rm{. }}~ (\ref{20a})-(\ref{20f}), (\ref{fla_c8}),(\ref{fla_p}),\nonumber
\end{flalign}
which aims to find the joint optimal mode selection, time assignment and  power allocation associated with MU $m$ for \textcolor[rgb]{0.00,0.00,0.00}{$m \in \{1,..., M\}$}. It is noticed that Problem $\textbf{P}_{0}$ is with discrete variable $\theta$, which is still not easy to tackle. In order to efficiently solve it, we decompose it into two subproblem $\textbf{P}_{1}$ and $\textbf{P}_{2}$ according to the working modes, i.e., the local computing mode and the fog offloading mode by fixing $\theta$. Then, we solve them separately and get some closed-form solutions \textcolor[rgb]{0.00,0.00,0.00}{related} to the two modes.

\subsubsection{Local Computing Mode}\label{local}
By setting $\theta$ = 1, MU $m$ works in the local computing mode, so Problem $\textbf{P}_{0}$ is simplified to be the following Problem $\textbf{P}_{1}$, i.e.,
\begin{flalign} \label{ref_comp1}
\textbf{P}_{1}: {\kern 2pt}\mathop {\min }\limits_{\tau_{\textrm{ipt}}, \tau_{\textrm{cpt}},
\rho_{(\textrm{local})} } \,
 & E_{\textrm{id}} + E_{\textrm{cpt}} - E_{\textrm{eh}}-E_{\textrm{s}}, \\
 {\rm{s}}{\rm{.t}}{\rm{. }}~ &   R_{\textrm{AP-u}}\ge R_{\textrm{th}}, \tag{23a}\label{22a}\nonumber \\ \quad
 \ {\rm{ }}~ &   \tau_{\textrm{cpt}}f_{\textrm{op}} \ge KR_{\textrm{AP-u}}T_{\textrm{b}}, \tag{23b}\label{22b} \nonumber \\ \quad
 \ {\rm{ }}~ &   \tau_{\textrm{ipt}} + \tau_{\textrm{cpt}} \leq T_{\textrm{b}}, \tau_{\textrm{ipt}}, \tau_{\textrm{cpt}}\in(0,T_{\textrm{b}}), \tag{23c}\label{22c}\nonumber \\ \quad
 \ {\rm{ }}~ &    \rho_{(\textrm{local})}\in(0,1). \tag{23d}\label{22d}\nonumber
\end{flalign}
By expanding the expressions, Problem $\textbf{P}_{1}$ is further equivalently rewritten as
\textcolor[rgb]{0.00,0.00,0.00}{\begin{small}
\begin{flalign} \label{ref_comp2}
\textbf{P}_{1\!-\!\textrm{A}}: {\kern 2pt}\!\!\!\mathop {\min }\limits_{\substack{\tau_{\textrm{ipt}}, \tau_{\textrm{cpt}},\\ \rho_{(\textrm{local})} }} \,
 & \!\!B\tau_{\textrm{ipt}}\log\!\left(\!1+\frac{\rho_{(\textrm{local})} P_\textrm{AP}\!\left|\bm{h_{\textrm{AP-u}}}^H\!\bm{w}\right|^2}{\sigma_{\textrm{n}}^2}\!\right) \!\left(KF_{0}\alpha M_{\textrm{c}}N_{0} \right. \nonumber \\
 &\left.\!\!\!\!\!\!\!\!\!\!\!\!\!\!\ln\!2+ \xi\right)\!- \eta\left(1\!-\rho_{(\textrm{local})} \right)P_\textrm{AP}\left|\bm{h_{\textrm{AP-u}}}^H \bm{w}\right|^2\tau_{\textrm{ipt}}-\iota-E_{\textrm{s}}, \\  \quad
 {\rm{s}}{\rm{.t}}{\rm{. }}~ &  \!\!B\frac{\tau_{\textrm{ipt}}}{T_{\textrm{b}}}\log\left(1+\frac{\rho_{(\textrm{local})} P_\textrm{AP}\left|\bm{h_{\textrm{AP-u}}}^H \bm{w}\right|^2}{\sigma_{\textrm{n}}^2}\right)\ge R_{\textrm{th}}, \tag{24a}\label{23a}\nonumber\\ \quad
 \ {\rm{ }}~ & \!\!\!\tau_{\textrm{cpt}}f_{\textrm{op}}\ge KB\tau_{\textrm{ipt}}\log\left(\!1+\frac{\rho_{(\textrm{local})} P_\textrm{AP}\left|\bm{h_{\textrm{AP-u}}}^H \bm{w}\right|^2}{\sigma_{\textrm{n}}^2}\!\right), \tag{24b}\label{23b}\nonumber\\ \quad
 \ {\rm{ }}~ &   \!\!\tau_{\textrm{ipt}} + \tau_{\textrm{cpt}} \leq T_{\textrm{b}}, \tau_{\textrm{ipt}}, \tau_{\textrm{cpt}}\in(0,T_{\textrm{b}}), \tag{24c}\label{23c}\nonumber \\ \quad
 \ {\rm{ }}~ &    \!\!\rho_{(\textrm{local})}\in(0,1). \tag{24d}\label{23d}\nonumber
\end{flalign}
\end{small}}
It is observed that variables $\tau_{\textrm{ipt}}$ and $\rho_{(\textrm{local})}$ are coupled together, so that Problem $\textbf{P}_{1-\textrm{A}}$ is non-convex and cannot be directly solved by using some standard convex optimization solution methods. Hence, by introducing a new slack variable $\varphi  = \rho_{(\textrm{local})} \tau_{\textrm{ipt}}$, we have that $\rho_{(\textrm{local})} = \frac{\varphi}{\tau_{\textrm{ipt}}}$.

Denote
$\textrm{C}_{\textrm{1}}=KF_{0}\alpha M_{\textrm{c}}N_{0}\ln\!2+\xi,$
$\textrm{C}_{\textrm{2}}=\eta P_\textrm{AP}\left|\bm{h_{\textrm{AP-u}}}^H\bm{w}\right|^2,$
and $f(\tau_{\textrm{ipt}},\varphi)=B\tau_{\textrm{ipt}}\log\left(1+\frac{\varphi P_\textrm{AP}\left|\bm{h_{\textrm{AP-u}}}^H \bm{w}\right|^2}{\tau_{\textrm{ipt}}\sigma_{\textrm{n}}^2}\right).$
So, $F(\tau_{\textrm{ipt}},\varphi)$ $=B\tau_{\textrm{ipt}}\log\left(1+\frac{\varphi P_\textrm{AP}\left|\bm{h_{\textrm{AP-u}}}^H \bm{w}\right|^2}{\tau_{\textrm{ipt}}\sigma_{\textrm{n}}^2}\right)\left(KF_{0}\alpha M_{\textrm{c}}N_{0}\ln\!2+\xi\right)-\eta$ $\left(\tau_{\textrm{ipt}}-\varphi \right)P_\textrm{AP}\left|\bm{h_{\textrm{AP-u}}}^H \bm{w}\right|^2\!=\! \textrm{C}_{\textrm{1}}f(\tau_{\textrm{ipt}},\varphi) -\textrm{C}_{\textrm{2}}(\tau_{\textrm{ipt}}-\varphi)$. Therefore, Problem $\textbf{P}_{1-\textrm{A}}$ can be rewritten to be
\begin{flalign} \label{ref_p1c}
\textbf{P}_{1-\textrm{B}}: {\kern 2pt}\mathop {\min }\limits_{\tau_{\textrm{ipt}},\varphi} \,
 &F(\tau_{\textrm{ipt}},\varphi)\!= \textrm{C}_{\textrm{1}}f(\tau_{\textrm{ipt}},\varphi)\!-\textrm{C}_{\textrm{2}}(\tau_{\textrm{ipt}}-\varphi)\!-\iota-E_{\textrm{s}} \\  \quad
 {\rm{s}}{\rm{.t}}{\rm{. }}~ &  f(\tau_{\textrm{ipt}},\varphi)\ge R_{\textrm{th}}T_{\textrm{b}}, \tag{25a}\label{24a} \\ \quad
 \ {\rm{ }}~ &   f(\tau_{\textrm{ipt}},\varphi) \leq \frac{\left(T_{\textrm{b}} - \tau_{\textrm{ipt}}\right)f_{\textrm{op}}}{K}, \tag{25b}\label{24b} \\ \quad
\ {\rm{ }}~ &  \tau_{\textrm{ipt}}\in(0,T_{\textrm{b}}),~~\varphi \in(0,T_{\textrm{b}}). \tag{25c}\label{24c}\nonumber
\end{flalign}
It can be seen that the first term of $F(\tau_{\textrm{ipt}},\varphi)$, i.e., $\textrm{C}_{\textrm{1}}f(\tau_{\textrm{ipt}},\varphi)$, is with the form of $y\log\left(1 +\frac{x}{y}\right)$, which is concave w.r.t. $x$ and $y$\cite{ref_convex}. \textcolor[rgb]{0.00,0.00,0.00}{That is,} $\textrm{C}_{\textrm{1}}f(\tau_{\textrm{ipt}},\varphi)$ is concave w.r.t $\tau_{\textrm{ipt}}$ and $\varphi$. Moreover, the second term of $F(\tau_{\textrm{ipt}},\varphi)$, i.e., $-\textrm{C}_{\textrm{2}}(\tau_{\textrm{ipt}}-\varphi)$, is linear with $\tau_{\textrm{ipt}}$ and $\varphi$. \textcolor[rgb]{0.00,0.00,0.00}{To} Problem $\textbf{P}_{1-\textrm{B}}$, because that the objective function is the minimum of a concave function w.r.t $\tau_{\textrm{ipt}}$ and $\varphi$, and constraint (\ref{24b}) is non-convex,  $\textbf{P}_{1-\textrm{B}}$ is still non-convex problem. Hence, we analyze and obtain some theoretical results as follows.

\begin{figure}[htb]
\centering
\includegraphics[width=0.44\textwidth]{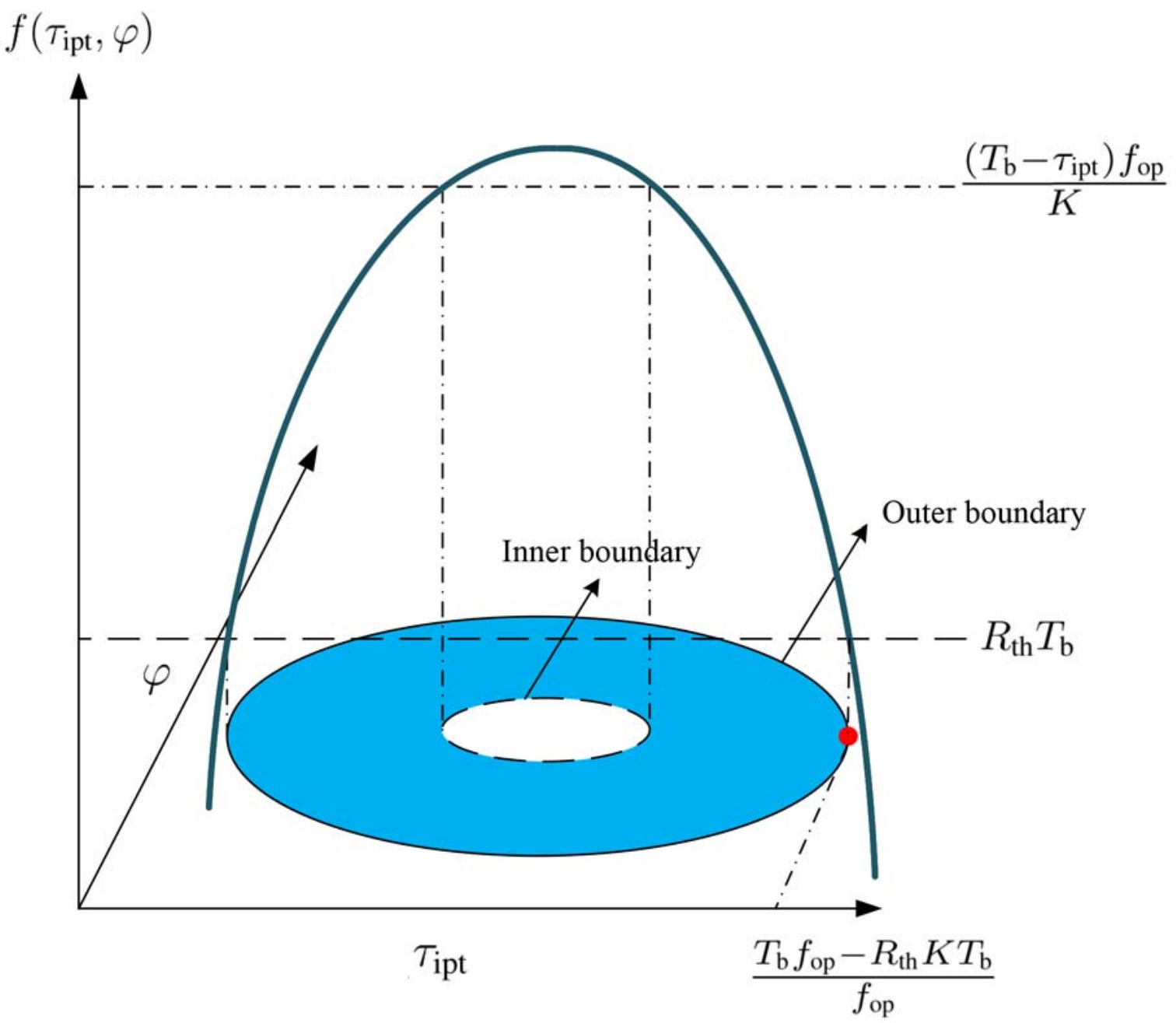}
\caption{Illustration of $\textbf{P}_{1-\textrm{B}}$}
\label{fig_s1}\vspace{-0.1 in}
\end{figure}

\textbf{  Proposition 1:}
\emph{Problem $\textbf{P}_{1-\textrm{B}}$ has feasible solutions only when} $R_{\textrm{th}} \leq \frac{\left(T_{\textrm{b}} - \tau_{\textrm{ipt}}\right)f_{\textrm{op}}}{KT_{\textrm{b}}}$.

\begin{IEEEproof}
From Constrain (\ref{24a}) and (\ref{24b}), one can see that $R_{\textrm{th}}T_{\textrm{b}} \leq f(\tau_{\textrm{ipt}},\varphi) \leq \frac{\left(T_{\textrm{b}} - \tau_{\textrm{ipt}}\right)f_{\textrm{op}}}{K}$, i.e., $R_{\textrm{th}} \leq \frac{\left(T_{\textrm{b}} - \tau_{\textrm{ipt}}\right)f_{\textrm{op}}}{KT_{\textrm{b}}}$. That is, when $R_{\textrm{th}} \leq \frac{\left(T_{\textrm{b}} - \tau_{\textrm{ipt}}\right)f_{\textrm{op}}}{KT_{\textrm{b}}}$, the intersection set of the two constraints is not empty, which is illustrated by Figure~\ref{fig_s1}. \textcolor[rgb]{0.00,0.00,0.00}{Therefore}, Proposition 1 is proved.
\end{IEEEproof}

Following Proposition 1, we obtain the following corollary.

\textcolor[rgb]{0.00,0.00,0.00}{\textbf{  Corollary 1:}
\emph{Problem $\textbf{P}_{1-\textrm{B}}$ has feasible \textcolor[rgb]{0.00,0.00,0.00}{solutions} only when}  $K \leq \frac{\left(T_{\textrm{b}} - \tau_{\textrm{ipt}}\right)f_{\textrm{op}}}{R_{\textrm{th}}T_{\textrm{b}}}$ \emph{for given} $R_{\textrm{th}}$ \emph{and} $f_{\textrm{op}}$, \emph{when} $f_{\textrm{op}} \geq \frac{R_{\textrm{th}}KT_{\textrm{b}}}{T_{\textrm{b}}-\tau_{\textrm{ipt}}}$ \emph{for given} $R_{\textrm{th}}$ \emph{and} $K$, \emph{and when} $\tau_{\textrm{ipt}} \leq \frac{T_{\textrm{b}}f_{\textrm{op}} - R_{\textrm{th}}KT_{\textrm{b}}}{f_{\textrm{op}}}$ \emph{for given} $R_{\textrm{th}}$, $f_{\textrm{op}}$ \emph{and} $K$.}

%

\textbf{  Lemma 1:}
\emph{The optimal $\tau_{\textrm{ipt}}$ for local computing is} $\tau^{*}_{\textrm{ipt(local)}} = \frac{T_{\textrm{b}}f_{\textrm{op}} - R_{\textrm{th}}KT_{\textrm{b}}}{f_{\textrm{op}}}$.

\begin{IEEEproof}
See Appendix A.
\end{IEEEproof}

\textbf{ Theorem 1:}
\emph{The optimal $\rho_{\textrm{(local)}}$ and $\tau_{\textrm{cpt}}$ for local computing mode are} $\rho_{\textrm{(local)}} ^{*} = \frac{\sigma_{\textrm{n}}^2}{P_\textrm{AP}\left|\bm{h_{\textrm{AP-u}}}^H \bm{w}\right|^2}\left(2^{\frac{R_{\textrm{th}}f_{\textrm{op}}}{B\left(f_{\textrm{op}} - KR_{\textrm{th}}\right)}} - 1\right)$ \emph{and} $\tau_{\textrm{cpt}}^{*}= \frac{KR_{\textrm{th}}T_{\textrm{b}}}{f_{\textrm{op}}}$, \emph{respectively.}
\begin{IEEEproof}
See Appendix B.
\end{IEEEproof}

\subsubsection{Fog Offloading Mode}\label{fog}
By setting $\theta$ = 0, \textcolor[rgb]{0.00,0.00,0.00}{MU} works in the fog offloading mode. In this case, Problem $\textbf{P}_{0}$ is simplified to be the following Problem $\textbf{P}_{2}$, i.e.,

\begin{flalign} \label{ref_offloading}
\textbf{P}_{2}: {\kern 2pt}&\!\!\!\!\!\!\mathop {\min }\limits_{\bm{\tau_{\textrm{offload}}}, \rho_{(\textrm{offload})}, P_{\textrm{u-f}}} \,
\!\!\!\!\!\!E_{\textrm{id}} + E_{\textrm{u-f}} - E_{\textrm{eh}}-E_{\textrm{s}}, \\
 {\rm{s}}{\rm{.t}}{\rm{. }}~\, &   R_{\textrm{u-f}}T_{\textrm{b}} \ge R_{\textrm{AP-u}}T_{\textrm{b}}, \tag{26a}\label{25a}\nonumber \\ \quad
\ {\rm{ }}~ &   \tau_{\textrm{ipt}} + \tau_{\textrm{u-f}} + \tau_{\textrm{fogcpt}} + \tau_{\textrm{f-u}} \leq T_{\textrm{b}}, 0 \preceq \bm{\tau_{\textrm{offload}}} \preceq T_{\textrm{b}}, \tag{26b}\label{25b}\nonumber \\ \quad
 \ {\rm{ }}~ &  \rho_{(\textrm{offload})}\in(0,1), \tag{26c}\label{25c}\nonumber \\ \quad
 \ {\rm{ }}~ &  (\ref{20a}), (\ref{20c}), (\ref{20d}), (\ref{20f}),\nonumber
\end{flalign}
in which Constraint (\ref{25a}) describes that the number of the bits transmitted from MU to FS should be no less than the bits transmitted to the MU. Similar to the process of Problem $\textbf{P}_{1}$. By introducing two new slack variable $\varpi = \rho_{(\textrm{offload})}\tau_{\textrm{ipt}}$, and $\lambda_{\textrm{u-f}} = \tau_{\textrm{u-f}}P_{\textrm{u-f}}$, Problem $\textbf{P}_{2}$ is transformed to be Problem $\textbf{P}_{2-\textrm{A}}$ in (\ref{ref_off511}).
\begin{figure*}[!ht]
\hrulefill
\textcolor[rgb]{0.00,0.00,0.00}{\begin{small}
\begin{flalign}\label{ref_off511}
\textbf{P}_{2-\textrm{A}}: {\kern 2pt}\!\!\!\!\!\!\!\!\!\mathop {\min }\limits_{\bm{\tau_{\textrm{offload}}}, \varpi, \lambda_{\textrm{u-f}}} \,
 \!\!\!\!\!\! &\xi B\tau_{\textrm{ipt}}\log\left(1+\frac{\varpi }{\tau_{\textrm{ipt}}}\frac{P_\textrm{AP}\left|\bm{h_{\textrm{AP-u}}}^H \bm{w}\right|^2}{\sigma_{\textrm{n}}^2}\right) + \lambda_{\textrm{u-f}}+ \varpi\eta 
 P_\textrm{AP}\left|\bm{h_{\textrm{AP-u}}}^H \bm{w}\right|^2 - \tau_{\textrm{ipt}}\eta P_\textrm{AP}\left|\bm{h_{\textrm{AP-u}}}^H \bm{w}\right|^2-\iota-E_{\textrm{s}}, \\  \quad
 {\rm{s}}{\rm{.t}}{\rm{. }}~ &  B\tau_{\textrm{ipt}}\log\left(1+\frac{\varpi }{\tau_{\textrm{ipt}}}\frac{P_\textrm{AP}\left|\bm{h_{\textrm{AP-u}}}^H \bm{w}\right|^2}{\sigma_{\textrm{n}}^2}\right)\ge R_{\textrm{th}}T_{\textrm{b}},  \tag{27a}\label{27a}\nonumber\\ \quad
 \ {\rm{ }}~ &  B\tau_{\textrm{u-f}}\log\left(1+\frac{\lambda_{\textrm{u-f}}}{\tau_{\textrm{u-f}}}\frac{\left|h_{\textrm{u-f}}\right|^2}{\sigma_{\textrm{s}}^2}\right)\ge 
 B\tau_{\textrm{ipt}}\log\left(1 +\frac{\varpi} {\tau_{\textrm{ipt}}}\frac{P_\textrm{AP}\left|\bm{h_{\textrm{AP-u}}}^H \bm{w}\right|^2}{\sigma_{\textrm{n}}^2}\right), \tag{27b}\label{27b}\nonumber\\ \quad
 \ {\rm{ }}~ &   \tau_{\textrm{fogcpt}}f_{\textrm{fogop}} \ge KB\tau_{\textrm{u-f}}\log\big(1+\frac{\lambda_{\textrm{u-f}}}{\tau_{\textrm{u-f}}}\frac{\left|h_{\textrm{u-f}}\right|^2}{\sigma_{\textrm{s}}^2}\!\big), \tag{27c}\label{27c}\nonumber \\ \quad
 \ {\rm{ }}~ &  B\tau_{\textrm{f-u}}\log\left(1+\frac{\left|h_{\textrm{f-u}}\right|^2P_{\textrm{f-u}}}{\sigma_{\textrm{f}}^2}\right) \geq 
 \beta B\tau_{\textrm{u-f}}\log\big(1+\frac{\lambda_{\textrm{u-f}}}{\tau_{\textrm{u-f}}}\frac{\left|h_{\textrm{u-f}}\right|^2}{\sigma_{\textrm{s}}^2}\big) , \tag{27d}\label{27d}\nonumber\\ \quad
 \ {\rm{ }}~ &   \frac{\lambda_{\textrm{u-f}}}{\tau_{\textrm{u-f}}}\leq P^{\textrm{(max)}}_{\textrm{u-f}}, \tag{27e}\label{27e}\nonumber \\ \quad
 \ {\rm{ }}~ &(\ref{25b}), (\ref{25c}),\nonumber
\end{flalign}
\end{small}}
\hrulefill
\end{figure*}

The objective of Problem $\textbf{P}_{2-\textrm{A}}$ function is a minimization of a concave function w.r.t $\tau_{\textrm{ipt}}$, $\varpi$, and $\lambda_{\textrm{u-f}}$. It is also difficult to solve due to the non-convexity of constraint sets (\ref{27a}) and (\ref{27b}).

\textbf{  Proposition 2:}
\emph{Problem $\textbf{P}_{2-\textrm{A}}$ has feasible \textcolor[rgb]{0.00,0.00,0.00}{solutions }only when} $R_{\textrm{th}} \leq B\frac{\tau_{\textrm{u-f}}}{T_{\textrm{b}}}\log\left(1+\frac{P^{\textrm{(max)}}_{\textrm{u-f}}\left|h_{\textrm{u-f}}\right|^2}{\sigma_{\textrm{s}}^2}\right)$.

\begin{IEEEproof}
From Constraint (\ref{27a}), (\ref{27b}) and (\ref{27e}) of Problem $\textbf{P}_{2-\textrm{A}}$, one can see that when $R_{\textrm{th}} \leq B\frac{\tau_{\textrm{u-f}}}{T_{\textrm{b}}}\log\left(1+\frac{P^{\textrm{(max)}}_{\textrm{u-f}}\left|h_{\textrm{u-f}}\right|^2}{\sigma_{\textrm{s}}^2}\right)$, the intersection set of the two constraints is not empty. Hence, Proposition 2 is proved.
\end{IEEEproof}

Following Proposition 2, we obtain the following corollary.

\textcolor[rgb]{0.00,0.00,0.00}{\textbf{  Corollary 2:}
\emph{Problem $\textbf{P}_{2-\textrm{A}}$ has feasible \textcolor[rgb]{0.00,0.00,0.00}{solutions} only when}  $P^{\textrm{(max)}}_{\textrm{u-f}}\!\geq\! \frac{\sigma_{\textrm{s}}^2}{\left|h_{\textrm{u-f}}\right|^2}\left(2^{\frac{R_{\textrm{th}}T_{\textrm{b}}}{B\tau_{\textrm{u-f}}}} - 1\right)$ \emph{for given} $R_{\textrm{th}}$ \emph{and} $h_{\textrm{u-f}}$, \emph{and when} $\tau_{\textrm{u-f}}\!\geq\! \frac{R_{\textrm{th}}T_{\textrm{b}}}{\log\left(1 +\frac{P^{\textrm{(max)}}_{\textrm{u-f}}\left|h_{\textrm{u-f}}\right|^2}{\sigma_{\textrm{s}}^2}\right)}$ \emph{for given} $R_{\textrm{th}}$, $h_{\textrm{u-f}}$ \emph{and} $P^{\textrm{(max)}}_{\textrm{u-f}}$.}


\textbf{  Proposition 3:}
\emph{The optimal $\tau_{\textrm{f-u}}$ and $\tau_{\textrm{fogcpt}}$ for fog offloading mode are} $\tau_{\textrm{f-u}}^{*} = \tfrac{\beta R_{\textrm{th}}T_{\textrm{b}}}{B\log\left(1+\tfrac{\left|h_{\textrm{f-u}}\right|^2P^{\textrm{(max)}}_{\textrm{f-u}}}{\sigma_{\textrm{f}}^2}\right)}$ \emph{and} $\tau_{\textrm{fogcpt}}^{*} = \frac{KR_{\textrm{th}}T_{\textrm{b}}}{f_{\textrm{fogop}}}$,
\emph{respectively.}

\begin{IEEEproof}
As is known, the energy required by transmitting and processing \textcolor[rgb]{0.00,0.00,0.00}{of data} is determined the number of bits, which means the increased bits cannot reduce the energy requirement. So, when  $R_{\textrm{AP-u}}$ and $R_{\textrm{u-f}}$ are equal to their threshold, $R_{\textrm{th}}$, and $R_{\textrm{f-u}}$, the energy requirement reaches its minimum value. On the other hand, according (\ref{14}) and (\ref{15}), one can see that the smaller $\tau_{\textrm{fogcpt}}$ and $\tau_{\textrm{f-u}}$ are, the larger  $\tau_{\textrm{ipt}}$ and $\tau_{\textrm{u-f}}$ are, which requires less energy for offloading energy or longer energy harvesting time. Therefore, in order to achieve the minimal energy requirement at the MU, $\tau_{\textrm{fogcpt}}$ and $\tau_{\textrm{f-u}}$ should be as small as possible. That is, the optimal $\tau_{\textrm{fogcpt}}$ and $\tau_{\textrm{f-u}}$ should be located \textcolor[rgb]{0.00,0.00,0.00}{close to} their lower boundaries. So, the optimal solution $\tau_{\textrm{fogcpt}}$ and $\tau_{\textrm{f-u}}$ can be given by $\tau_{\textrm{fogcpt}}^{*} = \frac{KR_{\textrm{th}}T_{\textrm{b}}}{f_{\textrm{fogop}}}$ and $\tau_{\textrm{f-u}}^{*} = \frac{\beta R_{\textrm{th}}T_{\textrm{b}}}{B\log\left(1+\frac{\left|h_{\textrm{f-u}}\right|^2P^{\textrm{(max)}}_{\textrm{f-u}}}{\sigma_{\textrm{f}}^2}\right)}$. Proposition 3 is therefore proved.
\end{IEEEproof}

With Proposition 3, Problem $\textbf{P}_{2-\textrm{A}}$ can be simplified and rewritten as $\textbf{P}_{2-\textrm{B}}$.
\textcolor[rgb]{0.00,0.00,0.00}{\begin{small}
\begin{flalign}\label{ref_off3}
\textbf{P}_{2-\textrm{B}}: {\kern 2pt}\!\!\!\!\!\!\!\!\!\!\!\!\mathop {\min }\limits_{\tau_{\textrm{ipt}}, \tau_{\textrm{u-f}}, \varpi, \lambda_{\textrm{u-f}}} \,
 \!\!\!\!\!& \xi B\tau_{\textrm{ipt}}\log\left(1+\frac{\varpi }{\tau_{\textrm{ipt}}}\frac{P_\textrm{AP}\left|\bm{h_{\textrm{AP-u}}}^H \bm{w}\right|^2}{\sigma_{\textrm{n}}^2}\right) + \varpi\eta P_\textrm{AP} \nonumber\\ \quad
 &\left|\bm{h_{\textrm{AP-u}}}^H \bm{w}\right|^2 - \tau_{\textrm{ipt}}\eta P_\textrm{AP}\left|\bm{h_{\textrm{AP-u}}}^H \bm{w}\right|^2+ \lambda_{\textrm{u-f}}-\iota-E_{\textrm{s}}, \\  \quad
 {\rm{s}}{\rm{.t}}{\rm{. }}~ & \tau_{\textrm{ipt}}\! +\!\tau_{\textrm{u-f}} \leq T_{\textrm{b}}\!-\! \frac{K\!R_{\textrm{th}}\!T_{\textrm{b}}}{f_{\textrm{fogop}}}\!-\!\frac{\beta R_{\textrm{th}}T_{\textrm{b}}}{B\log\!\left(1\! +\! \frac{\left|h_{\textrm{f-u}}\right|^2P^{\textrm{(max)}}_{\textrm{f-u}}}{\sigma_{\textrm{f}}^2}\!\right) }, \nonumber \\ \quad
 &\tau_{\textrm{ipt}}, \tau_{\textrm{u-f}}, \in(0,T_{\textrm{b}}), \tag{28a}\label{28a}\nonumber \\ \quad
 \ {\rm{ }}~ &(\ref{27a}), (\ref{27b}), (\ref{27e}), (\ref{25c}). \nonumber
\end{flalign}
\end{small}}

The first term of the objective function in Problem $\textbf{P}_{2-\textrm{B}}$ is concave w.r.t $\tau_{\textrm{ipt}}$ and $\varpi$ with the form of $y\log\left(1 +\frac{x}{y}\right)$. The second and the third terms are the linear function w.r.t $\varpi$ and $\tau_{\textrm{ipt}}$, respectively. \textcolor[rgb]{0.00,0.00,0.00}{The} fourth term is the linear function w.r.t $\lambda_{\textrm{u-f}}$. As a result, the objective function is a concave function w.r.t $\tau_{\textrm{ipt}}$, $\varpi$ and $\lambda_{\textrm{u-f}}$. However, because of the non-convexity of constraint sets (\ref{27a}) and (\ref{27b}), Problem $\textbf{P}_{2-\textrm{B}}$ also cannot be solved directly with \textcolor[rgb]{0.00,0.00,0.00}{the} standard convex optimization methods. Thus, we further analyze and deal with it as follows.

By using a similar proof of Lemma 1, it can be prove that the optimal solution to Problem $\textbf{P}_{2-\textrm{B}}$ is also located at the boundary of its constraint set. The outer boundary is obtained when the equalities of the inequality constraints (\ref{27a}), (\ref{27b}) and (\ref{28a}) hold, i.e., $B\tau_{\textrm{ipt}}\log\left(1+\frac{\varpi }{\tau_{\textrm{ipt}}}\frac{P_\textrm{AP}\left|\bm{h_{\textrm{AP-u}}}^H \bm{w}\right|^2}{\sigma_{\textrm{n}}^2}\right)= R_{\textrm{th}}T_{\textrm{b}}$, $B\tau_{\textrm{u-f}}\log\left(1+\frac{\left|h_{\textrm{u-f}}\right|^2P_{\textrm{u-f}}}{\sigma_{\textrm{s}}^2}\right) = R_{\textrm{th}}T_{\textrm{b}}$, and $\tau_{\textrm{ipt}}+\tau_{\textrm{u-f}}=\mathfrak{T_{\textrm{b}}}$, respectively, in which $\mathfrak{T_{\textrm{b}}} = T_{\textrm{b}}- \frac{KR_{\textrm{th}}T_{\textrm{b}}}{f_{\textrm{fogop}}} - \frac{\beta R_{\textrm{th}}T_{\textrm{b}}}{B\log\left(1 + \frac{\left|h_{\textrm{f-u}}\right|^2P^{\textrm{(max)}}_{\textrm{f-u}}}{\sigma_{\textrm{f}}^2}\right) }$. Therefore, each of $\varpi$, $\tau_{\textrm{u-f}}$, and $\lambda_{\textrm{u-f}}$ can be regarded as a function w.r.t $\tau_{\textrm{ipt}}$. So, the objective function can be transformed to be $\vartheta(\tau_{\textrm{ipt}})$ as shown in (\ref{ref_off4}).
\begin{small}
\begin{flalign}\label{ref_off4}
\vartheta(\tau_{\textrm{ipt}})=& \xi R_{\textrm{th}}T_{\textrm{b}} + \frac{(\mathfrak{T_{\textrm{b}}}-\tau_{\textrm{ipt}})\sigma_{\textrm{s}}^2}{\left|h_{\textrm{u-f}}\right|^2}\left(2^{\frac{R_{\textrm{th}}T_{\textrm{b}}}{B(\mathfrak{T_{\textrm{b}}}-\tau_{\textrm{ipt}})}} - 1\right)-\tau_{\textrm{ipt}}\eta \nonumber \\ \quad
& \!\!\!\!\!\!\left(P_\textrm{AP}\left|\bm{h_{\textrm{AP-u}}}^H \bm{w}\right|^2 - \sigma_{\textrm{n}}^2\left(2^{\frac{R_{\textrm{th}}T_{\textrm{b}}}{B\tau_{\textrm{ipt}}}} - 1\right) \right)-\iota-E_{\textrm{s}}.
\end{flalign}
\end{small}

\textbf{  Lemma 2:}
\emph{$\vartheta(\tau_{\textrm{ipt}})$ is convex w.r.t $\tau_{\textrm{ipt}}$. }

\begin{IEEEproof}
See Appendix C.
\end{IEEEproof}

\textbf{ Theorem 2:}
\emph{Let}
\textcolor[rgb]{0.00,0.00,0.00}{\begin{small}
\begin{equation} \label{eq:2}
\left\{ \begin{aligned}
&\alpha=\mathop {\textrm{arg}}\limits_{\tau_{\textrm{ipt}}\in(0, T_{\textrm{b}})}\Bigg\{\tau_{\textrm{ipt}}\mid\frac{\sigma_{\textrm{s}}^2}{\left|h_{\textrm{u-f}}\right|^2} \left(2^{\frac{R_{\textrm{th}}T_{\textrm{b}}}{B\left(\mathfrak{T_{\textrm{b}}} -\tau_{\textrm{ipt}}\right)}}\left(\textrm{ln}2\,\tfrac{R_{\textrm{th}}T_{\textrm{b}}}{B\left(\mathfrak{T_{\textrm{b}}} -\tau_{\textrm{ipt}}\right)}\right.\right.  \\ \quad
&\,\,\,\,\,\,\,\,\,\,\,\,\left.\left.-1\right)+1\right)+\eta \sigma_{\textrm{n}}^2\left(2^{\tfrac{R_{\textrm{th}}T_{\textrm{b}}}{B\tau_{\textrm{ipt}}}}\left(1- \textrm{ln}2\frac{R_{\textrm{th}}T_{\textrm{b}}}{B\tau_{\textrm{ipt}}}\right)-1\right)\\ \quad
&\,\,\,\,\,\,\,\,\,\,\,\,\,-\eta P_\textrm{AP}\left|\bm{h_{\textrm{AP-u}}}^H \bm{w}\right|^2-\iota-E_{\textrm{s}}=0\Bigg\},\\
&\gamma = \frac{\sigma_{\textrm{n}}^2}{P_\textrm{AP}\left|\bm{h_{\textrm{AP-u}}}^H \bm{w}\right|^2}\big(2^{\frac{R_{\textrm{th}}T_{\textrm{b}}}{B\alpha}} - 1\big),\\
&\delta = \mathfrak{T_{\textrm{b}}}-\alpha,\,\,\varrho = \frac{\sigma_{\textrm{s}}^2}{\left|h_{\textrm{u-f}}\right|^2}\big(2^{\frac{R_{\textrm{th}}T_{\textrm{b}}}{B(\mathfrak{T_{\textrm{b}}}-\alpha)}} - 1\big).
\end{aligned} \right.
\end{equation}
\end{small}}
\emph{If}
$\varrho \leq P_{\textrm{u-f}}^{\textrm{(max)}}$, \emph{the optimal \textcolor[rgb]{0.00,0.00,0.00}{solutions are} } $\tau_{\textrm{ipt}}^{*}=\alpha, \rho_{(\textrm{offload})}^{*}=\gamma, \tau_{\textrm{u-f}}^{*}=\delta, P_{\textrm{u-f}}^{*}=\varrho$; \emph{\textcolor[rgb]{0.00,0.00,0.00}{otherwise},}  \emph{the optimal solution is}
\textcolor[rgb]{0.00,0.00,0.00}{\begin{small}
\begin{equation} \label{eq:3}
\left\{ \begin{aligned}
                  &\tau_{\textrm{ipt}}^{*} =  \mathfrak{T_{\textrm{b}}}-\frac{R_{\textrm{th}}T_{\textrm{b}}}{B\log\big(1 +\frac{\left|h_{\textrm{u-f}}\right|^2P^{\textrm{(max)}}_{\textrm{u-f}}}{\sigma_{\textrm{s}}^2}\big)},\\
                  &\rho_{(\textrm{offload})}^{*} = \frac{\sigma_{\textrm{n}}^2}{P_\textrm{AP}\left|\bm{h_{\textrm{AP-u}}}^H \bm{w}\right|^2}\left(2^{\frac{R_{\textrm{th}}T_{\textrm{b}}}{B\tau_{\textrm{ipt}}^{*}}} - 1\right),\\
                  &\tau_{\textrm{u-f}}^{*} = \frac{R_{\textrm{th}}T_{\textrm{b}}}{B\log\big(1+\frac{\left|h_{\textrm{u-f}}\right|^2P^{\textrm{(max)}}_{\textrm{u-f}}}{\sigma_{\textrm{s}}^2}\big)},\\
                  &P_{\textrm{u-f}}^{*} = P^{\textrm{(max)}}_{\textrm{u-f}}.
                          \end{aligned} \right.
                          \end{equation}
                          \end{small}
}
\begin{IEEEproof}
See Appendix D.
\end{IEEEproof}

\textit{\textbf{Step 2}: Optimization of $\theta$}

With the closed-form or semi-closed-form solutions to the two modes derived in \textcolor[rgb]{0.00,0.00,0.00}{\textbf{Step 1}}, \textbf{Step 2} is able to calculate the minimal energy requirement (i.e., $E_{\textrm{u(local)}}^{*}$ and $E_{\textrm{u(offload)}}^{*}$)  with very low computational complexity. Therefore, for \textcolor[rgb]{0.00,0.00,0.00}{each MU}, the optimal mode selection can be determined by
\begin{equation}
\theta^{*} =\left\{
             \begin{array}{lr}
             1,  \,\,\,\,\textrm{if}\,\,E_{\textrm{u(local)}}^{*}\leq E_{\textrm{u(offload)}}^{*} \\
             0, \,\,\,\,\textrm{otherwise}.
             \end{array}
\right.
\end{equation}
Therefore, \textcolor[rgb]{0.00,0.00,0.00}{the minimal energy requirement of each MU can be given by (\ref{eqz22})}, \textcolor[rgb]{0.00,0.00,0.00}{which is a piecewise function depended on the mode selection and in which } $\varsigma=\frac{\sigma_{\textrm{s}}^2}{\left|h_{\textrm{u-f}}\right|^2}\left(2^{\frac{R_{\textrm{th}}T_{\textrm{b}}}{B(\mathfrak{T_{\textrm{b}}} -\tau_{\textrm{ipt}}^{*})}} - 1\right)$.
\begin{figure*}[b]
\hrulefill
\textcolor[rgb]{0.00,0.00,0.00}{\begin{small}
\begin{flalign}
\label{eqz22}
&E_{\textrm{u}}^*=\mathop {\min }
  \{E_{\textrm{u(local)}}^{*}, E_{\textrm{u(offload)}}^{*}\}\nonumber\\ \quad
&=\left\{
             \begin{array}{lr}
             \left(KF_{0}\alpha M_{\textrm{c}}N_{0}\ln\!2 + \xi\right)R_{\textrm{th}}T_{\textrm{b}} - \eta P_{\textrm{AP}}\left|\bm{h_{\textrm{AP-u}}}^H \bm{w}\right|^2\left(T_{\textrm{b}} -  \frac{KR_{\textrm{th}}T_{\textrm{b}}}{f_{\textrm{op}}}\right)\left(1-\frac{\sigma_{\textrm{n}}^2}{P_\textrm{AP}\left|\bm{h_{\textrm{AP-u}}}^H \bm{w}\right|^2}\left(2^{\frac{R_{\textrm{th}}f_{\textrm{op}}}{B\left(f_{\textrm{op}} - KR_{\textrm{th}}\right)}} - 1\right)\right)-\iota-E_{\textrm{s}}, \\
             {\kern 290pt} {\rm if}  \, E_{\textrm{u(local)}}^{*}\leq E_{\textrm{u(offload)}}^{*}, \\
             \xi R_{\textrm{th}}T_{\textrm{b}} +\frac{\sigma_{\textrm{s}}^2}{\left|h_{\textrm{u-f}}\right|^2} (\mathfrak{T_{\textrm{b}}}-\tau_{\textrm{ipt}}^{*}) \left(2^{\frac{R_{\textrm{th}}T_{\textrm{b}}}{B(\mathfrak{T_{\textrm{b}}} -\tau_{\textrm{ipt}}^{*})}} - 1\right)+\eta \sigma_{\textrm{n}}^2\left(2^{\frac{R_{\textrm{th}}T_{\textrm{b}}}{B\tau_{\textrm{ipt}}^{*}}} - 1\right)-\tau_{\textrm{ipt}}^{*}\eta P_\textrm{AP}\left|\bm{h_{\textrm{AP-u}}}^H \bm{w}\right|^2-\iota-E_{\textrm{s}}, \\
             {\kern 290pt} {\rm if}  \, E_{\textrm{u(local)}}^{*}> E_{\textrm{u(offload)}}^{*}\,\, {\rm and} \,\, P_\textrm{{u-f}}^{\textrm{(max)}}\geq \varsigma, \\
             \left(\frac{P^{\textrm{(max)}}_{\textrm{u-f}}}{B\log\left(1 +\frac{\left|h_{\textrm{u-f}}\right|^2P^{\textrm{(max)}}_{\textrm{u-f}}}{\sigma_{\textrm{s}}^2}\right)} + \xi\right)R_{\textrm{th}}T_{\textrm{b}} - \eta P_{\textrm{AP}}\left|\bm{h_{\textrm{AP-u}}}^H \bm{w}\right|^2\left(1-\frac{\sigma_{\textrm{n}}^2}{P_\textrm{AP}\left|\bm{h_{\textrm{AP-u}}}^H \bm{w}\right|^2}\left(2^{\frac{R_{\textrm{th}}T_{\textrm{b}}}{B\tau_{\textrm{ipt}}^{*}}} - 1\right)\right)\tau_{\textrm{ipt}}^{*}-\iota-E_{\textrm{s}}, \\
             {\kern 290pt} {\rm if}  \, E_{\textrm{u(offload)}}^{*}> E_{\textrm{u(local)}}^{*}\,\,  {\rm and} \,\,P_\textrm{{u-f}}^{\textrm{(max)}}< \varsigma. \\
             \end{array}
\right.
\end{flalign}
\end{small}}
\end{figure*}

\textbf{\emph{II. Optimization of $\Psi$}}

With the obtained optimal result, i.e., $\{\bm{\tau^{*}_{\textrm{local}}}, \bm{\tau^{*}_{\textrm{offload}}}, \bm{\rho^{*}}, P^{*}_{\textrm{u-f}}, \theta^{^*}\}$, we present a \textcolor[rgb]{0.00,0.00,0.00}{scheme} to optimize $\Psi$. Before that, we analyze the computational complexity to find the global optimal $\Psi^*$ via combinatorial optimization scheme. To determine the optimal $\Psi^*$, it is equal to solve the problem of assigning $M$ tasks to $M$ different workers, which is with the complexity of $\mathcal{O}(M!)$. It is too complex, especially for a relatively large \textcolor[rgb]{0.00,0.00,0.00}{size of} $M$. More importantly, to determine the global optimal $\Psi^*$, the CSI of $M$ time blocks must be known, which is not practical, as it is really difficult to accurately estimate and predict the CSI of future $M$ time blocks within current time block. To make the user scheduling in time domain more practical, we present a user scheduling scheme, \textcolor[rgb]{0.00,0.00,0.00}{which is shown in \textbf{Algorithm 1} in details}, where for each time block, the unscheduled sensor who requires the smallest energy is scheduled.
\begin{algorithm}[h!]
\caption{The user scheduling scheme}
\label{alg:u}
\begin{algorithmic}[1]
\STATE Initialize MU number $\mathcal{M}=\{1,...,M\}$, $\mathcal{M}_{s}=\emptyset$.
\STATE \textbf{While}
\STATE  \quad $m$=arg $\mathop {\textrm{min}}\limits_{i \in \mathcal{M}}$ $\{E_{\textrm{u}}^{(i)}\}$,

\STATE  \quad Move $m$ from $\mathcal{M}$ into $\mathcal{M}_{s}$,
\STATE  \quad Until $\mathcal{M}=\emptyset$.
\STATE \textbf{End}
\end{algorithmic}
\end{algorithm}

\textcolor[rgb]{0.00,0.00,0.00}{According to the closed-form and semi-closed-form expressions above, the computation complexity of local computing mode is $\mathcal{O}(1)$ based on the closed-form solution. The computation complexity of fog offloading mode is $\mathcal{O}(\textrm{log}n)$ based on the bisection method, where $n$ denotes the array length of the time block duration $T_{\textrm{b}}$. The computation complexity of the user scheduling scheme of $M$ users is $\mathcal{O}(M)$. Therefore, the computation complexity of our proposed design is about $\mathcal{O}(M\textrm{log}n)$.}
\section{Discussion}\label{discuss}
In Section \ref{solution}, we obtain some closed-form and semi-closed-form expressions associated with the optimal configurations to minimize the multi-user energy requirement for the consider the \textcolor[rgb]{0.00,0.00,0.00}{fog computing-assisted PS SWIPT system}. In order to provide simpler results for engineers, this section will analyze the relationships between the system deployment and the mode selection, by which one \textcolor[rgb]{0.00,0.00,0.00}{can} easily determine the deployment area of the MU and the optimal mode selection without calculating the minimal required energy associated with the two modes.
\textcolor[rgb]{0.00,0.00,0.00}{\subsection{The deployment area versus distance $d_{\textrm{AP-u}}$}}
\textcolor[rgb]{0.00,0.00,0.00}{At first, we discuss deployment area of the MUs. Denote the distance between HAP and MU as $d_{\textrm{AP-u}}$ and the maximal feasible
distance\footnote{\textcolor[rgb]{0.00,0.00,0.00}{The maximal feasible distance means that when $d_{\textrm{AP-u}}\leq d_{\textrm{AP-u}}^{\textrm{(max)}}$, the MU can work; otherwise, the MU cannot work.}} between HAP and MU as $d_{\textrm{AP-u}}^{\textrm{(max)}}$, respectively. By regarding the location of the HAP as a reference and taking the effect of pass loss fading into account, according to (1), $d_{\textrm{AP-u}}^{\textrm{(max)}}$ can be calculated by $d_{\textrm{AP-u}}^{\textrm{(max)}}=10^{\tfrac{L(\textrm{dB})^{\textrm{(max)}}+28-20\log f_{c}}{n}}$ in terms of $L(\textrm{dB})^{\textrm{(max)}}$, where $L(\textrm{dB})^{\textrm{(max)}}$ is the path loss related to $d_{\textrm{AP-u}}^{\textrm{(max)}}$, so that one can get $d_{\textrm{AP-u}}^{\textrm{(max)}}$ by equivalently getting $L(\textrm{dB})^{\textrm{(max)}}$.}

\textcolor[rgb]{0.00,0.00,0.00}{\textbf{ Proposition 4:}
\emph{The MU's maximal path loss in the available deployment area,} $L(\textrm{dB})^{\textrm{(max)}}$, \emph{can be obtained by} $L(\textrm{dB})^{\textrm{(max)}} = \textrm{max}\{L(\textrm{dB})^{\textrm{(max)}}_\textrm{(local)},L(\textrm{dB})^{\textrm{(max)}}_\textrm{(offload)}\}$, \emph{i.e.,}
\begin{equation}\label{34}
L(\textrm{dB})^{\textrm{(max)}} =\left\{
             \begin{array}{lr}
             \frac{f_{\textrm{op}}(KAR_{\textrm{th}}+\xi R_{\textrm{th}}-\iota-E_{\textrm{s}}) +\sigma_\textrm{n}^2C\eta(f_{\textrm{op}}-KR_{\textrm{th}})}{\eta (f_{\textrm{op}}-KR_{\textrm{th}})P_{\textrm{AP}}}, \\
             {\kern 60pt} {\rm if} \, L(\textrm{dB})^{\textrm{(max)}}_\textrm{(local)}\geq L(\textrm{dB})^{\textrm{(max)}}_\textrm{(offload)}, \\
             \frac{(P^{\textrm{(max)}}_{\textrm{u-f}}+\xi F)R_{\textrm{th}}T_{\textrm{b}}+\eta\sigma_\textrm{n}^2DEF-\iota-E_{\textrm{s}}}{\eta EFP_{\textrm{AP}}},
             {\kern 7pt} {\rm otherwise},
             \end{array}
\right.
\end{equation}
\emph{where} $A\!=\!F_{0}\alpha M_{\textrm{c}}\!N_{0}\!\ln\!2, C\!=\!2^{\frac{R_{\textrm{th}}\!f_{\textrm{op}}}{B\left(f_{\textrm{op}}\!-\! KR_{\textrm{th}}\right)}}\!-\!1$, $D\!=\!2^{\frac{R_{\textrm{th}}\!T_{\textrm{b}}}{CE}}\!-\!1$, $E\!=\!\mathfrak{T_{\textrm{b}}}\!-\!\frac{R_{\textrm{th}}T_{\textrm{b}}}{B\log\left(1\!+\! \frac{\left|h_{\textrm{u-f}}\right|^2P^{\textrm{(max)}}_{\textrm{u-f}}}{\sigma_{\textrm{s}}^2}\right)}$, \emph{and} $F\!=\!B\log\left(1\!+\! \frac{\left|h_{\textrm{u-f}}\right|^2P^{\textrm{(max)}}_{\textrm{u-f}}}{\sigma_{\textrm{s}}^2}\right)$.}
\textcolor[rgb]{0.00,0.00,0.00}{\begin{IEEEproof}
See Appendix E.
\end{IEEEproof}}

\textcolor[rgb]{0.00,0.00,0.00}{To get more concise result, we consider the high signal-to-noise ratio (SNR) scenario and obtain
the following corollary.}

\textcolor[rgb]{0.00,0.00,0.00}{\textbf{ Corollary 3:}
\emph{In high SNR case, i.e.,} $P_{\textrm{AP}}\gg\sigma_\textrm{n}^2$, \emph{the MU's maximal path loss in the available deployment area,} $L(\textrm{dB})^{\textrm{(max)}}_{\textrm{SNR}}$, \emph{can be simplified to} $L(\textrm{dB})^{\textrm{(max)}}_{\textrm{SNR}} = \textrm{max}\{L(\textrm{dB})^{\textrm{(max)}}_\textrm{SNRlo},L(\textrm{dB})^{\textrm{(max)}}_\textrm{SNRof}\}$, \emph{i.e.,}
\begin{equation}\label{35}
L(\textrm{dB})^{\textrm{(max)}}_{\textrm{SNR}} =\left\{
             \begin{array}{lr}
             \frac{f_{\textrm{op}}(KAR_{\textrm{th}}+\xi R_{\textrm{th}} -\iota-E_{\textrm{s}})}{\eta (f_{\textrm{op}}-KR_{\textrm{th}})P_{\textrm{AP}}}, \\
             {\kern 30pt} {\rm if} \, L(\textrm{dB})^{\textrm{(max)}}_\textrm{SNRlo}\geq L(\textrm{dB})^{\textrm{(max)}}_\textrm{SNRof}, \\
             \frac{(P^{\textrm{(max)}}_{\textrm{u-f}}+\xi F)R_{\textrm{th}}T_{\textrm{b}}-\iota-E_{\textrm{s}}}{\eta EFP_{\textrm{AP}}},
             {\kern 12pt} {\rm otherwise}.
             \end{array}
\right.
\end{equation}}
\textcolor[rgb]{0.00,0.00,0.00}{\begin{IEEEproof}
By taking $(P_{\textrm{AP}}/\sigma_\textrm{n}^2)\rightarrow \infty$ into account, (\ref{34}) can be simplified to be (\ref{35}). Therefore, Corollary 3 is proved.
\end{IEEEproof}}
\textcolor[rgb]{0.00,0.00,0.00}{Note that $(f_{\textrm{op}}-K)$ should be larger than 0. That is, in high SNR case, it is required that $K< f_{\textrm{op}}$ for performing local computing.}
\textcolor[rgb]{0.00,0.00,0.00}{\subsection{Mode selection versus the computational complexity $K$}
In this section, we shall discuss the quantitative relationship between the parameter $K$ and mode selection. When the two modes require the same amount of energy, there is only a unique solution w.r.t $K$, which means
there exists a value of $K$ associated with the mode selection. Such $K$ is called the reference $K$, denoted as $K_0$. When $K$ is smaller than $K_0$, local computing mode is a better option; otherwise, when $K$ larger than $K_0$, the fog offloading mode should be selected. $K_0$ can be calculated in term of the following Proposition 5.}

\textcolor[rgb]{0.00,0.00,0.00}{\textbf{ Proposition 5:}
\emph{If the computing capability of the FS is sufficient, i.e.,} ${f_{\textrm{fogop}}}\gg K_0R_{\textrm{th}}T_{\textrm{b}}$, \emph{the computational complexity threshold,} $K_0$, \emph{can be expressed by} (\ref{eqnk1}), \emph{in which} $G = P_\textrm{AP}\left|\bm{h_{\textrm{AP-u}}}^H \bm{w}\right|^2$ and $H = \frac{R_{\textrm{th}}T_{\textrm{b}}}{F}$.
\begin{IEEEproof}
By taking $E_{\textrm{u(local)}}$ to be equal to $E_{\textrm{u(offload)}}$, Proposition 5 can be derived.
\end{IEEEproof}}

\newcounter{mytempeqncnt}
\begin{figure*}[!t]
\normalsize
\setcounter{mytempeqncnt}{\value{equation}}
\setcounter{equation}{35}
\begin{equation}
\label{eqnk1}
\textcolor[rgb]{0.00,0.00,0.00}{K_0\!=\!\frac{f_{\textrm{op}}\!\!\left(\!\!A\ln\!2\, R_{\textrm{th}}T_{\textrm{b}}f_{\textrm{op}}\!+\!\eta G\ln\!2\,R_{\textrm{th}}T_{\textrm{b}}\!+\!\eta\sigma_n^2
\ln\!2\,R_{\textrm{th}}T_{\textrm{b}}\!+\!{\mathcal{W}}\!\left(\!\!-\frac{\eta\sigma_n^2\ln\!2\,R_{\textrm{th}}T_{\textrm{b}}\mathrm{e}^{ -\frac{\left( Af_{\textrm{op}}+\eta G+\eta\sigma_n^2\right)\ln\!2\,R_{\textrm{th}}T_{\textrm{b}}}{GH}}}{GH}\!\right)\!GH\! +\!\ln\!2\,R_{\textrm{th}}H\!\right)}{R_{\textrm{th}}\!\left(\!A\ln\!2\,R_{\textrm{th}}T_{\textrm{b}}f_{\textrm{op}}\! +\!\eta G\ln\!2\,R_{\textrm{th}}T_{\textrm{b}}\!+\!\eta\sigma_n^2
\ln\!2\,R_{\textrm{th}}T_{\textrm{b}}\!+\!{\mathcal{W}}\!\left(\!\!-\frac{\eta\sigma_n^2\ln\!2\,R_{\textrm{th}}T_{\textrm{b}}\mathrm{e}^{ -\frac{\left( Af_{\textrm{op}}+\eta G+\eta\sigma_n^2\right)\ln\!2\,R_{\textrm{th}}T_{\textrm{b}}}{GH}}}{GH}\!\right)\!GH\!\right)}}
\end{equation}
\setcounter{equation}{\value{mytempeqncnt}}
\hrulefill
\vspace*{4pt}
\end{figure*}
\textcolor[rgb]{0.00,0.00,0.00}{\subsection{Mode selection versus the data compression ratio $\beta$}
In this sub section, we discuss the quantitative relationship between parameter $\beta$ and mode selection with other parameters fixed. Similarly, when the two modes have the same energy requirement, there is only a unique solution w.r.t $\beta$, which means there exists a value of $\beta$ associated with the mode selection. Such $\beta$ is called the reference $\beta$, denoted to be $\beta_0$. When $\beta$ is smaller than $\beta_0$, fog offloading mode is a better option; otherwise, when $\beta$ larger than $\beta_0$, the local computing mode should be selected. $\beta_0$ can be calculated in terms of the following Proposition 6.}

\textcolor[rgb]{0.00,0.00,0.00}{\textbf{ Proposition 6:}
\emph{The data compressed ratio threshold,} $\beta_0$, \emph{can be expressed by (\ref{fla_z}), in which}
$I=KAR_{\textrm{th}}T_{\textrm{b}}-\eta G\left(T_{\textrm{b}}-\frac{KR_{\textrm{th}}T_{\textrm{b}}}{f_{\textrm{op}}}\right)$ $\left(1-\frac{\sigma_{\textrm{n}}^2C}{G}\right)-P^{\textrm{(max)}}_{\textrm{u-f}}H$,
$J=T_{\textrm{b}}-H-\frac{KR_{\textrm{th}}T_{\textrm{b}}}{f_{\textrm{fogop}}}$, and $L=\frac{B\log\left( 1 + \frac{\left|h_{\textrm{f-u}}\right|^2P^{\textrm{(max)}}_{\textrm{f-u}}}{\sigma_{\textrm{f}}^2}\right)}{ R_{\textrm{th}}T_{\textrm{b}}}$
\begin{IEEEproof}
By taking $E_{\textrm{u(local)}}$ to be equal to $E_{\textrm{u(offload)}}$, Proposition 6 can be derived.
\end{IEEEproof}}

\newcounter{mytemp}
\begin{figure*}[!t]
\normalsize
\setcounter{mytemp}{\value{equation}}
\setcounter{equation}{36}
\begin{equation}
\label{fla_z}
\textcolor[rgb]{0.00,0.00,0.00}{\beta_0=L\left(G+\frac{I\ln\!2\,R_{\textrm{th}}T_{\textrm{b}}}{\eta G\ln\!2\,R_{\textrm{th}}T_{\textrm{b}}+\eta\sigma_n^2\ln\!2\,R_{\textrm{th}}T_{\textrm{b}}+{\mathcal{W}}\left(-\frac{\eta\sigma_n^2\ln\!
2\,R_{\textrm{th}}T_{\textrm{b}}\mathrm{e}^{-\frac{\eta\left(G+\sigma_n^2\right)\ln\!2\,R_{\textrm{th}}T_{\textrm{b}}}{IJ}}}{IJ}\right)IJ}\right)}
\end{equation}
\setcounter{equation}{\value{mytemp}}
\hrulefill
\vspace*{4pt}
\end{figure*}

\section{Application in a frame by frame system}\label{frame}
The results in section III and IV were obtained for a given time frame. That is, for \textcolor[rgb]{0.00,0.00,0.00}{each MU}, in its scheduled time block, i.e., the $m$-th $T_{\textrm{b}}$, the mode selection, power splitting ratio and time assignment can be optimally configured to minimize the \textcolor[rgb]{0.00,0.00,0.00}{energy} requirement of MU $m$, for \textcolor[rgb]{0.00,0.00,0.00}{$m \in\{1, 2,..., M\}$}. Considering that in practice, the energy storage in MU's battery also impacts the system performance. Therefore, in this section, we discuss how to run our proposed multi-user system in a frame by frame continuous manner.

Denote $E^{(m)}_{\textrm{s}}[l]$ as the stored energy \textcolor[rgb]{0.00,0.00,0.00}{in its battery} at MU $m$ at the beginning of the $l$-th time frame and $E^{(m)}_{\textrm{u}}[l]$ as the required energy associated with the $l$-th time frame.

Obviously, when $E^{(m)}_{\textrm{s}}[l]<E^{(m)}_{\textrm{u}}[l]$, the system can not work without harvesting energy from the HAP. Therefore, in this case, the $l$-th time block with interval $T_{\textrm{b}}$ should be only used for MU $m$ to harvest energy. That is, $\tau_\textrm{ipt}$ = $T_{\textrm{b}}$ and $\rho$ = 0. Consequently, the energy storage at the end of $l$-th time block can be given by
\begin{equation}
E^{(m)}_{\textrm{s}}[l] = \left\{
             \begin{array}{lr}
             E^{(m)}_{\textrm{s}}[l]+E^{(m)}_{\textrm{eh}}[l],\\
             {\kern 35pt} {\rm if}  \,E^{(m)}_{\textrm{u}}[l] < 0 \,\,{\rm and}\,\, E^{(m)}_{\textrm{u}}[l] < E^{(m)}_{\textrm{s}}[l], \\
             E^{(m)}_{\textrm{s}}[l]+E^{(m)}_{\textrm{u}}[l],\\
             {\kern 35pt} {\rm if}  \, E^{(m)}_{\textrm{u}}[l] < 0 \,\,{\rm and}\,\, E^{(m)}_{\textrm{u}}[l] \geq E^{(m)}_{\textrm{s}}[l], \\
             E^{(m)}_{\textrm{s}}[l]+E^{(m)}_{\textrm{u}}[l],
             {\kern 70pt} {\rm \textcolor[rgb]{0.00,0.00,0.00}{otherwise,}}
             \end{array}
\right.
\end{equation}
where $l$=1, 2, 3, $\cdot\cdot\cdot$. With $E^{(m)}_{\textrm{s}}[l]$, the proposed multi-user scheduling scheme and the corresponding optimized configurations can be applied in the $l$-th time frame.

\section{SIMULATION RESULTS}\label{results}
In this section, we provide some numerical results to discuss the performance of \textcolor[rgb]{0.00,0.00,0.00}{our considered fog computing-assisted PS SWIPT system}. At first, the single MU system is simulated for a given time block to discuss the performance of the two information processing modes, and then, the multi-user system is simulated for a given time frame in order to discuss the performance of the proposed scheduling scheme. Finally, the system is simulated in a frame-by-frame continuous operation scenario in order to discuss the long-time system performance.

\textcolor[rgb]{0.00,0.00,0.00}{In the simulations, the channel coefficients were generated following Rician fading model and the rest parameters were configured according to \cite{ref_re13}, \cite{ref_circuit} and \cite{ref_f0}. For clarity, the detail parameter settings are provided in Table I\footnote{\textcolor[rgb]{0.00,0.00,0.00}{Note that the logic operations per second is not equivalent to the CPU cycles per second. According to \cite{ref_44}, at room temperature, the Landauer limit is $N_0\ln\!2\approx3\times10^{-21}J$. Thus, it is estimated that the computing rate fop of the computer is able to reach $1/(3\times100\times10^{-21}J\times0.1)\approx10^{18}$ logic operations per second. The frequency of FS can be given by $f_{\textrm{FS}}=f_{\textrm{lopt}}\times N_{\textrm{core}}\times N_{\textrm{CPU}}$, where $f_{\textrm{lopt}}$ is the floating-point operations per second, $N_{\textrm{core}}$ is the number of core count per CPU, and $N_{\textrm{CPU}}$ is the number of CPU. Thus, in this paper, it is assumed that there are $10^4$ logic operations per floating-point operation and 4 CPU with 8 core of each one.}}, which \textcolor[rgb]{0.00,0.00,0.00}{did not change} in the sequel unless otherwise specified.} It is noticed that each point in the figures of this section was obtained by averaging $10^3$ channel realizations.

\begin{table}[htb]
\centering
\caption{Simulation Parameters}
\begin{tabular}{l|c|c}
  \hline
  \hline
  Parameters & Notation &Values \\
  \hline
  \hline
  The HAP's power budget & $P_\textrm{AP}$  & 1 watt \\
  The HAP's antenna number & $N_A$  & 8 \\
  The system bandwidth & $B$  &  2 MHz \\
  The system noise power & $\sigma^{2}_{\textrm{n}}$, $\sigma^{2}_{\textrm{s}}$, $\sigma^{2}_{\textrm{f}}$  & -140 dBm \\
  The Rician factor  & $K_\textrm{rice}$  & 3.5 dB \\
  The transmission frequency & $f_c$  & 915 MHz \\
  The distance power loss coefficient & $N$  & 22 \\
  The \textcolor[rgb]{0.00,0.00,0.00}{minimal} required rate & $R_{\textrm{th}}$  & 20 Kb/s \\
  The time block duration & $T_{\textrm{b}}$  & 1 sec \\
  The task's \textcolor[rgb]{0.00,0.00,0.00}{computational} complexity & $K$  & $10^4$ oper/bit \\
  The energy harvesting efficiency & $\eta$  & 0.6 \\
  The decoding required energy factor & $\xi$  & $10^{-10}$ J/bit \\
  The MU's power budget & $P^{\textrm{(max)}}_\textrm{u-f}$  & $10^{-3}$ watt \\
  The MU's operational capability  & $f_{\textrm{op}}$  & $10^9$ oper/s \\
  The FS's operational capability  & $f_{\textrm{fogop}}$  & $10^{15}$ oper/s \\
  The time-dependent immaturity factor & $M_\textrm{c}$  & $10^4$ \\
  The circuit activity factor & $\alpha$  & 0.1${\sim}$0.2 \\
  The circuit fanout number  & $F_0$  & 3${\sim}$4 \\
  The result scaling factor & $\beta$  & $10^{-2}$ \\
  \hline
\end{tabular}
\label{parametertable}
\end{table}

\subsection{System performance versus $K$ for a given time block}
\begin{figure}[htb]
\centering
\includegraphics[width=0.435\textwidth]{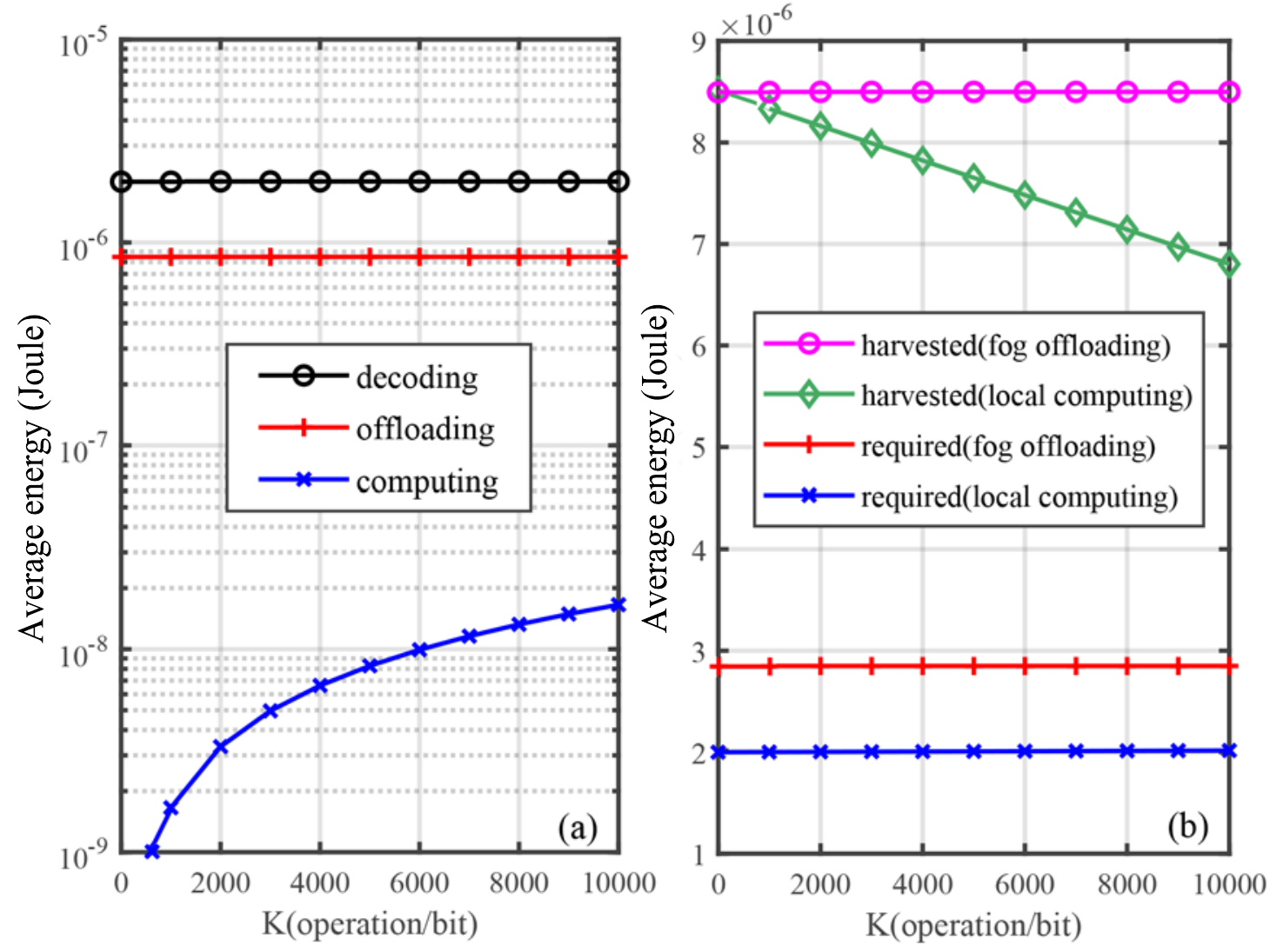}
\caption{\textcolor[rgb]{0.00,0.00,0.00}{(a) The average required energy for decoding, offloading and computing per block versus $K$, (b) The average harvested and required energy per block of the two modes versus $K$}}
\label{fig_tu1}\vspace{-0.1 in}
\end{figure}

First, we discuss the effect of $K$ on the energy requirement at MU $m$. Fig. \ref{fig_tu1} shows the harvested and required energy versus $K$, where $d_\textrm{AP-u}$ = 10 m and $d_\textrm{u-f}$ = 8 m. It can be seen that the energy required for decoding and offloading do not change with the increment of $K$, but that for computing increases linearly versus $K$. Moreover, the energy harvested in the fog offloading mode does not change versus $K$, but that harvested in local computing mode deceases linearly versus $K$, which is affected by $\tau_{\textrm{ipt(local)}}$ that closely related to $K$. As $K$ increases, $\tau_{\textrm{ipt(local)}}$ decreases linearly, and then the energy harvested decreases proportionally.
\begin{figure}[htb]
\centering
\includegraphics[width=0.435\textwidth]{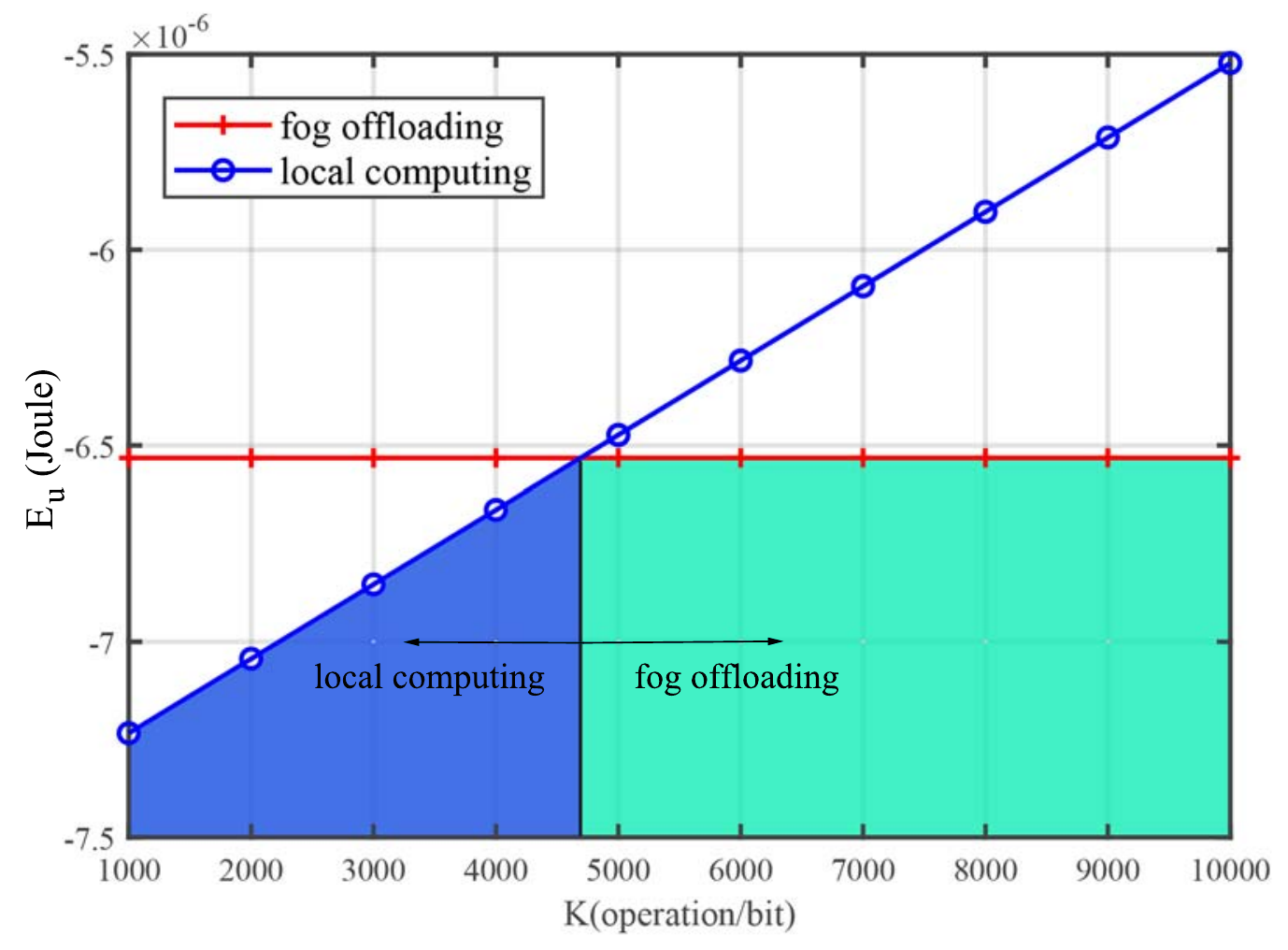}
\caption{\textcolor[rgb]{0.00,0.00,0.00}{The average energy requirement $E_{\textrm{u}}$ of the two modes versus $K$, where $E_{\textrm{u}}$ is defined in (\ref{lcenergy}) and (\ref{lcenergy1})}}
\label{fig_three1}\vspace{-0.1 in}
\end{figure}

Fig. \ref{fig_three1} shows the minimal energy requirement at MU $m$ of the local computing mode and the fog offloading mode versus $K$. It can be seen that there is an intersection point between the two lines associated with the two modes at $K=$ 4800 operations per bit. This indicates that there exists a certain value of $K$, when $K$ is less than the certain value (i.e., the intersection point), the local computing is a better option; \textcolor[rgb]{0.00,0.00,0.00}{otherwise}, the fog computing mode is better. This result is consistent with the analysis in \textcolor[rgb]{0.00,0.00,0.00}{\textbf{Proposition 5}}.

\subsection{System performance versus distances and locations for a given time block}
\subsubsection{The minimal required energy versus $d_\textrm{AP-u}$ with fixed $d_\textrm{u-f}$}

\begin{figure}[htb]
\centering
\includegraphics[width=0.37\textwidth]{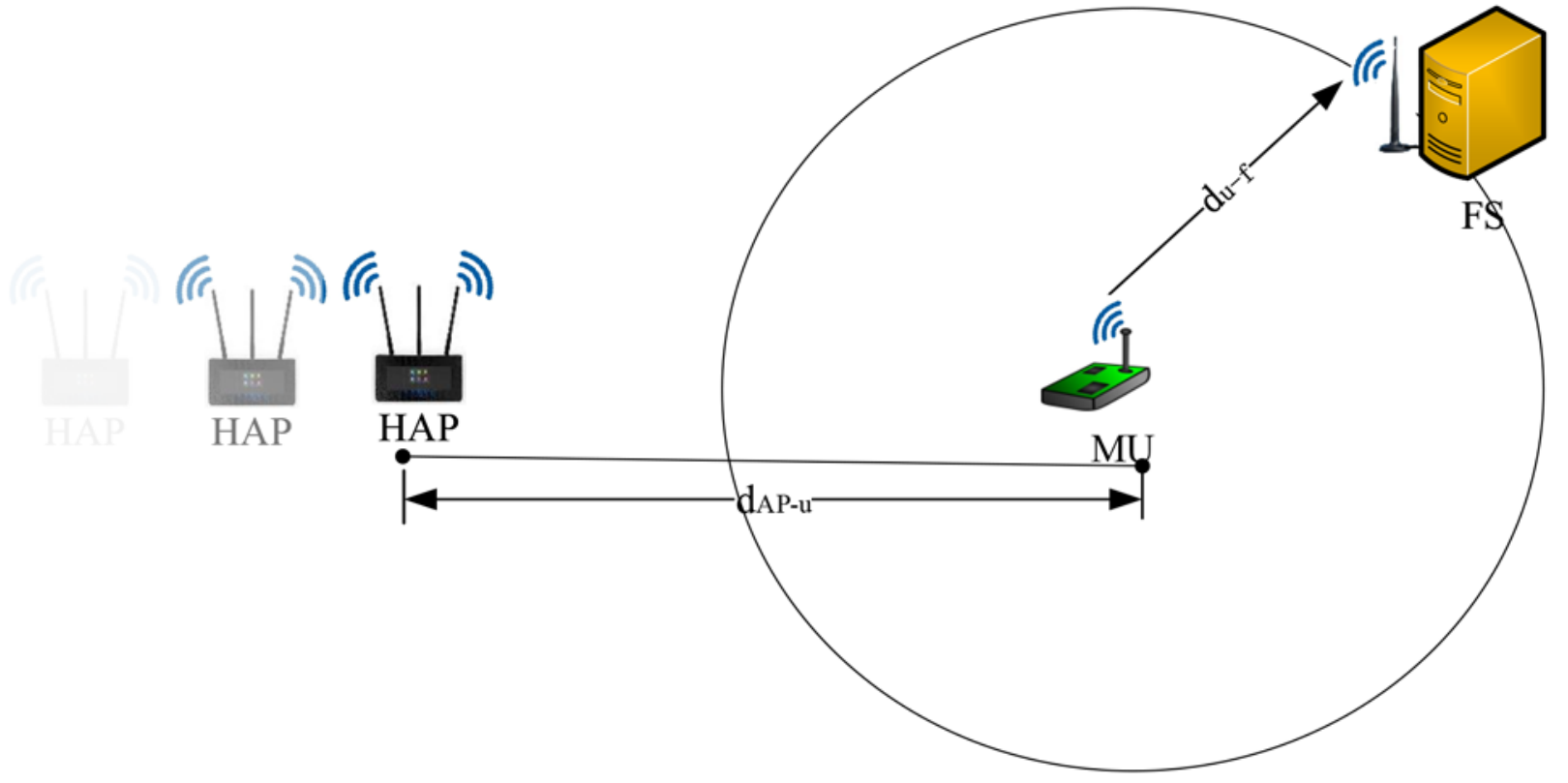}
\caption{\textcolor[rgb]{0.00,0.00,0.00}{Illustration of the simulation scenarios, where $d_{\textrm{AP-u}}$ is changed with a fixed $d_{\textrm{u-f}}$}}
\label{fig_ds10fixed}\vspace{-0.1 in}
\end{figure}

\begin{figure}[htb]
\centering
\includegraphics[width=0.435\textwidth]{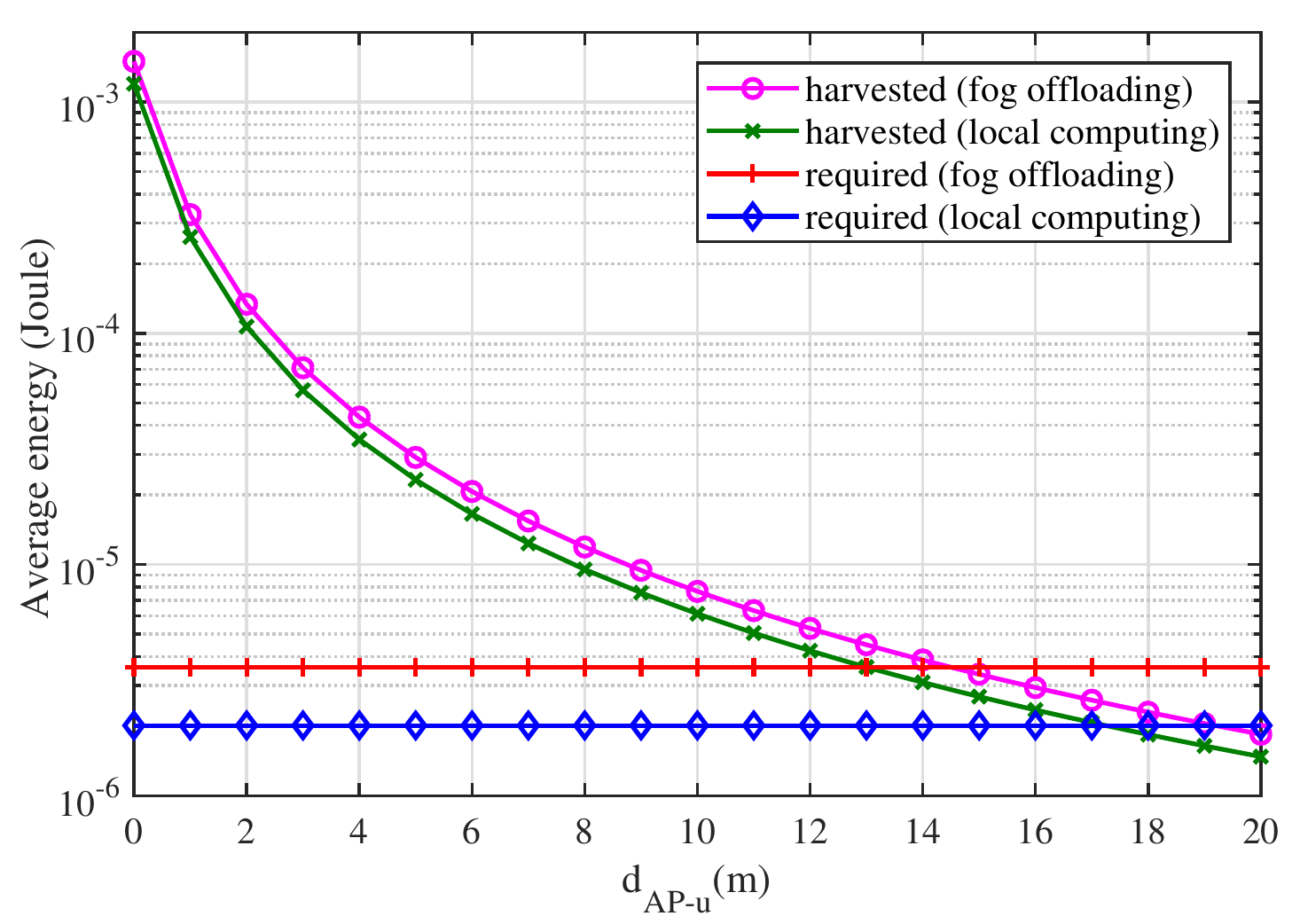}
\caption{\textcolor[rgb]{0.00,0.00,0.00}{The harvested energy and required energy per block of the two modes} versus $d_\textrm{AP-u}$}
\label{fig_distance}\vspace{-0.1 in}
\end{figure}

Fig. \ref{fig_distance} shows the average harvested and required energy at MU $m$ versus $d_\textrm{AP-u}$, where the distance between MU $m$ and the FS is fixed, i.e., $d_\textrm{u-f}$ = 8m, and \textcolor[rgb]{0.00,0.00,0.00}{the HAP is moved away from MU $m$} as illustrated in Fig. \ref{fig_ds10fixed}. From Fig. \ref{fig_distance}, one can see that with the increment of $d_\textrm{AP-u}$, the required energy of the two modes almost do not change but the harvested energy of both two modes decrease, because the required energy is independent of $d_\textrm{AP-u}$, but the harvested energy closely depends on the path loss fading of the wireless channels.
\begin{figure}[htb]
\centering
\includegraphics[width=0.435\textwidth]{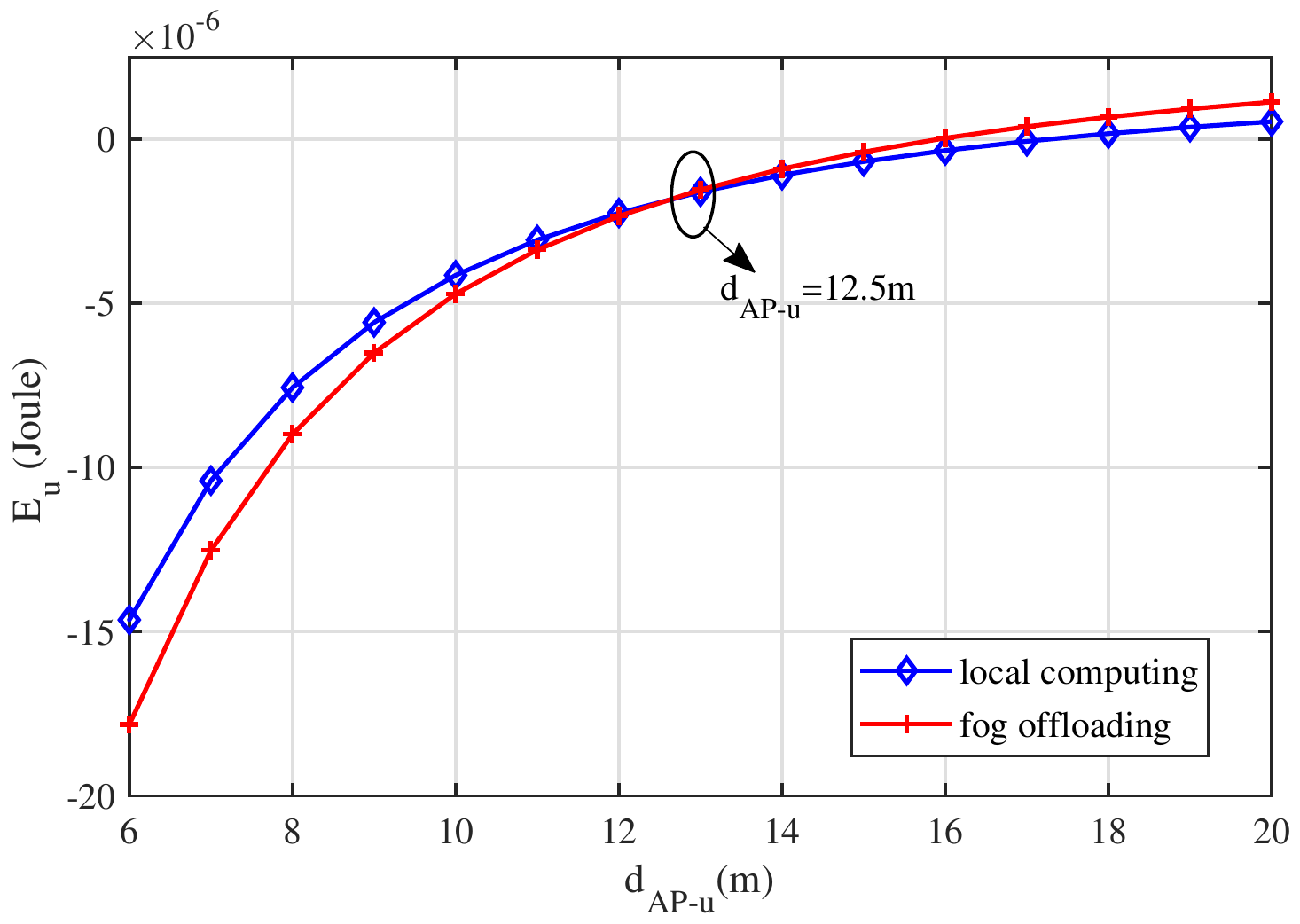}
\caption{\textcolor[rgb]{0.00,0.00,0.00}{The average energy requirement $E_{\textrm{u}}$ of the two modes versus  $d_\textrm{AP-u}$, where $E_{\textrm{u}}$ is defined in (\ref{lcenergy}) and (\ref{lcenergy1})}}
\label{fig_u-f}\vspace{-0.1 in}
\end{figure}

Fig. \ref{fig_u-f} shows the minimal energy requirement of the two modes at MU $m$ versus $d_\textrm{AP-u}$ with the same settings as Fig. \ref{fig_distance}, where the distance between MU $m$ and the FS is fixed as $d_\textrm{u-f}$ = 8m, \textcolor[rgb]{0.00,0.00,0.00}{and the HAP is moved away from MU $m$}. It is seen that there is an intersection point between the two curves at $d_\textrm{AP-u}$ = 12.5 m. This implies that when $d_\textrm{AP-u}$ is less than a certain value, the local computing mode is a better choice; \textcolor[rgb]{0.00,0.00,0.00}{otherwise,} the fog computing mode should be selected. This observation is consistent to the analysis in \textcolor[rgb]{0.00,0.00,0.00}{\textbf{Proposition 4}}.

\subsubsection{Fixed the distance between HAP and FS}
\begin{figure}[htb]
\centering
\includegraphics[width=0.46\textwidth]{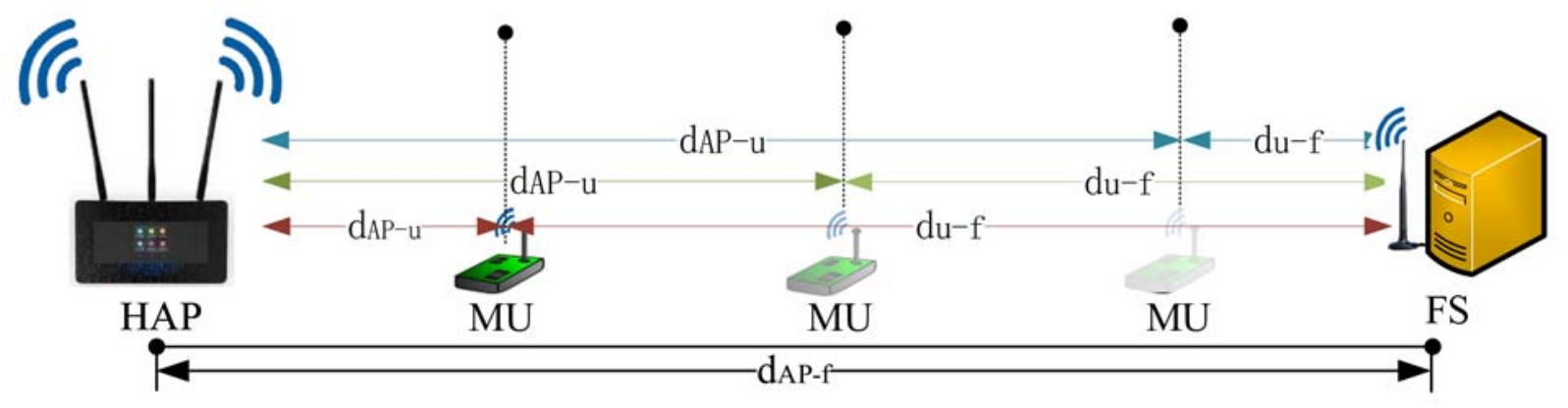}
\caption{Illustration of the HAP, MU $m$, and FS locations}
\label{fig_apfogposition}\vspace{-0.1 in}
\end{figure}

\begin{figure}[htb]
\centering
\includegraphics[width=0.435\textwidth]{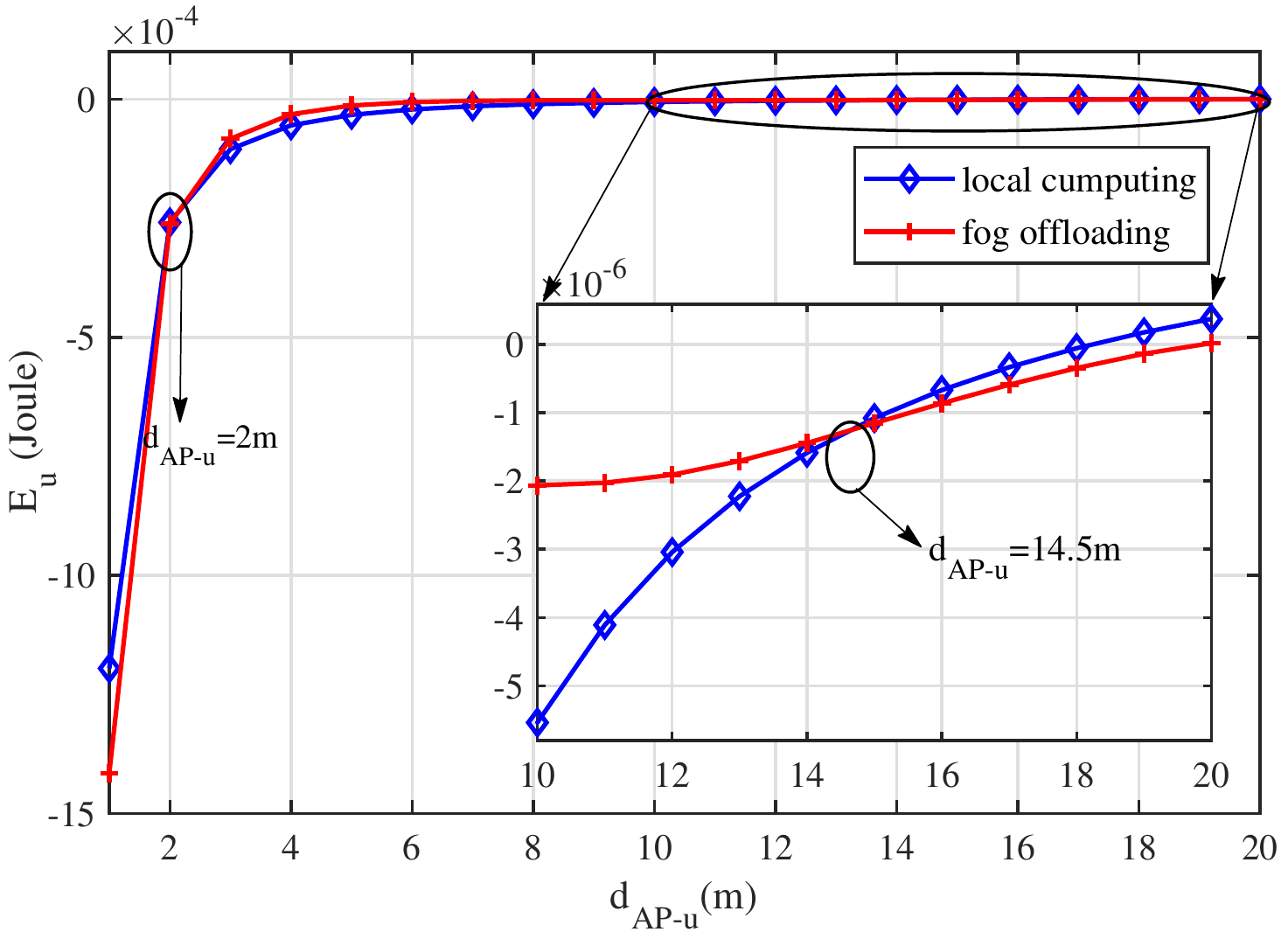}
\caption{\textcolor[rgb]{0.00,0.00,0.00}{The average energy requirement $E_{\textrm{u}}$ of the two modes versus $d_\textrm{AP-u}$ and $d_\textrm{u-f}$, where $E_{\textrm{u}}$ is defined in (\ref{lcenergy}) and (\ref{lcenergy1})}}
\label{fig_apfog}\vspace{-0.1 in}
\end{figure}

Fig. \ref{fig_apfog} shows the minimal energy requirement of the two modes at MU $m$ versus $d_\textrm{AP-u}$ and $d_\textrm{u-f}$, where the distance between the HAP and the FS is fixed, i.e., $d_\textrm{AP-f}=$ 20m, and MU $m$ moves along the line between HAP and FS, as illustrated in Fig.\ref{fig_apfogposition}. In Fig. \ref{fig_apfog}, there exists two intersection points between the two curves at $d_\textrm{AP-u}$ = 2m and 14.5m. This indicates that there are two thresholds of $d_\textrm{AP-f}$. When $d_\textrm{AP-f}$ is smaller than the lower threshold, i.e., $d_\textrm{AP-u}$ = 2m, or larger than the higher threshold, i.e., $d_\textrm{AP-u}$=14.5m, the fog offloading mode is the better choice; \textcolor[rgb]{0.00,0.00,0.00}{otherwise}, the local computing mode is the better option.

\begin{figure}[h!]
\centering
\includegraphics[width=0.435\textwidth]{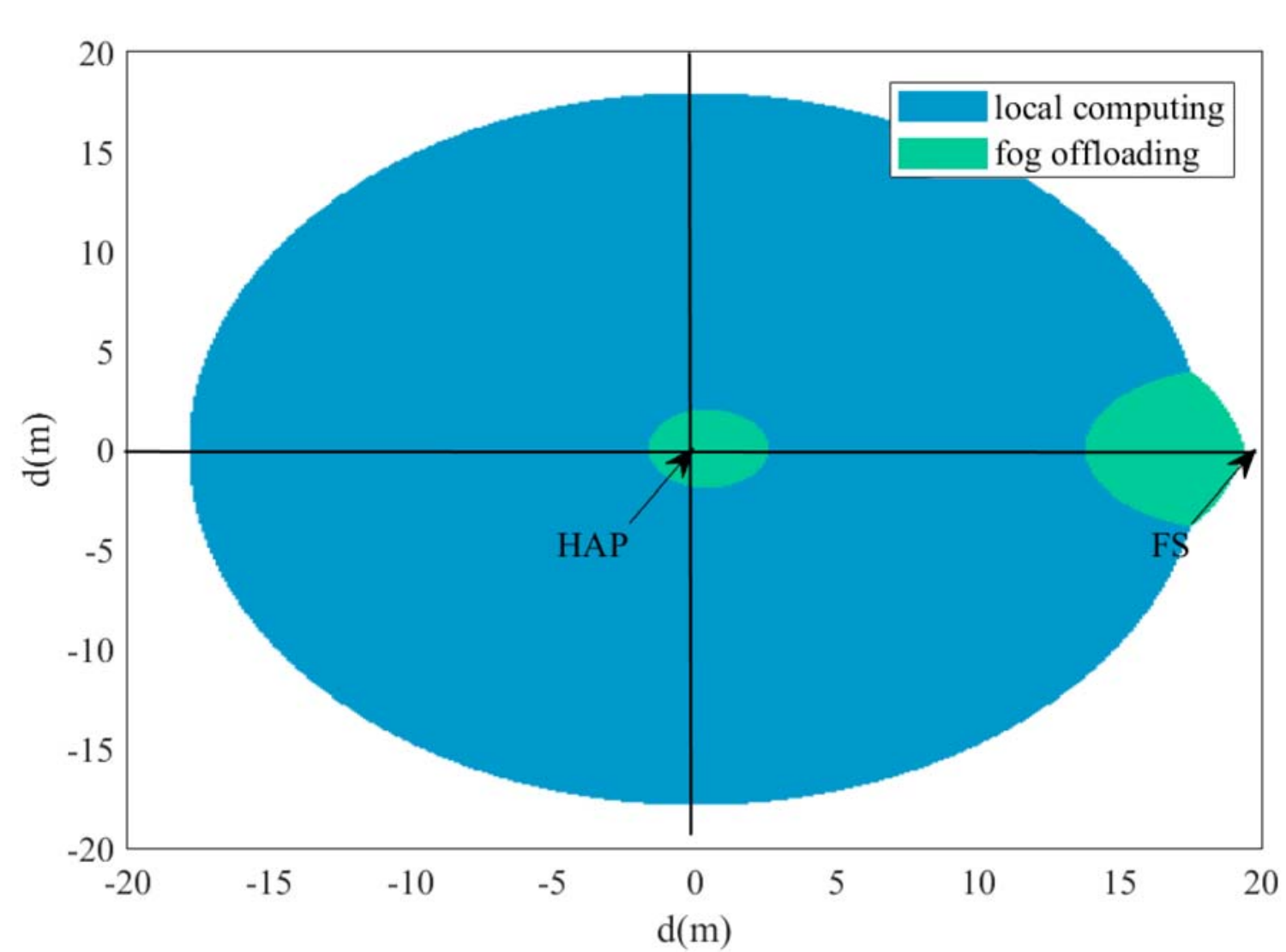}
\caption{\textcolor[rgb]{0.00,0.00,0.00}{Comparison of the two modes versus MU's location}}
\label{fig_buju}\vspace{-0.1 in}
\end{figure}

Motivated by Fig. \ref{fig_apfog}, the mode selection is discussed in Fig. \ref{fig_buju} over a two-dimensional coordinate plane in order to provide deeper insights, where the HAP is positioned at the original point (i.e., (0, 0)) and the FS is positioned at (0, 20). \textcolor[rgb]{0.00,0.00,0.00}{MU $m$ can be located at arbitrary point} on the plane. The minimal energy requirement of the two modes is compared. In the green area, the fog offloading mode should be selected and in the blue area, the local computing mode is a better choice. In the white region, the system has to firstly accumulate the energy until the stored energy is sufficient to work. \textcolor[rgb]{0.00,0.00,0.00}{The reason may be that when MU $m$ is positioned closed to the FS, less time is required to offload the task, leading to more time to perform energy harvesting, so the fog offloading mode is better. When MU $m$ is far from the FS (around 0 m area), it is relatively close to the HAP, in this case, MU $m$ can receive stronger power, so that it is able to harvest more energy, leading to the fog offloading is a better option.}

\begin{figure}[h!]
\centering
\includegraphics[width=0.45\textwidth]{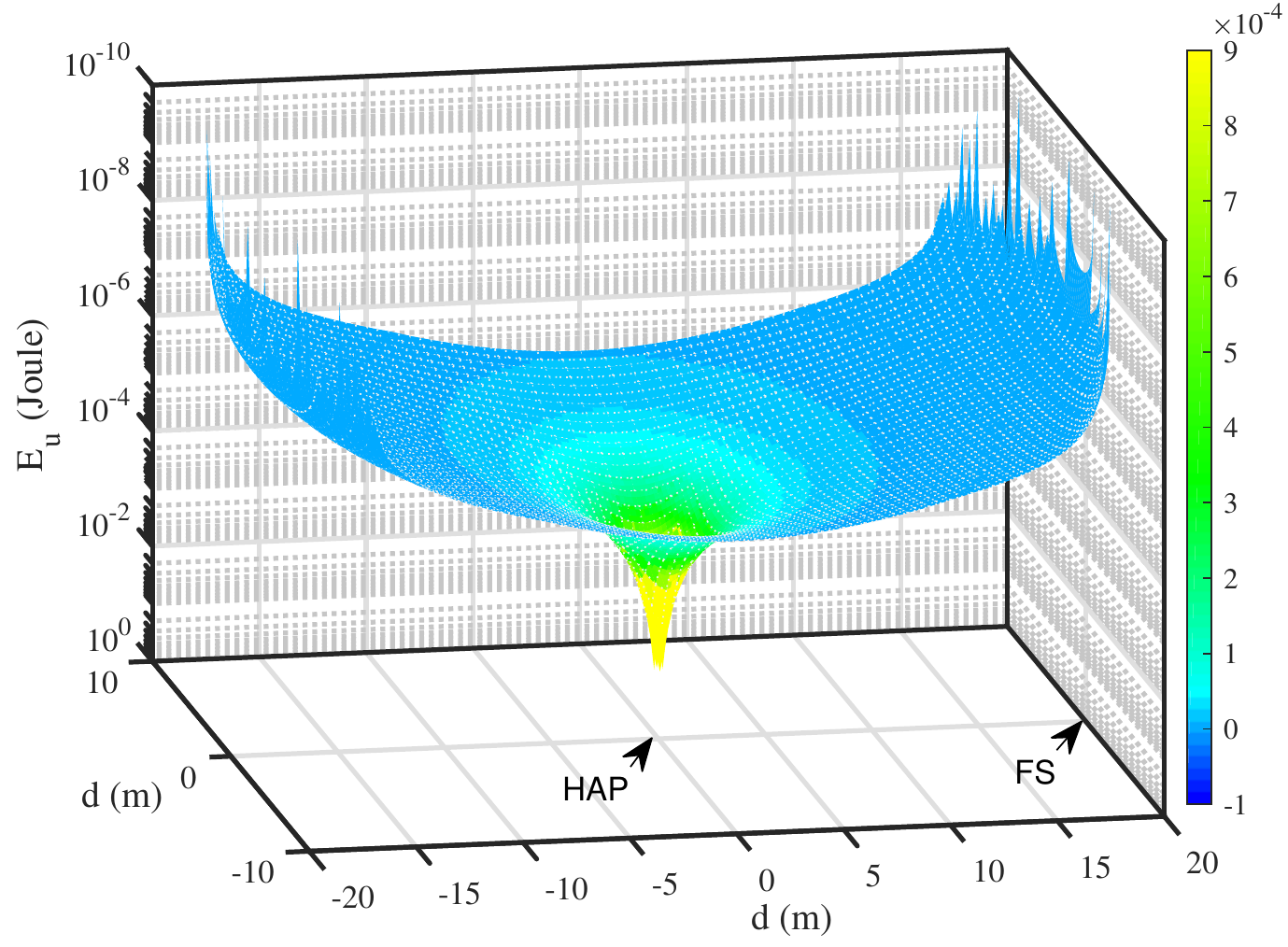}
\caption{\textcolor[rgb]{0.00,0.00,0.00}{The average energy requirement $E_{\textrm{u}}$ of our proposed method with the optimal mode selection}}
\label{fig_menggubao}
\end{figure}

The minimal energy requirement of MU $m$ associated with the two modes is also plotted in a 3-D figure as shown in Fig. \ref{fig_menggubao}. It can be seen that when MU $m$ is closely positioned to the HAP, it requires relatively low power to meet the information \textcolor[rgb]{0.00,0.00,0.00}{transmission} and computing requirements.

\subsection{System performance versus $P_{\textrm{AP}}$ for a given time block}
\begin{figure*}
\centering
\vspace{-0.1 in}
\subfigure[$P_{\textrm{AP}}$ = 0.5 watt]{
\begin{minipage}[b]{0.23\textwidth}
\centering
\includegraphics[width=1.7in]{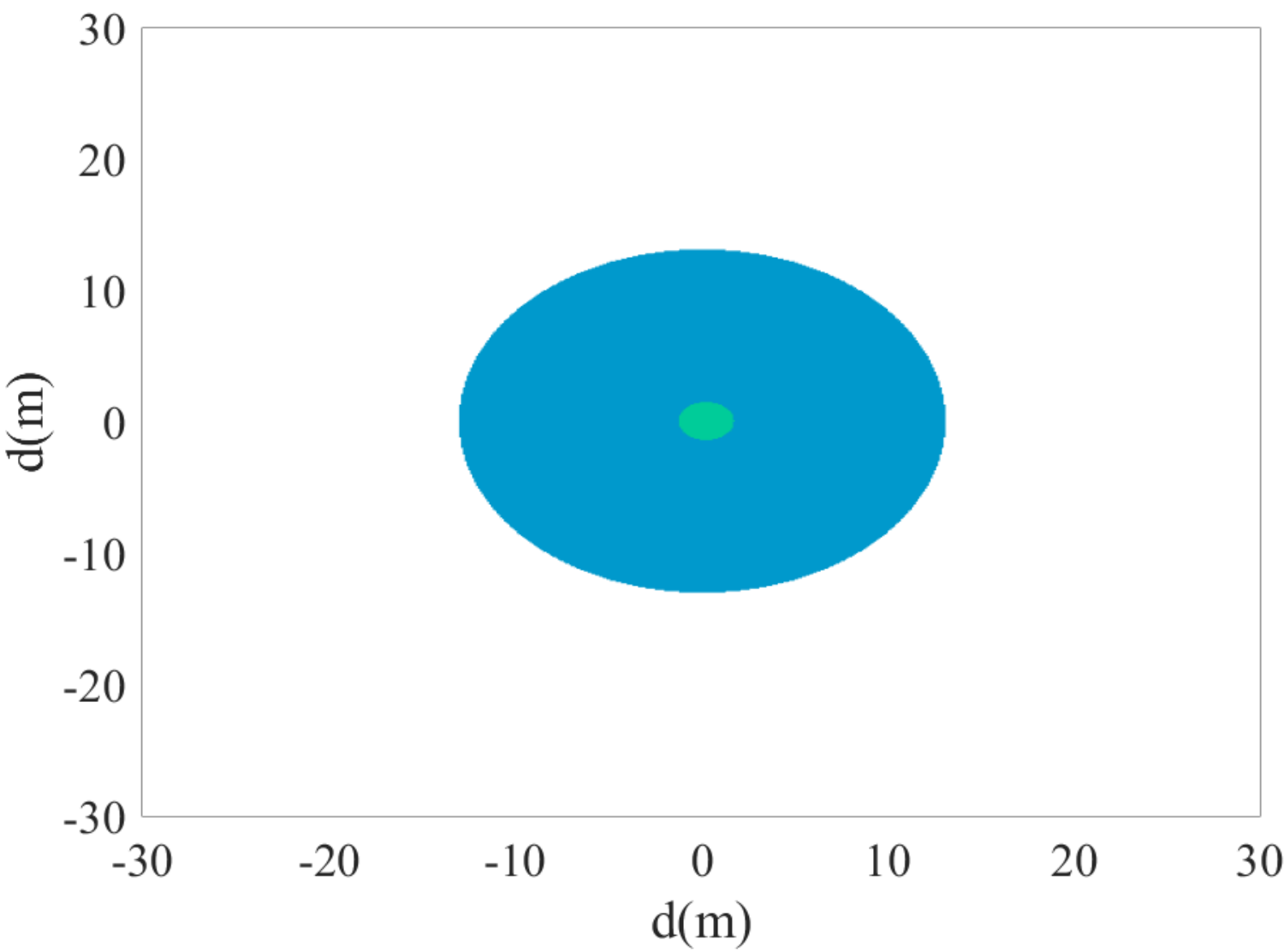}
\label{fig_pap05}
\end{minipage}}
\hspace{0.01\linewidth}
\subfigure[$P_{\textrm{AP}}$ = 1.0 watt]{
\begin{minipage}[b]{0.23\textwidth}
\centering
\includegraphics[width=1.7in]{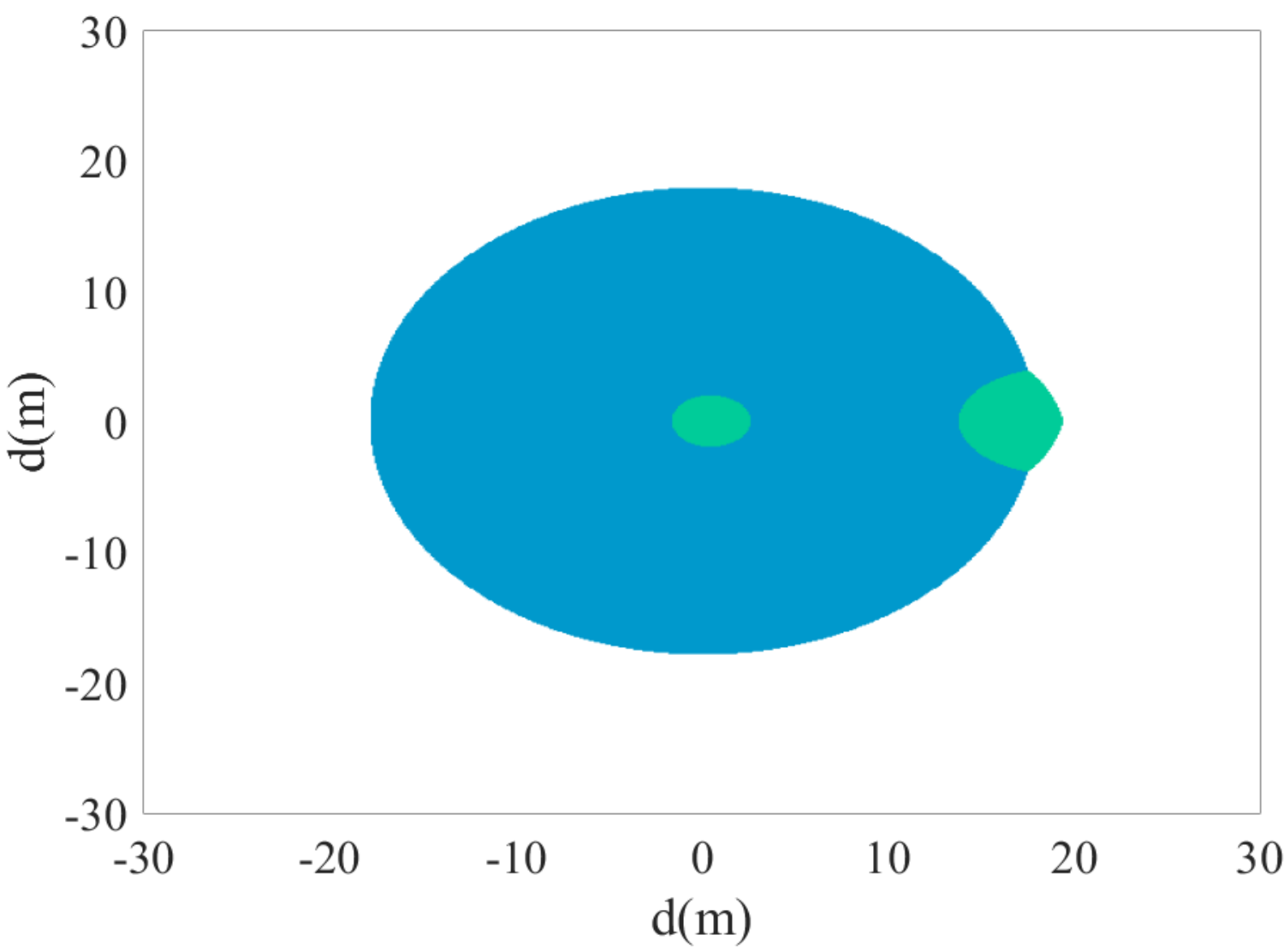}
\label{fig_pap10}
\end{minipage}}%
\hspace{0.01\linewidth}%
\subfigure[$P_{\textrm{AP}}$ = 1.5 watt]{
\begin{minipage}[b]{0.23\textwidth}
\centering
\includegraphics[width=1.7in]{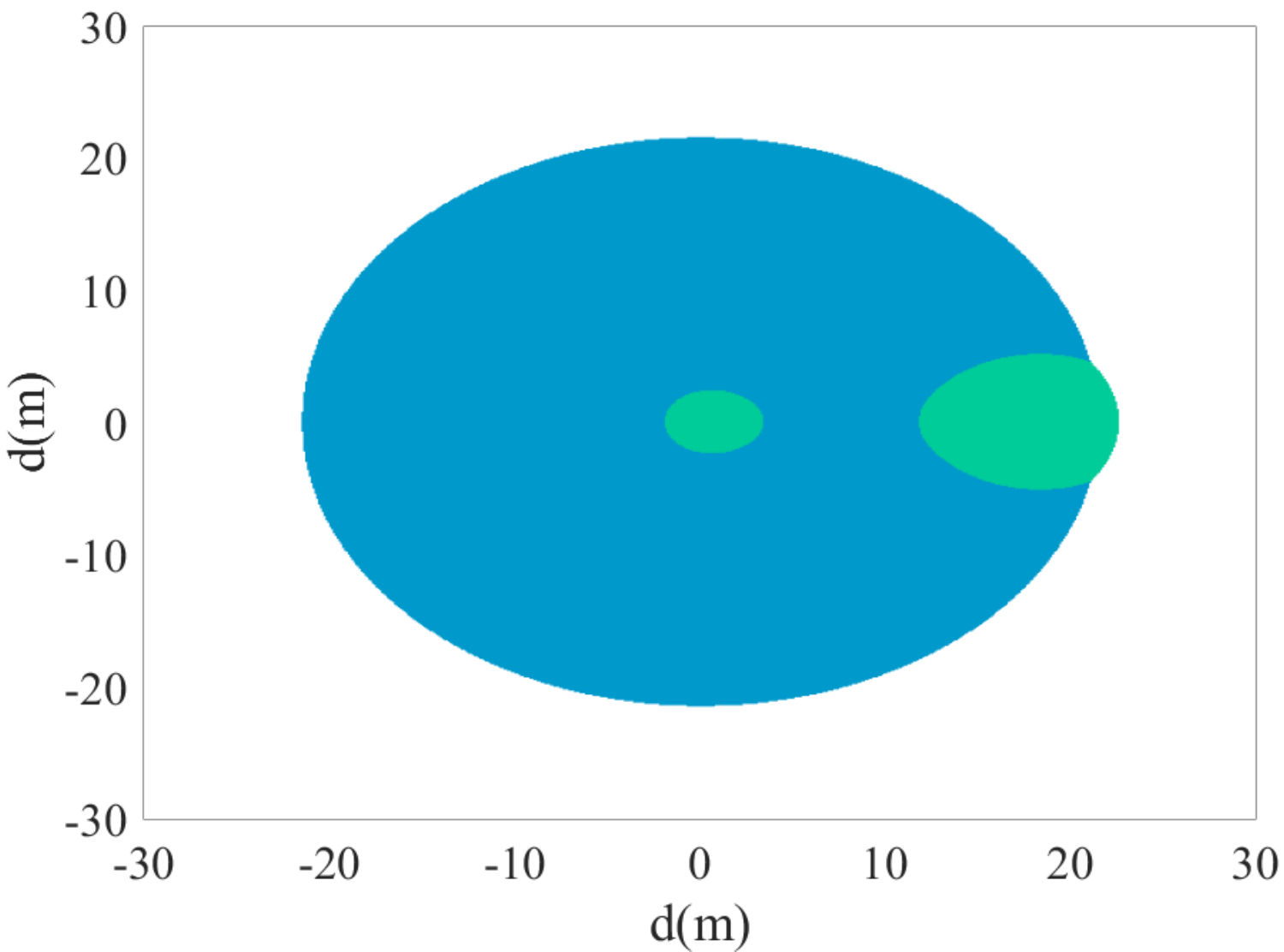}
\label{fig_pap15}
\end{minipage}}
\hspace{0.01\linewidth}%
\subfigure[$P_{\textrm{AP}}$ = 2.0 watt]{
\begin{minipage}[b]{0.23\textwidth}
\centering
\includegraphics[width=1.7in]{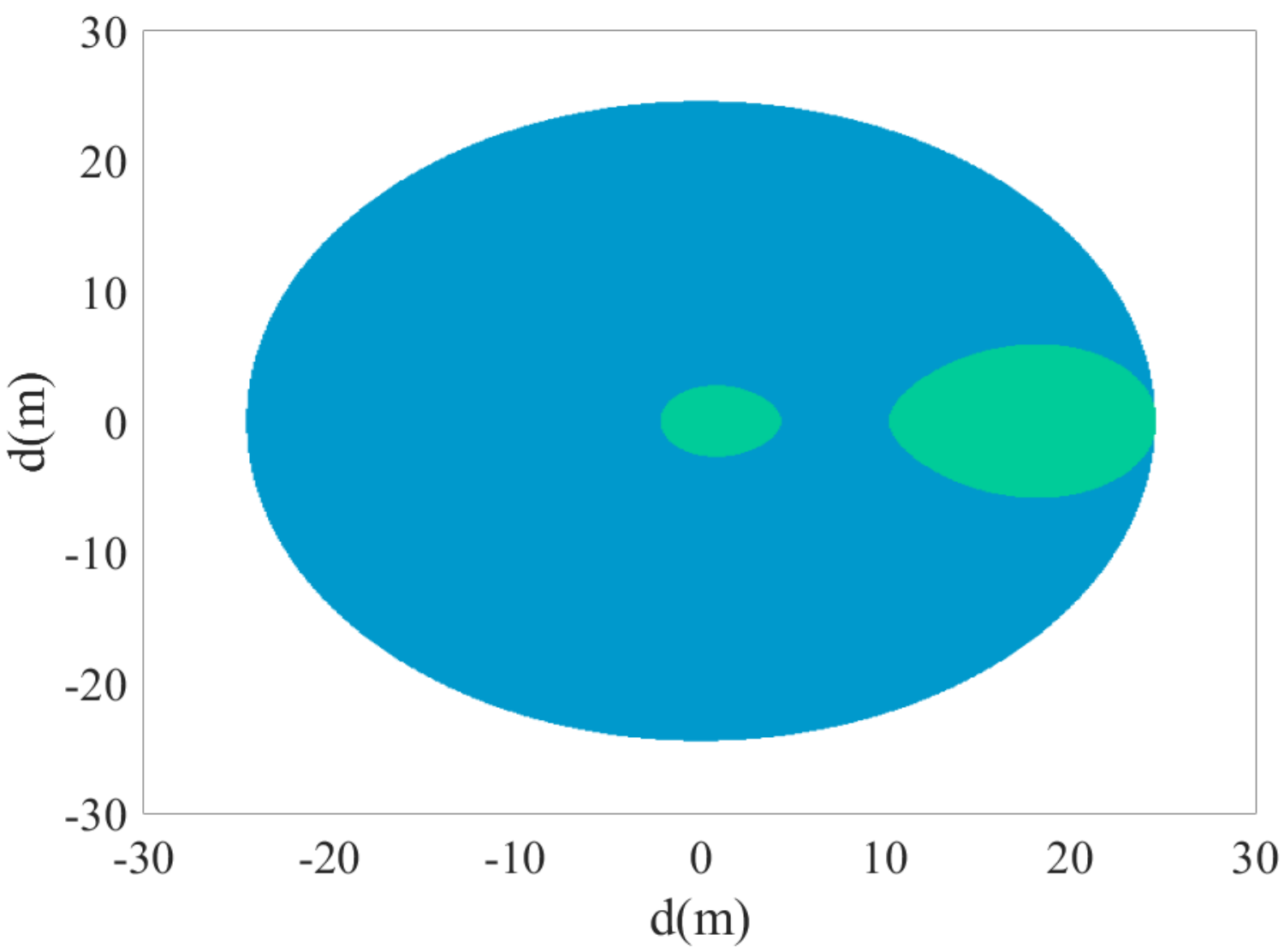}
\label{fig_pap20}
\end{minipage}}%
\hspace{0.01\linewidth}%
\subfigure[$P_{\textrm{AP}}$ = 2.5 watt]{
\begin{minipage}[b]{0.23\textwidth}
\centering
\includegraphics[width=1.7in]{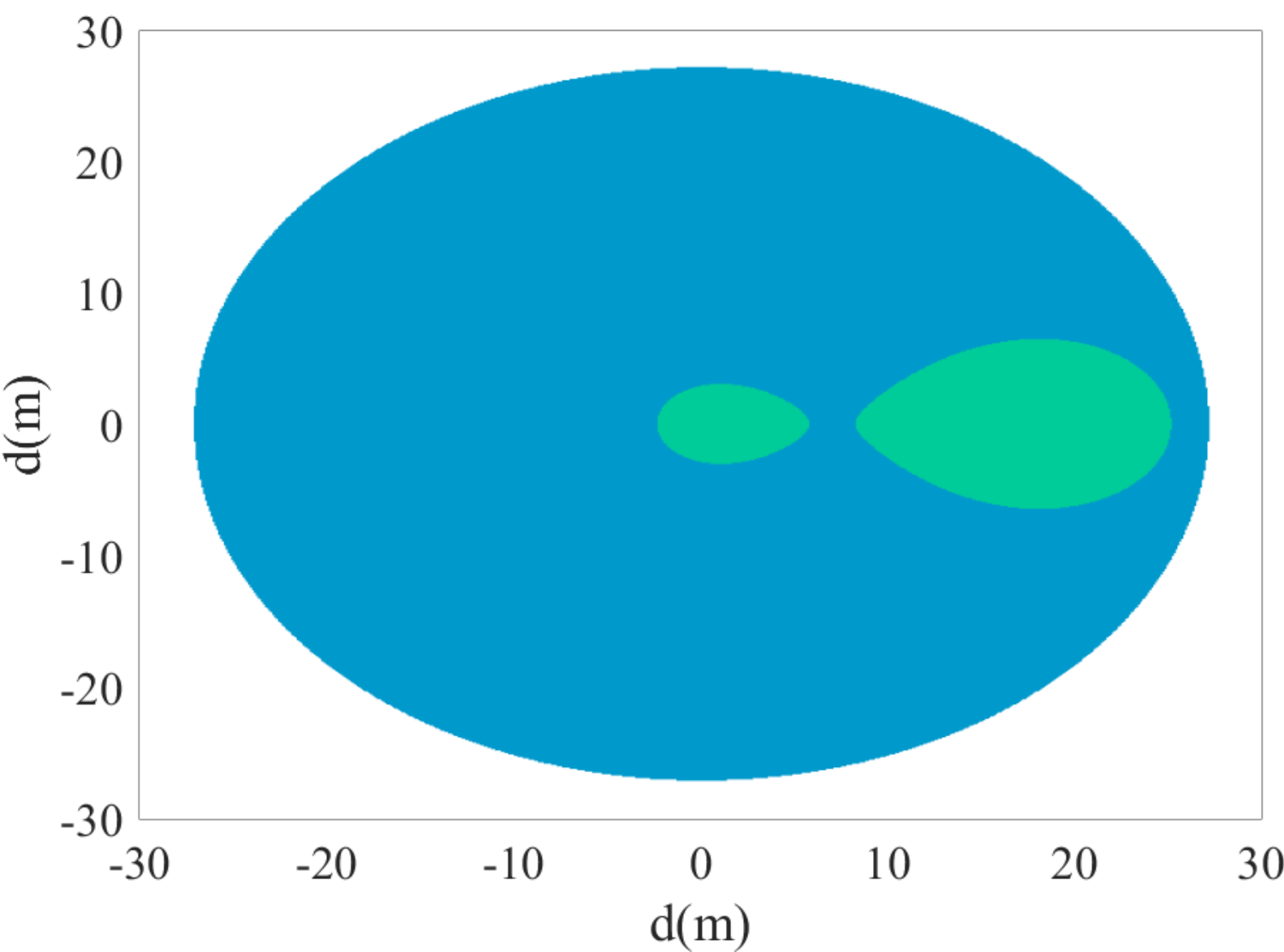}
\label{fig_pap25}
\end{minipage}}%
\hspace{0.01\linewidth}%
\subfigure[$P_{\textrm{AP}}$ = 3.0 watt]{
\begin{minipage}[b]{0.23\textwidth}
\centering
\includegraphics[width=1.7in]{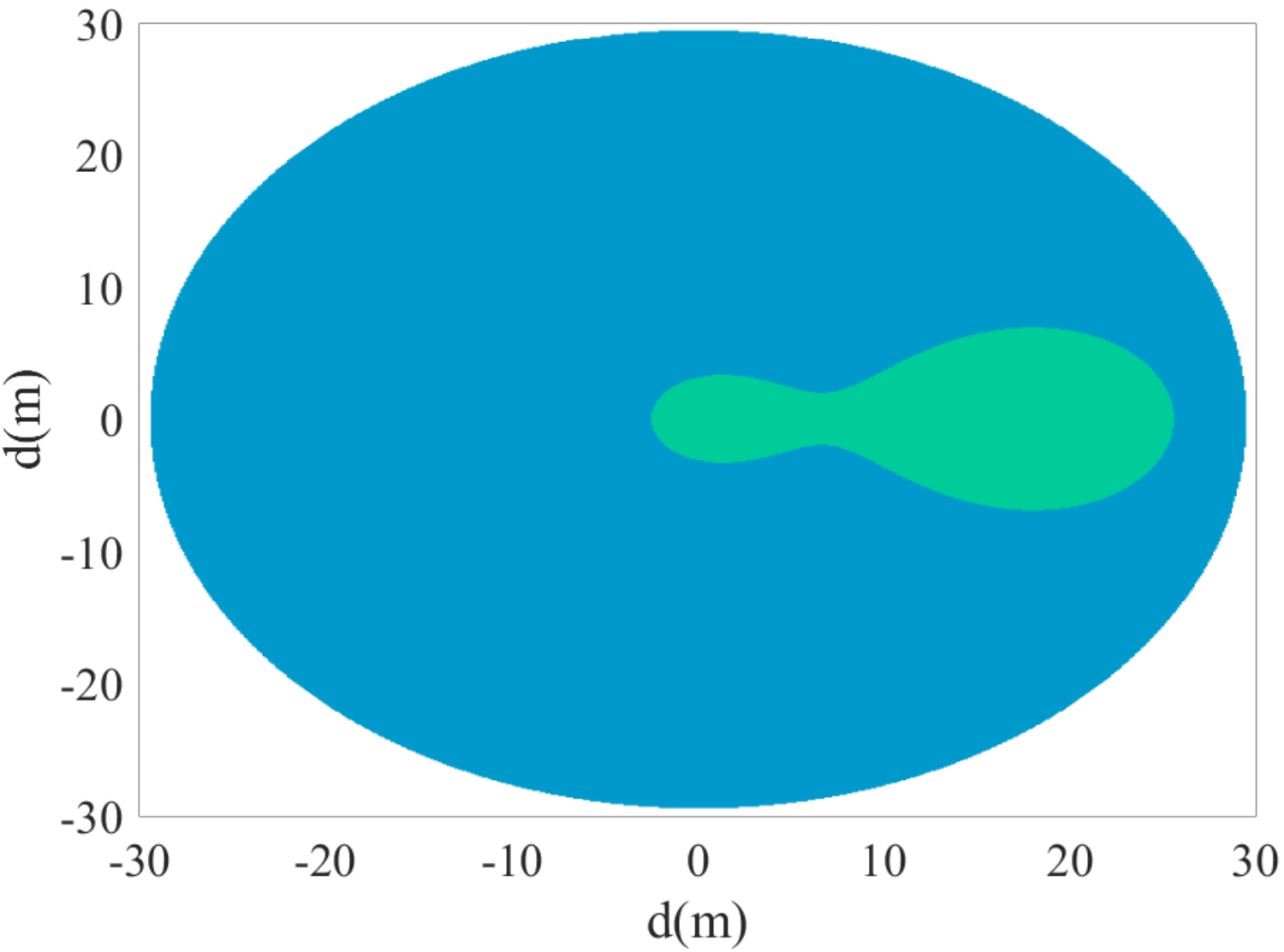}
\label{fig_pap30}
\end{minipage}}
\hspace{0.01\linewidth}
\subfigure[$P_{\textrm{AP}}$ = 5.0 watt]{
\begin{minipage}[b]{0.23\textwidth}
\centering
\includegraphics[width=1.7in]{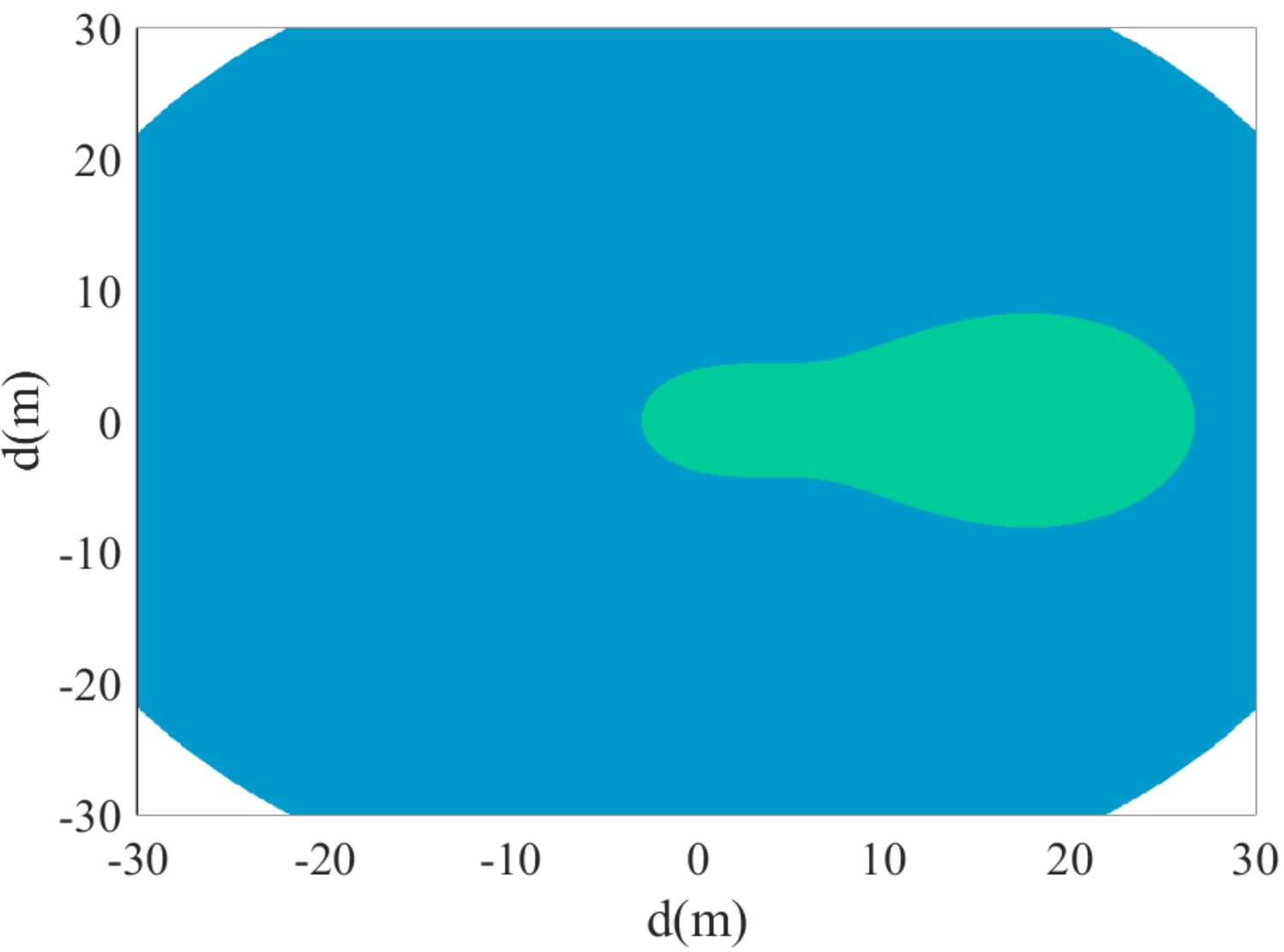}
\label{fig_pap50}
\end{minipage}}%
\hspace{0.01\linewidth}%
\subfigure[$P_{\textrm{AP}}$ = 10.0 watt]{
\begin{minipage}[b]{0.23\textwidth}
\centering
\includegraphics[width=1.7in]{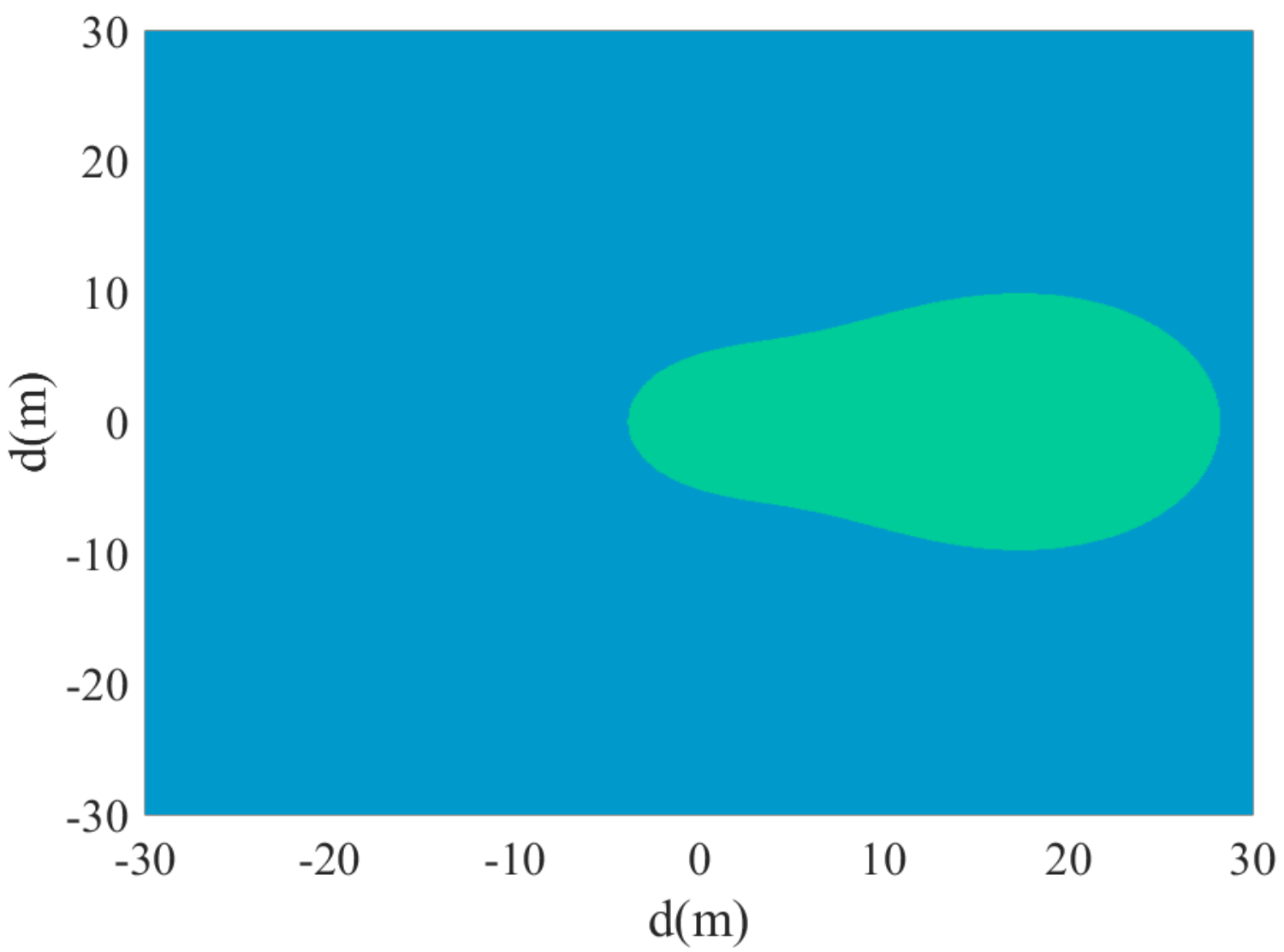}
\label{fig_pap100}
\end{minipage}}%
\hspace{0.01\linewidth}%
\caption{\textcolor[rgb]{0.00,0.00,0.00}{Comparisons of the two modes} versus $P_{\textrm{AP}}$}
\label{fig_pap}
\end{figure*}

With the same settings as Fig. \ref{fig_buju}, we discuss the effects of $P_{\textrm{AP}}$ on the mode selection over the 2-D coordinate plane, where $P_{\textrm{AP}}$ is increased from 0.5 Watt (as shown in Fig. \ref{fig_pap05}) to 10.0 Watt (as shown in Fig. \ref{fig_pap100}). From these figures, one can see that: 1) The feasible work area of the system expands around the HAP with the increment of $P_{\textrm{AP}}$; 2) The green areas (where the fog offloading mode is superior to the local computing one) become larger and larger with the increment of $P_{\textrm{AP}}$. This is because that when MU $m$ is positioned closed to the HAP, it can receive more power. When MU $m$ is positioned closed to the FS, less time is required to offload the task, leading to more time to perform energy harvesting.

\subsection{System performance versus $\beta$ for a given block}
\begin{figure}[h]
\centering
\includegraphics[width=0.435\textwidth]{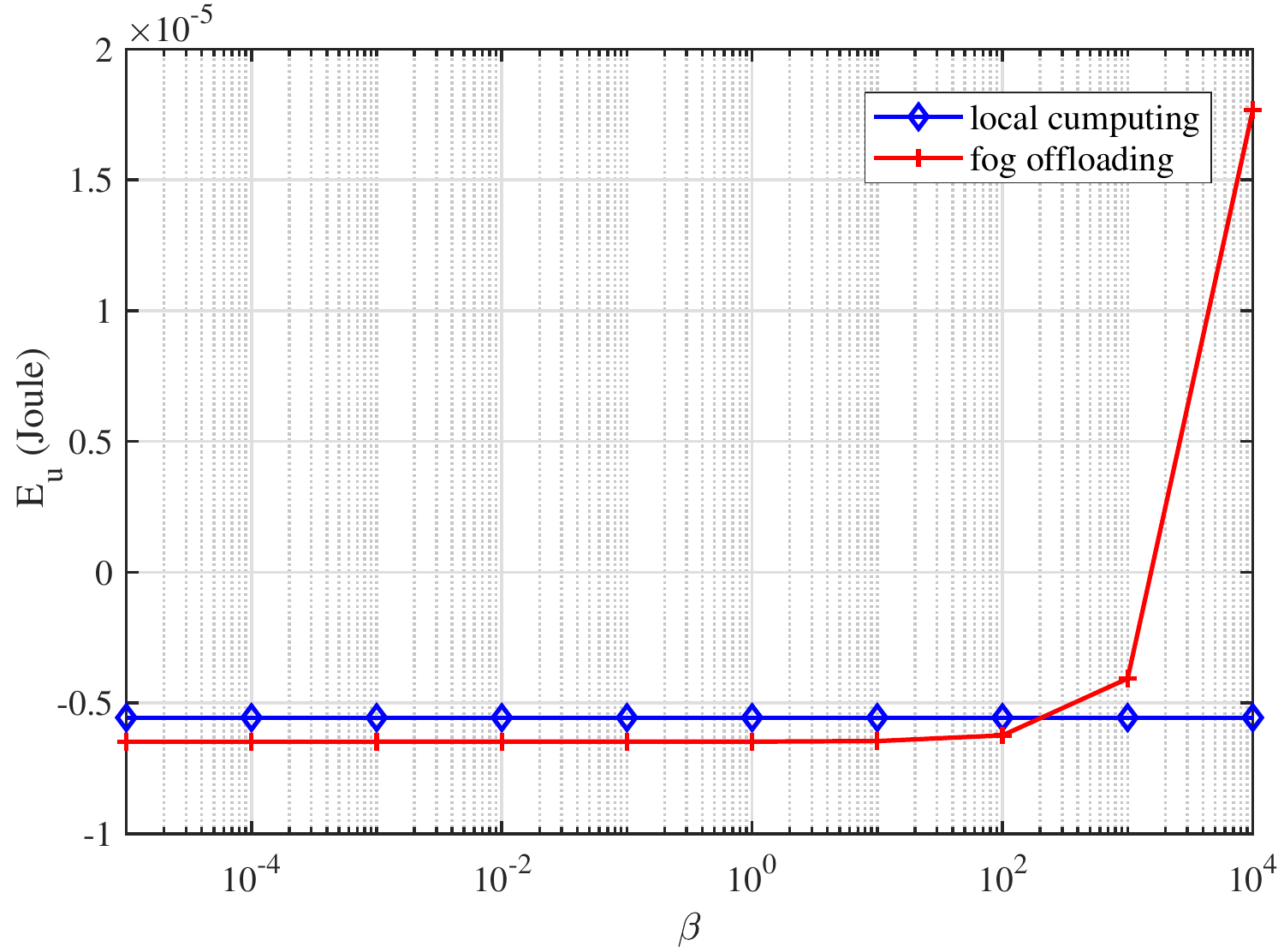}
\caption{\textcolor[rgb]{0.00,0.00,0.00}{The average energy requirement $E_{\textrm{u}}$ of the two modes versus $\beta$, where $E_{\textrm{u}}$ is defined in (\ref{lcenergy}) and (\ref{lcenergy1})}}
\label{fig_beta}\vspace{-0.1 in}
\end{figure}
The effect of the data compressed ratio, $\beta$, on the energy requirement of the two modes at MU $m$ is shown in Fig. \ref{fig_beta}, where $d_\textrm{AP-u}$ = 10 m and $d_\textrm{u-f}$ = 8 m. One can see that the required energy of the local computing mode does not change with the increment of $\beta$, while that of the fog offloading mode changes significantly. It is also observed that there exists an intersection point between the two curves at $\beta$ = $10^2$. When $\beta$ is less than $10^2$, the local computing mode is a better choice; \textcolor[rgb]{0.00,0.00,0.00}{otherwise}, the fog offloading mode should be selected. The reason is that the bigger the size of the calculated results, the more the time required to feedback them to the MU $m$, resulting in less time to harvest energy. This result is consistent with the analysis in \textcolor[rgb]{0.00,0.00,0.00}{\textbf{Proposition 6}}.

\begin{figure}
\centering
\includegraphics[width=0.45\textwidth]{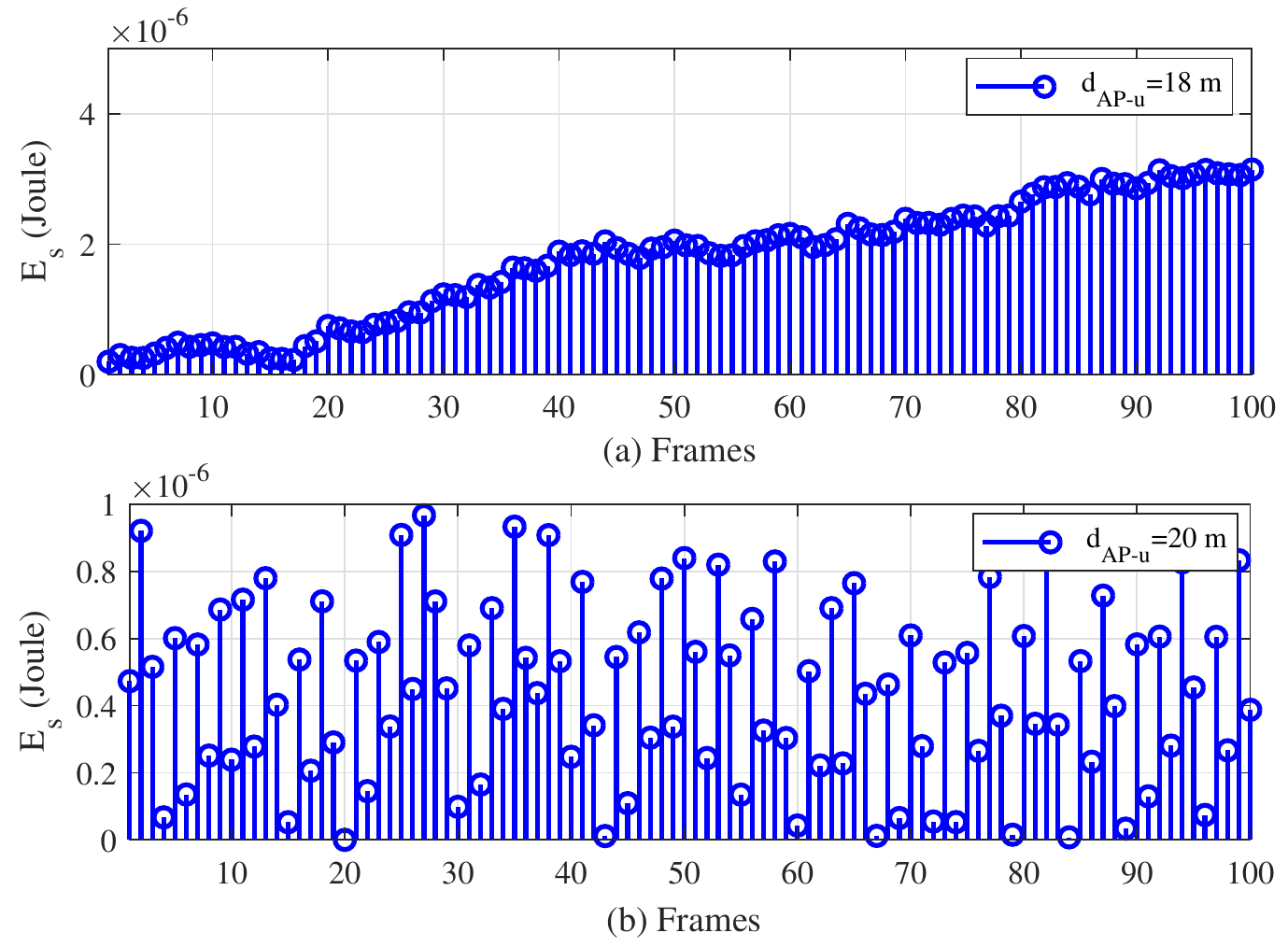}
\caption{\textcolor[rgb]{0.00,0.00,0.00}{(a) The energy storage in the battery for $d_\textrm{AP-u}$=18 m and (b) the energy storage in the battery for $d_\textrm{AP-u}$=20 m}}
\label{fig_tu51}\vspace{-0.1 in}
\end{figure}
\begin{figure}
\centering
\includegraphics[width=0.45\textwidth]{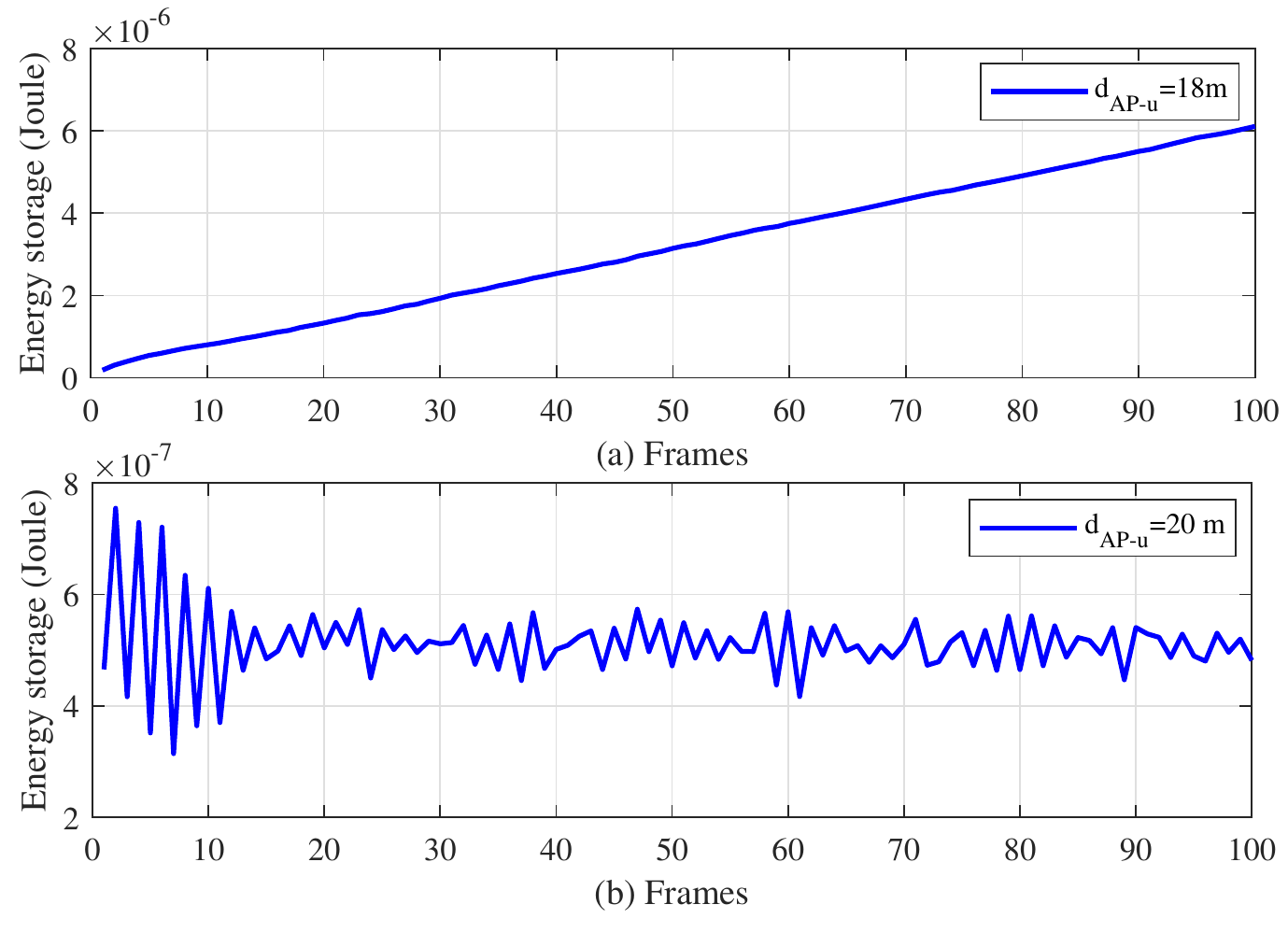}
\caption{\textcolor[rgb]{0.00,0.00,0.00}{(a) The average energy storage in the battery for $d_\textrm{AP-u}$=18 m and (b) the energy storage in the battery for $d_\textrm{AP-u}$=20 m}}
\label{fig_tu61}\vspace{-0.1 in}
\end{figure}

\begin{figure*}
\centering
\vspace{-0.2 in}
\subfigure{
\begin{minipage}[b]{0.33\textwidth}
\centering
\includegraphics[width=2.4in]{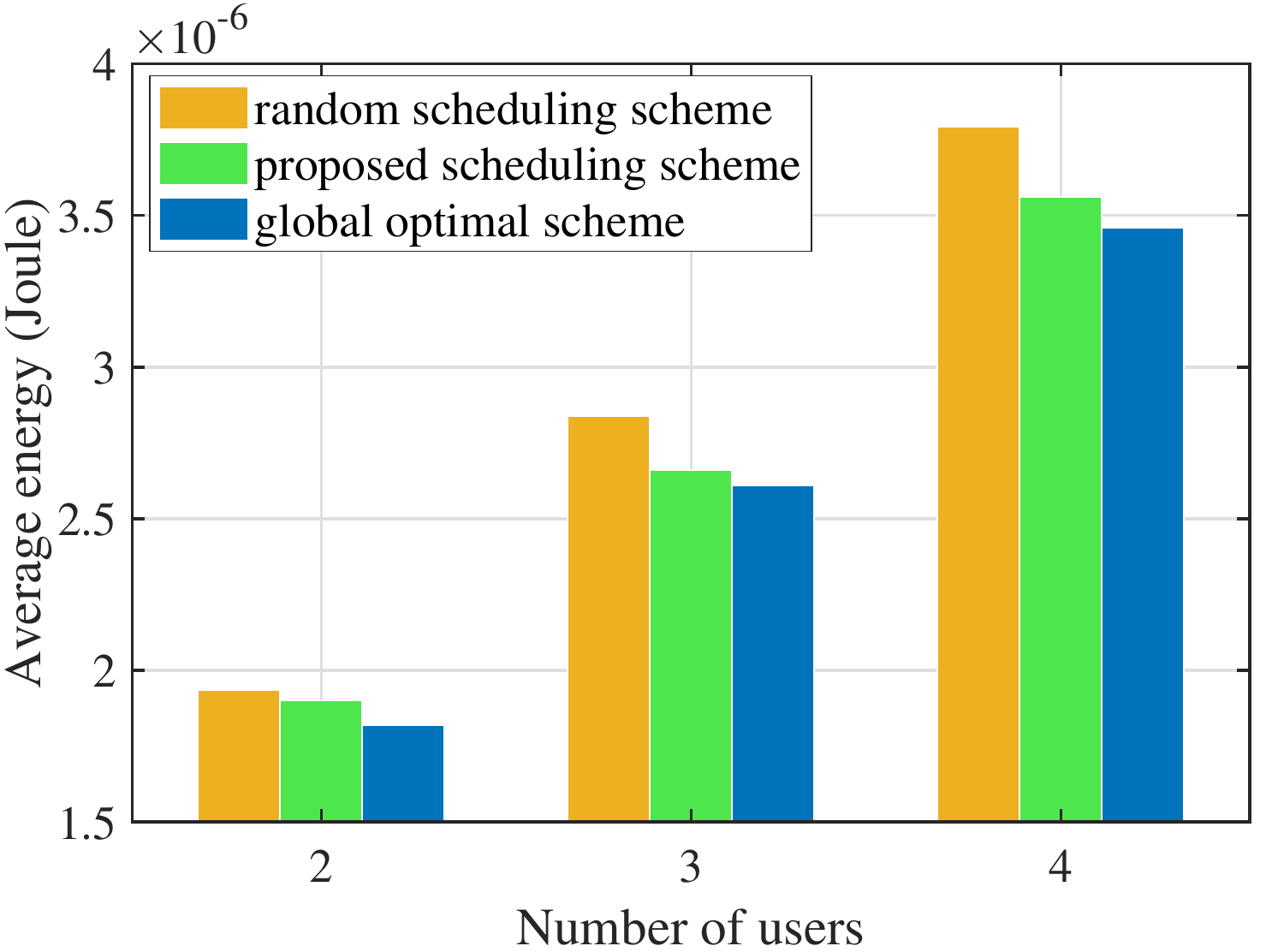}
\caption{\textcolor[rgb]{0.00,0.00,0.00}{The average energy} requirement versus $M$}
\label{fig_multireq}
\end{minipage}}%
\subfigure{
\begin{minipage}[b]{0.33\textwidth}
\centering
\includegraphics[width=2.4in]{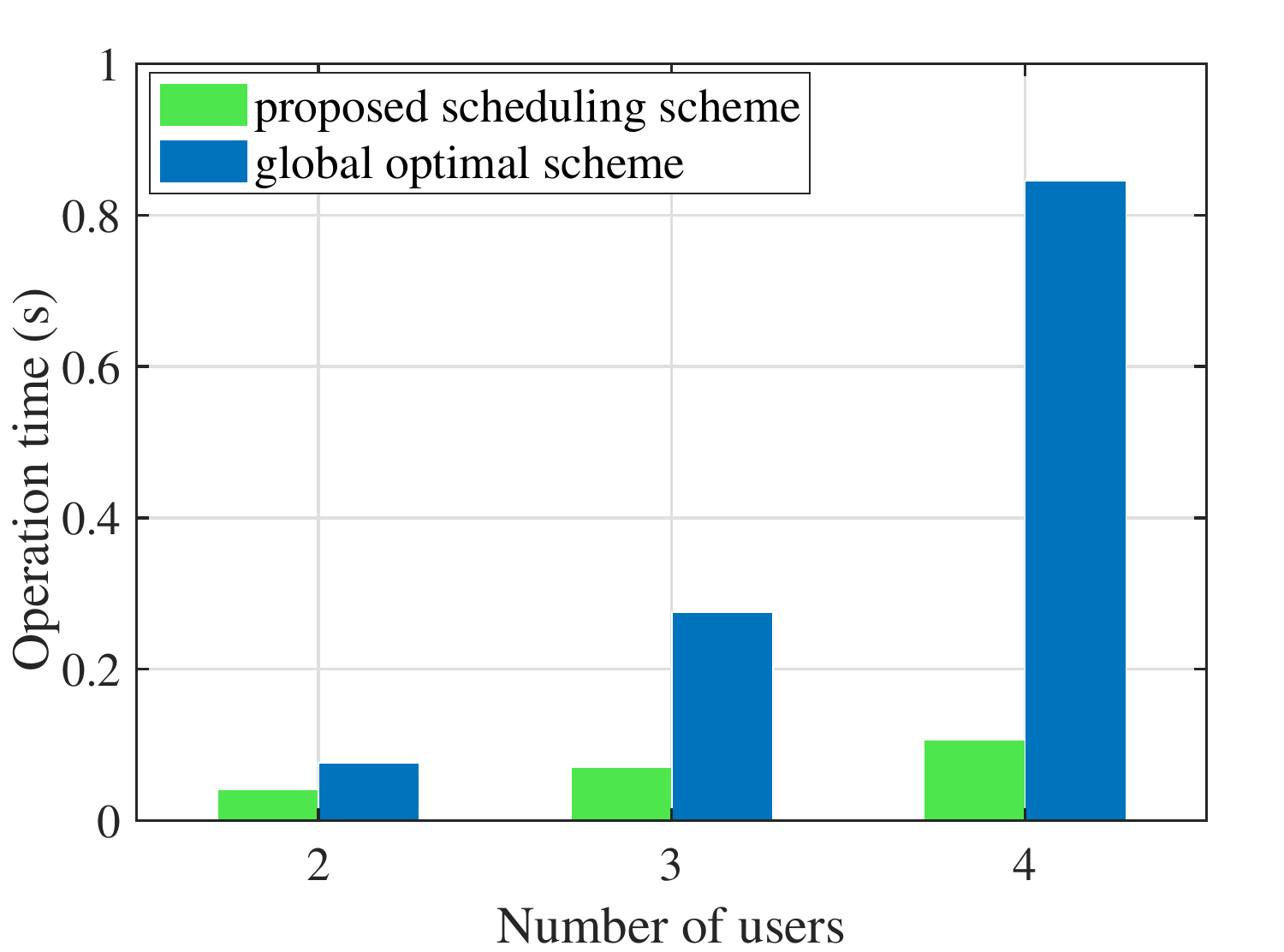}
\caption{\textcolor[rgb]{0.00,0.00,0.00}{The average operation} time versus $M$}
\label{fig_timecomplex}
\end{minipage}}%
\subfigure{
\begin{minipage}[b]{0.33\textwidth}
\centering
\includegraphics[width=2.4in]{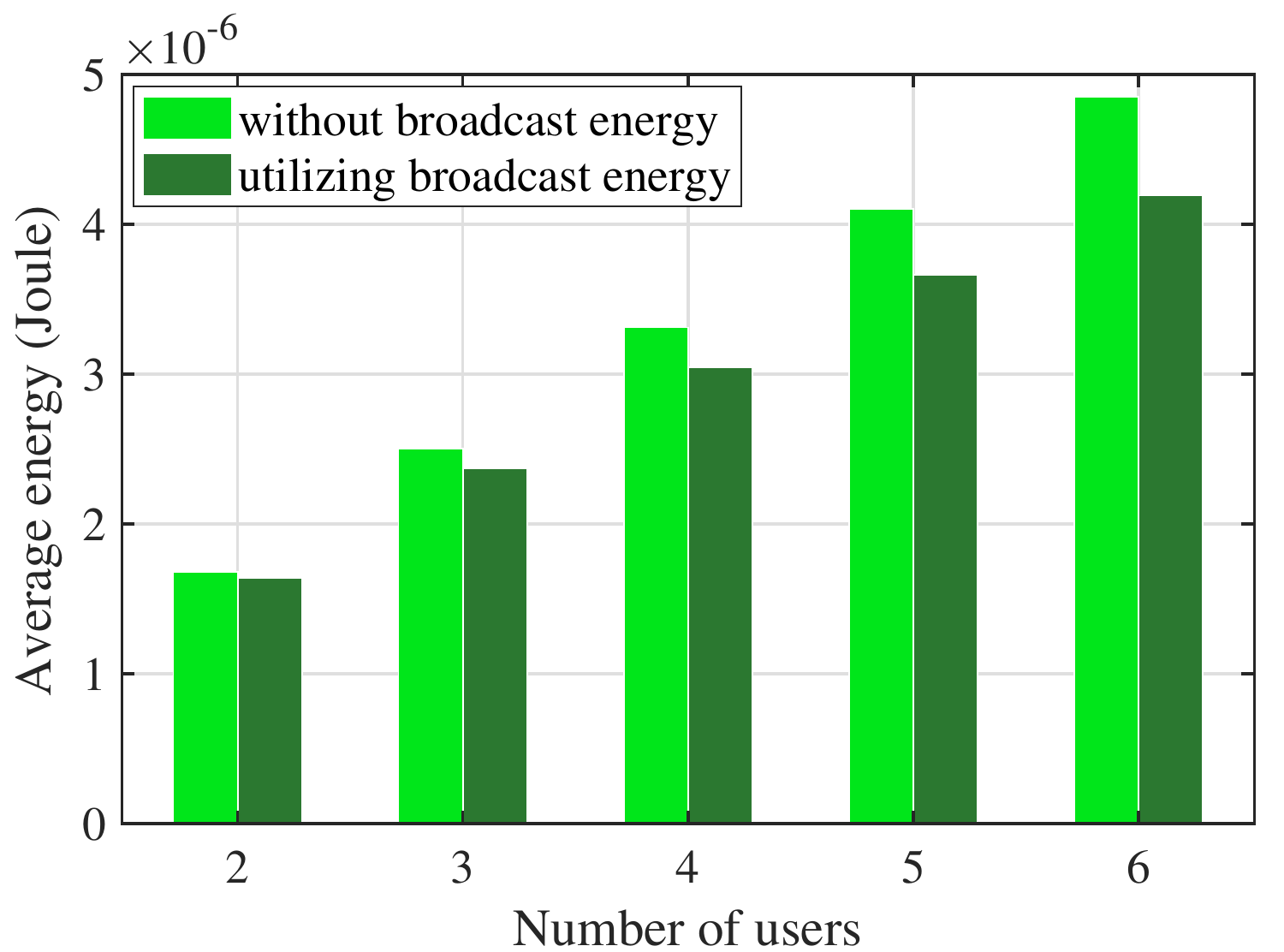}
\caption{\textcolor[rgb]{0.00,0.00,0.00}{The average energy} requirement versus $M$}
\label{fig_broad}
\end{minipage}}
\vspace{-0.2 in}
\end{figure*}

\subsection{System performance versus $E_{\textrm{s}}$ in a frame-by-frame applications}

Fig. \ref{fig_tu51} shows the energy storage level of MU $m$ versus 100 continuous time frames with different $d_\textrm{AP-u}$, i.e., $d_\textrm{AP-u}$ = 18m and 20m. As is shown in Fig. \ref{fig_tu51}(a), since the distance between the HAP and MU $m$ is relatively short and the amount of harvested energy is sufficient for completing the processes including the information decoding and local computing or data offloading, MU $m$ storages the harvested energy every frame. So, it is able to process the data in all time frames. However, with $d_\textrm{AP-u}$ = 20m as shown in Fig. \ref{fig_tu51}(b), MU $m$ has to harvest energy by consuming several time blocks due to the weak channel condition.

Fig.\ref{fig_tu61} plots the average energy storage level versus 100 continuous time frames with different $d_\textrm{AP-u}$, i.e., $d_\textrm{AP-u}$ = 18m and 20m. Each frame is averaged over $10^3$ channel realizations. It is observed that the average stored energy is almost linearly increasing for $d_\textrm{AP-u}$ = 18m while oscillatory varies for $d_\textrm{AP-u}$=20 m. These observations are consistent with those in Fig. \ref{fig_tu51}(a) and Fig. \ref{fig_tu51}(b).

\subsection{System Performance with multiple MUs}

Fig. \ref{fig_multireq} compares the average minimal energy requirement of our proposed scheduling scheme with two benchmark schemes, i.e., the random scheduling scheme and the global optimal scheduling scheme, versus the number of sensors, where the global optimal scheduling scheme is realized by exhaustive search method. It is seen that our proposed scheduling scheme requires less energy than the random scheduling scheme while only has a small gap with the global optimal scheduling scheme. This validates the effective of our proposed scheduling scheme.

Fig. \ref{fig_timecomplex} compares the average operation time of our proposed scheduling scheme with the global optimal scheduling scheme. It is seen that the operation time of our proposed scheduling scheme is significantly less than that of the global optimal scheduling scheme. In other words, by using our presented multi-user scheduling scheme, the approximate optimal result can be achieved with low computational complexity. \textcolor[rgb]{0.00,0.00,0.00}{Besides, compared to the exhaustive search method, the more the sensor number is, the less time is needed by our proposed scheduling scheme.}

Fig. \ref{fig_broad} show \textcolor[rgb]{0.00,0.00,0.00}{the effect of parameter $\iota$ on the system performance.} It can be seen that via collecting the energy in the signals transmitted by previous nodes, the total required power can be  further decreased.

\subsection{\textcolor[rgb]{0.00,0.00,0.00}{System Performance with inaccurate CSI}}
\begin{figure}
\centering
\includegraphics[width=0.435\textwidth]{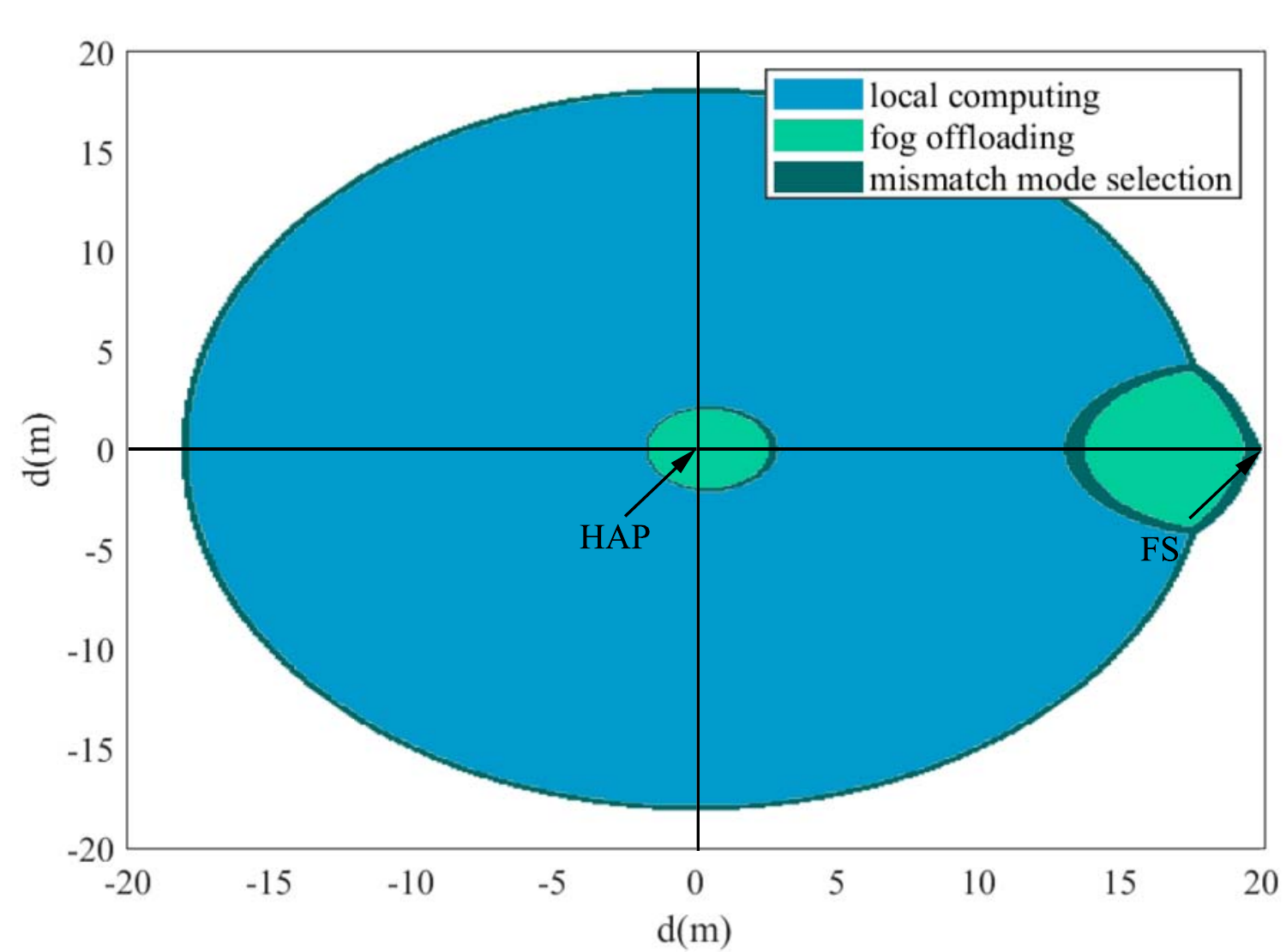}
\caption{\textcolor[rgb]{0.00,0.00,0.00}{The mode selection area with inaccurate CSI}}
\label{fig_error}\vspace{-0.1 in}
\end{figure}
\textcolor[rgb]{0.00,0.00,0.00}{With the same settings used in Fig. \ref{fig_buju}, we discuss the effect of the inaccurate CSI on the MU's mode selection. The result is shown in Fig. \ref{fig_error}, where the deep area represents the mismatch mode selection with $\epsilon=5\%$. $\epsilon$ represents the error factor of the CSI, and $\epsilon=5\%$ means that the inaccurate CSI has $5\%$ deviation with the accurate CSI. Compared with Fig. \ref{fig_buju}, it is seen that there are some mismatch areas of the mode selection in Fig. \ref{fig_error}, which indicates that the inaccurate CSI affects the results of the optimal mode selection.}

\begin{figure}
\centering
\includegraphics[width=0.435\textwidth]{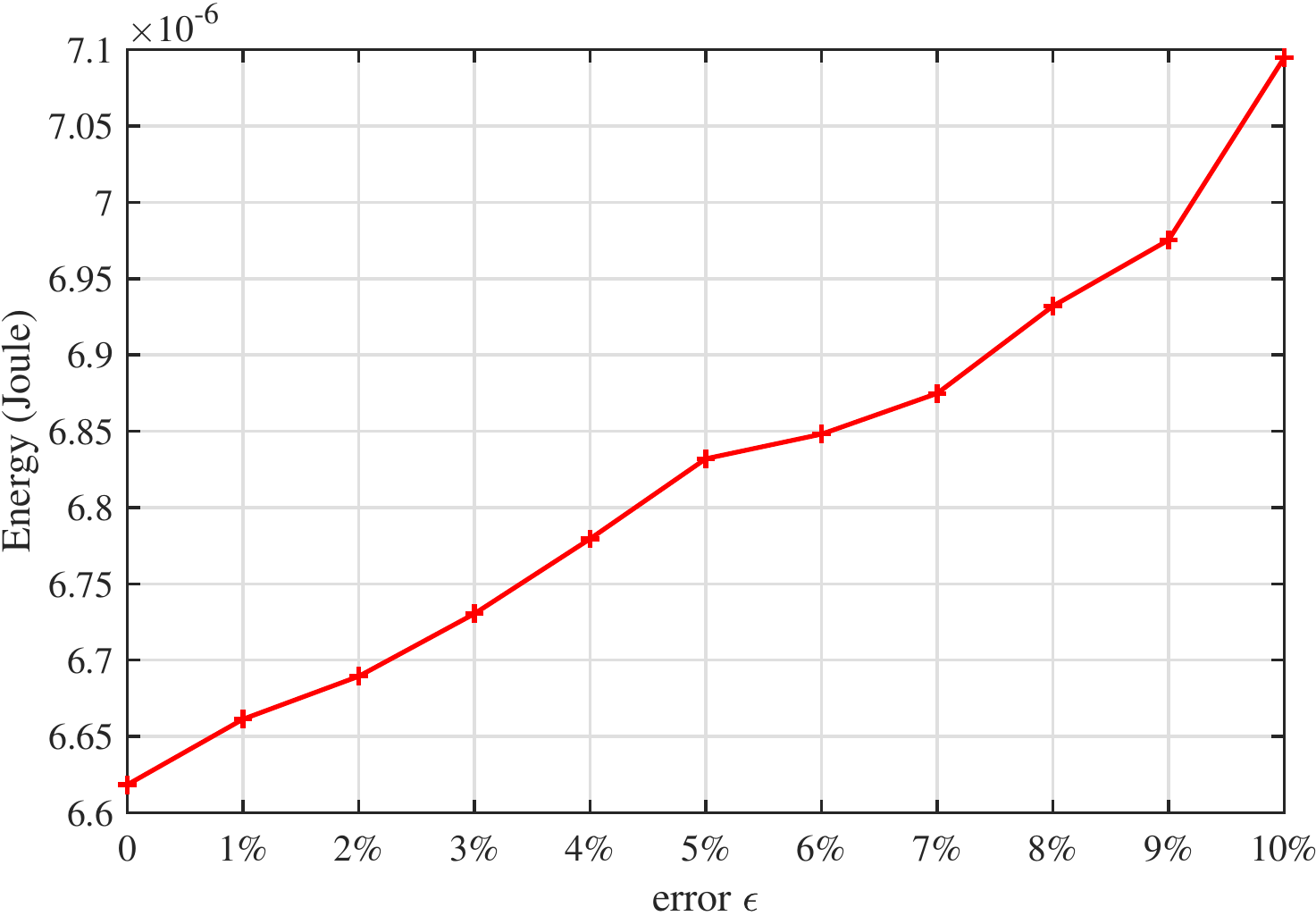}
\caption{\textcolor[rgb]{0.00,0.00,0.00}{The average minimal system energy requirement versus the error factor $\epsilon$ of the inaccurate CSI}}
\label{fig_error3}\vspace{-0.1 in}
\end{figure}
\textcolor[rgb]{0.00,0.00,0.00}{Fig. \ref{fig_error3} shows the average minimal required energy of the system versus the error factor $\epsilon$. It is observed that with the increment of $\epsilon$, the average minimal required energy is increased, and the bigger $\epsilon$ is, the more energy is required, which means the error factor $\epsilon$ affects the system. In Fig. \ref{fig_error3}, it shows that when $\epsilon$ is up to 10$\%$, the required energy is increased 7$\%$.}

\section{Conclusion} \label{conclusion}
This paper studied a multi-user fog computing-assisted SWIPT system with PS receiver architectures. Both the local computing mode and the fog offloading mode of MU were investigated. The closed-form and semi-closed-form expressions for the optimal configurations were derived, and the efficient user scheduling scheme was proposed. Simulation results showed that, for each MU, when the sensor is located close to the HAP and the FS, the fog offloading mode is the better choice; \textcolor[rgb]{0.00,0.00,0.00}{otherwise}, the local computing mode should be selected. The system performance in a frame-by-frame manner was also simulated, which showed that using the energy stored in the battery and that harvested from the signals transmitted by previous scheduled sensors can further decrease the system energy requirement.

\begin{appendices}
\section{The Proof of Lemma 1}
The objective function of Problem $\textbf{P}_{1-\textrm{B}}$, $F(\tau_{\textrm{ipt}}, \varphi)$, is a concave function, so that the minimal value of it must be at the outer boundary of its feasible set. Moreover, because of the concavity of the left hand side of Constraint (26a) and (26b), combining Proposition 1, the outer boundary of the feasible set of the $F(\tau_{\textrm{ipt}}, \varphi)$ satisfies that $f(\tau_{\textrm{ipt}}, \varphi) = R_{\textrm{th}}T_{\textrm{b}}$, which is presented by the black line on the ($\tau_{\textrm{ipt}}$, $f(\tau_{\textrm{ipt}}, \varphi)$) plane as shown in Fig. 3. When the points on the binary, i.e., $\{\tau_{\textrm{ipt}}, \varphi\}$ satisfies $f(\tau_{\textrm{ipt}}, \varphi) = R_{\textrm{th}}T_{\textrm{b}}$, $F(\tau_{\textrm{ipt}}, \varphi)$ achieves its minimal value. This is equivalent to that the minimal value of the $F(\tau_{\textrm{ipt}}, \varphi)$ is obtained when $B\tau_{\textrm{ipt}}\log\left(1+\frac{\varphi P_\textrm{AP}\left|\bm{h_{\textrm{AP-u}}}^H \bm{w}\right|^2}{\tau_{\textrm{ipt}}\sigma_{\textrm{n}}^2}\right)= R_{\textrm{th}}T_{\textrm{b}}$.
Therefore, the objective function of Problem $\textbf{P}_{1-\textrm{B}}$ can be transformed into as
\begin{flalign} \label{ref_th}
F(\tau_{\textrm{ipt}},\varphi)=\textrm{C}_{\textrm{1}}R_{\textrm{th}}T_{\textrm{b}} -\textrm{C}_{\textrm{2}}(\tau_{\textrm{ipt}} - \varphi)-\iota-E_{\textrm{s}}. \tag{A.1}
\end{flalign}
Then, the suitable $\{\tau_{\textrm{ipt}}, \varphi\}$ at the outer boundary $f(\tau_{\textrm{ipt}}, \varphi) = R_{\textrm{th}}T_{\textrm{b}}$ has to be determined to make $F(\tau_{\textrm{ipt}}, \varphi)$ reach the minimal value. According to the boundary condition $B\tau_{\textrm{ipt}}\log\left(1+\frac{\varphi P_\textrm{AP}\left|\bm{h_{\textrm{AP-u}}}^H \bm{w}\right|^2}{\tau_{\textrm{ipt}}\sigma_{\textrm{n}}^2}\right)= R_{\textrm{th}}T_{\textrm{b}}$, and (\ref{ref_th}), we have that $\varphi=\tfrac{\tau_{\textrm{ipt}}\sigma_{\textrm{n}}^2}{P_\textrm{AP}\left|\bm{h_{\textrm{AP-u}}}^H \bm{w}\right|^2}\left(2^{\tfrac{R_{\textrm{th}}T_{\textrm{b}}}{B\tau_{\textrm{ipt}}}} - 1\right)$.
Then, by substituting it into the objective function, the objective function is transformed into as
\begin{small}
\begin{flalign} \label{ref_comp6}
\mathfrak{f}(\tau_{\textrm{ipt}})&=\textrm{C}_{\textrm{1}}R_{\textrm{th}}T_{\textrm{b}}\underbrace{-\textrm{C}_{\textrm{2}}\tau_{\textrm{ipt}} \left(1-\tfrac{\sigma_{\textrm{n}}^2}{P_\textrm{AP}\left|\bm{h_{\textrm{AP-u}}}^H \bm{w}\right|^2}\overbrace{\left(2^{\tfrac{R_{\textrm{th}}T_{\textrm{b}}}{B\tau_{\textrm{ipt}}}}- 1\right)}^{(a)}\right)}_{(b)}\nonumber\\ \quad
&-\iota-E_{\textrm{s}},\tag{A.2}
\end{flalign}
\end{small}
which becomes a single-variable function w.r.t $\tau_{\textrm{ipt}}$. It can be proved that the $\mathfrak{f}(\tau_{\textrm{ipt}})$ is a monotonically decreasing function w.r.t the first constant term but a monotonically increasing function w.r.t. the second term. Nevertheless, in the second term, the increasing rate w.r.t $\tau_{\textrm{ipt}}$ of $(b)$ part in (\ref{ref_comp6}) is smaller than the increasing rate w.r.t $\tau_{\textrm{ipt}}$ of $(a)$ in (\ref{ref_comp6}). So, $\mathfrak{f}(\tau_{\textrm{ipt}})$ is a decreasing function w.r.t $\tau_{\textrm{ipt}}$. As a result, when $\tau_{\textrm{ipt}}$ reach the upper boundary, i.e., $\tau_{\textrm{ipt}} = \frac{T_{\textrm{b}}f_{\textrm{op}} - R_{\textrm{th}}KT_{\textrm{b}}}{f_{\textrm{op}}}$, $\mathfrak{f}(\tau_{\textrm{ipt}})$ arrives at its minimal valve. Note that as shown in Fig. 3, $\tau_{\textrm{ipt}} = \frac{T_{\textrm{b}}f_{\textrm{op}} - R_{\textrm{th}}KT_{\textrm{b}}}{f_{\textrm{op}}}$ is obtained in the case that the inequality of Constraint (26a) and (26b) adopt equal signs simultaneously, namely, the dynamic inner boundary meet the outer boundary $\frac{\left(T_{\textrm{b}} - \tau_{\textrm{ipt}}\right)f_{\textrm{op}}}{K}=R_{\textrm{th}}T_{\textrm{b}}$. Therefore, Lemma 1 is proved.
\section{The Proof of Theorem 1}
Following lemma 1, when $\{\tau_{\textrm{ipt}}, \varphi\}$ is at the outer boundary $f(\tau_{\textrm{ipt}}, \varphi) = R_{\textrm{th}}T_{\textrm{b}}$, the objective function  $F(\tau_{\textrm{ipt}}, \varphi)$ achieves its minimal value.  $\varphi=\frac{\tau_{\textrm{ipt}}\sigma_{\textrm{n}}^2}{P_\textrm{AP}\left|\bm{h_{\textrm{AP-u}}}^H \bm{w}\right|^2}\left(2^{\frac{R_{\textrm{th}}T_{\textrm{b}}}{B\tau_{\textrm{ipt}}}} - 1\right)$ and $\tau_{\textrm{ipt}} = $ $\tfrac{T_{\textrm{b}}f_{\textrm{op}} - R_{\textrm{th}}KT_{\textrm{b}}}{f_{\textrm{op}}}$, combining with $\rho = \frac{\varphi}{\tau_{\textrm{ipt}}}$, we have that $\rho_{\textrm{(local)}} ^{*} = $ $ \tfrac{\sigma_{\textrm{n}}^2}{P_\textrm{AP}\left|\bm{h_{\textrm{AP-u}}}^H \bm{w}\right|^2}\left(2^{\tfrac{R_{\textrm{th}}f_{\textrm{op}}}{B\left(f_{\textrm{op}} - KR_{\textrm{th}}\right)}} - 1\right)$. Moreover, when $ \tau_{\textrm{cpt}}<\tfrac{KR_{\textrm{th}}T_{\textrm{b}}}{f_{\textrm{op}}}$, $F(\tau_{\textrm{ipt}}, \varphi)$ has a smaller value with a larger $\tau_{\textrm{ipt}}$. So, $\tau_{\textrm{cpt}}^{*}=\tfrac{KR_{\textrm{th}}T_{\textrm{b}}}{f_{\textrm{op}}}$. Therefore, Theorem 1 is proved.
\section{The Proof of Lemma 2}
By deriving the second-order deviation of function $\vartheta(\tau_{\textrm{ipt}})$, we find it is always larger than zero in (C.3), where for clarity, we denote that $a=\frac{\sigma_{\textrm{s}}^2}{\left|h_{\textrm{u-f}}\right|^2}$, $b=\frac{R_{\textrm{th}}\!T_{\textrm{b}}}{B}$, $c=\eta P_\textrm{AP}\left|\bm{h_{\textrm{AP-u}}}^H \bm{w}\right|^2$, and $ d=\frac{\sigma_{\textrm{n}}^2}{P_\textrm{AP}\left|\bm{h_{\textrm{AP-u}}}^H \bm{w}\right|^2}$.
\begin{figure*}[htb]
\label{eeeqz}
\begin{flalign}
\vartheta(\tau_{\textrm{ipt}})&=\xi R_{\textrm{th}}T_{\textrm{b}}+a\left(\mathfrak{T_{\textrm{b}}}-\tau_{\textrm{ipt}}\right)\left(2^{\frac{b}{\mathfrak{T_{\textrm{b}}}-\tau_{\textrm{ipt}}}}-1\right)- c\tau_{\textrm{ipt}}\left(1-d\left(2^{\frac{b}{\tau_{\textrm{ipt}}}}-1\right)\right)
\nonumber\\ \quad&=\xi R_{\textrm{th}}T_{\textrm{b}}+a\mathfrak{T_{\textrm{b}}}2^{\frac{b}{\mathfrak{T_{\textrm{b}}}-\tau_{\textrm{ipt}}}}-a\mathfrak{T_{\textrm{b}}} -a\tau_{\textrm{ipt}}2^{\frac{b}{\mathfrak{T_{\textrm{b}}}-\tau_{\textrm{ipt}}}}+a\tau_{\textrm{ipt}}-c\tau_{\textrm{ipt}}+ cd\tau_{\textrm{ipt}}2^{\tfrac{b}{\tau_{\textrm{ipt}}}}-cd\tau_{\textrm{ipt}},\tag{C.1}\nonumber\\\quad
\vartheta'(\tau_{\textrm{ipt}})
&=a\mathfrak{T_{\textrm{b}}}2^{\frac{b}{\mathfrak{T_{\textrm{b}}}-\tau_{\textrm{ipt}}}}\frac{b\textrm{ln}2}{\left(\mathfrak{T_{\textrm{b}}}-\tau_{\textrm{ipt}}\right)^2} - a2^{\frac{b}{\mathfrak{T_{\textrm{b}}} -\tau_{\textrm{ipt}}}}- a\tau_{\textrm{ipt}}2^{\frac{b}{\mathfrak{T_{\textrm{b}}}-\tau_{\textrm{ipt}}}}\frac{b\textrm{ln}2}{\left(\mathfrak{T_{\textrm{b}}} -\tau_{\textrm{ipt}}\right)^2}+a-c +cd2^{\frac{b}{\tau_{\textrm{ipt}}}}- cd\tau_{\textrm{ipt}}2^{\frac{b}{\tau_{\textrm{ipt}}}}\tfrac{b\textrm{ln}2}{\tau_{\textrm{ipt}}^2}-cd
\nonumber\\\quad
&=a\left(\mathfrak{T_{\textrm{b}}}-\tau_{\textrm{ipt}}\right)2^{\frac{b}{\mathfrak{T_{\textrm{b}}}-\tau_{\textrm{ipt}}}}\frac{b\textrm{ln}2}{\left(\mathfrak{T_{\textrm{b}}}- \tau_{\textrm{ipt}}\right)^2}-a2^{\frac{b}{\mathfrak{T_{\textrm{b}}}-\tau_{\textrm{ipt}}}}+a-c-cd\left(2^{\frac{b}{\tau_{\textrm{ipt}}}}\left(1 -\frac{b\textrm{ln}2}{\tau_{\textrm{ipt}}}\right)-1\right)
\nonumber\\\quad
&=a\left(2^{\frac{b}{\mathfrak{T_{\textrm{b}}}-\tau_{\textrm{ipt}}}}\left(\frac{b\textrm{ln}2}{\mathfrak{T_{\textrm{b}}}-\tau_{\textrm{ipt}}}-1\right)+1\right) +cd\left(2^{\frac{b}{\tau_{\textrm{ipt}}}}\left(1-\frac{b\textrm{ln}2}{\tau_{\textrm{ipt}}}\right)-1\right)-c,\tag{C.2}\nonumber\\\quad
\vartheta''(\tau_{\textrm{ipt}})&=a\left(2^{\frac{b}{\mathfrak{T_{\textrm{b}}}-\tau_{\textrm{ipt}}}}\frac{b\textrm{ln}2}{\left(\mathfrak{T_{\textrm{b}}} -\tau_{\textrm{ipt}}\right)^2}\left(\frac{b\textrm{ln}2}{\mathfrak{T_{\textrm{b}}}-\tau_{\textrm{ipt}}}\!-\!1\right) +2^{\frac{b}{\mathfrak{T_{\textrm{b}}}-\tau_{\textrm{ipt}}}}\left(\frac{b\textrm{ln}2}{\left(\mathfrak{T_{\textrm{b}}}\!-\!\tau_{\textrm{ipt}}\right)^2}\right)\right) -cd\left(2^{\frac{b}{\tau_{\textrm{ipt}}}}\frac{b\textrm{ln}2}{\tau_{\textrm{ipt}}^2}\left(1\!-\!\frac{b\textrm{ln}2}{\tau_{\textrm{ipt}}}\right)
-2^{\frac{b}{\tau_{\textrm{ipt}}}}\frac{b\textrm{ln}2}{\tau_{\textrm{ipt}}^2}\right)
\nonumber\\ \quad &=a2^{\frac{b}{\mathfrak{T_{\textrm{b}}}-\tau_{\textrm{ipt}}}}\frac{\left(b\textrm{ln}2\right)^2}{\left(\mathfrak{T_{\textrm{b}}}-\tau_{\textrm{ipt}}\right)^3}+ cd2^{\frac{b}{\tau_{\textrm{ipt}}}}\frac{\left(b\textrm{ln}2\right)^2}{\tau_{\textrm{ipt}}^3}>0.
\tag{C.3}\label{eeeqz1}
\end{flalign}
\hrulefill
\end{figure*}

$\vartheta''(\tau_{\textrm{ipt}})>0$ always hold, so, Lemma 2 can be proved.
\section{The proof of Theorem 2}
\begin{figure}[h]
\centering
\includegraphics[width=0.32\textwidth]{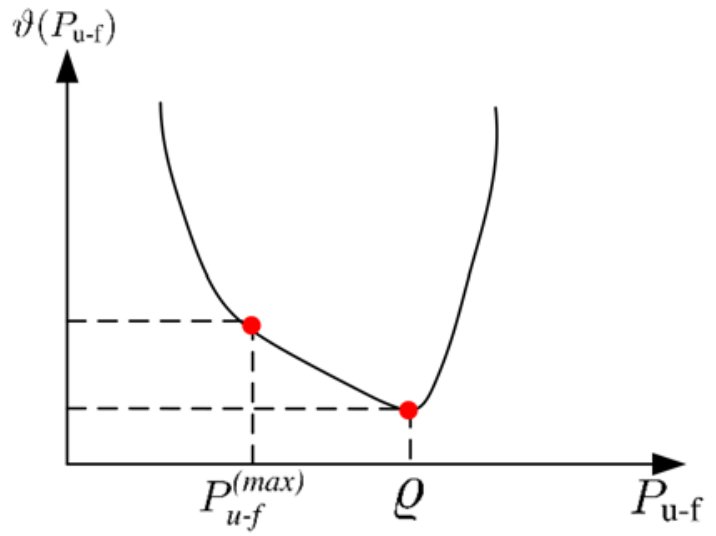}
\caption{Illustration of Theorem 2}
\label{s2}\vspace{-0.1 in}
\end{figure}

According to Lemma 2, $\tau^*_{\textrm{ipt}}$ can be obtained by setting $\frac{\partial\vartheta(\tau_{\textrm{ipt}})}{\partial\tau_{\textrm{ipt}}}=0$, i.e. $\alpha$ in (31). Although such an equation has no explicit expression of $\tau^*_{\textrm{ipt}}$, its numerical result can be obtained by using the bisection method. Once $\tau^*_{\textrm{ipt}}$ is obtained, by substituting it into the equations, i.e., $B\tau_{\textrm{ipt}}\log\left(1+\frac{\varpi }{\tau_{\textrm{ipt}}}\frac{P_\textrm{AP}\left|\bm{h_{\textrm{AP-u}}}^H \bm{w}\right|^2}{\sigma_{\textrm{n}}^2}\right)= R_{\textrm{th}}T_{\textrm{b}}$, $B\tau_{\textrm{u-f}}\log\left(1+\frac{\left|h_{\textrm{u-f}}\right|^2P_{\textrm{u-f}}}{\sigma_{\textrm{s}}^2}\right) = R_{\textrm{th}}T_{\textrm{b}}$, and $\tau_{\textrm{ipt}}+\tau_{\textrm{u-f}}=\mathfrak{T_{\textrm{b}}}$, respectively, the optimal $\rho_{(\textrm{offload})}^{*}$, $\tau_{\textrm{u-f}}^{*}$ and $P_{\textrm{u-f}}^{*}$ can be derived as shown by $\gamma$, $\delta$, $\varrho$ in (31), respectively. However, such optimal solution only can be achieved when $\varrho \leq P_{\textrm{u-f}}^{\textrm{(max)}}$. But for the case that $\varrho > P_{\textrm{u-f}}^{\textrm{(max)}}$, the optimal $P_{\textrm{f-u}}$ cannot reach $\varrho$ as illustrated in Fig. \ref{s2}.

In this case, $P_{\textrm{u-f}}^*=P_{\textrm{u-f}}^{\textrm{(max)}}$, because by combining Lemma 2 with $\tau_{\textrm{ipt}}=\mathfrak{T_{\textrm{b}}}-\frac{R_{\textrm{th}}T_{\textrm{b}}}{B \log\left(1+\frac{\left|h_{\textrm{u-f}}\right|^2P_{\textrm{u-f}}}{\sigma_{\textrm{s}}^2}\right)}$, it can be proved that $\vartheta(P_{\textrm{u-f}})$ is convex w.r.t $P_{\textrm{u-f}}$. Further, the optimal $\tau^{*}_{\textrm{ipt}}$, $\rho_{(\textrm{offload})}^{*}$, and $\tau_{\textrm{u-f}}^{*}$ can be derived by substituting $P_{\textrm{u-f}}^{\textrm{(max)}}$ \textcolor[rgb]{0.00,0.00,0.00}{into} the equations $B\tau_{\textrm{ipt}}\log\left(1+\frac{\varpi }{\tau_{\textrm{ipt}}}\frac{P_\textrm{AP}\left|\bm{h_{\textrm{AP-u}}}^H \bm{w}\right|^2}{\sigma_{\textrm{n}}^2}\right)= R_{\textrm{th}}T_{\textrm{b}}$, $B\tau_{\textrm{u-f}}\log\left(1+\frac{\left|h_{\textrm{u-f}}\right|^2P_{\textrm{u-f}}}{\sigma_{\textrm{s}}^2}\right) = R_{\textrm{th}}T_{\textrm{b}}$, and $\tau_{\textrm{ipt}}+\tau_{\textrm{u-f}}=\mathfrak{T_{\textrm{b}}}$.
Theorem 2 is thus proved.
\textcolor[rgb]{0.00,0.00,0.00}{\section{The proof of Proposition 4}
From (33), it is observed that $E_\textrm{id}=\xi R_{\textrm{th}}T_{\textrm{b}}$ and $E_\textrm{cpt}=KAR_{\textrm{th}}T_{\textrm{b}}$ are independent of $L(\textrm{dB})$, where $A$ = $F_{0}\alpha M_{\textrm{c}}N_{0}\ln\!2$, and $E_\textrm{eh}$ can be expressed as a function of $L(\textrm{dB})$, i.e.,
\begin{flalign}\label{ll}
  E_\textrm{eh}(L(\textrm{dB}))=&\eta \left(T_{\textrm{b}}- \frac{KR_{\textrm{th}}T_{\textrm{b}}}{f_{\textrm{op}}}\right)
  \left(1-\frac{\sigma_{\textrm{n}}^2C}{P_\textrm{AP} L(\textrm{dB})}\right)P_{\textrm{AP}} \nonumber \\ \quad
  &L(\textrm{dB})+\iota, \tag{D.1}
\end{flalign}
where $C$ = $2^{\frac{R_{\textrm{th}}f_{\textrm{op}}}{B\left(f_{\textrm{op}} - KR_{\textrm{th}}\right)}} - 1$. Since only when $E_\textrm{id} +E_\textrm{cpt}\leq E_\textrm{eh}+E_{\textrm{s}}$, the system can work. $L(\textrm{dB})$ must be smaller than a threshold such that $E_\textrm{id} +E_\textrm{cpt}\leq E_\textrm{eh}+E_{\textrm{s}}$ holds. It is a fact that $L(\textrm{dB})^{\textrm{(max)}}$ makes the equality of $E_\textrm{id} +E_\textrm{cpt}\leq E_\textrm{eh}+E_{\textrm{s}}$ be true.
Therefore, by setting $\xi R_{\textrm{th}}T_{\textrm{b}}+K AR_{\textrm{th}}T_{\textrm{b}}=\eta P_{\textrm{AP}}L(\textrm{dB})^{\textrm{(max)}}_\textrm{(local)}\left(T_{\textrm{b}}- \frac{KR_{\textrm{th}}T_{\textrm{b}}}{f_{\textrm{op}}}\right)\left(1-\frac{\sigma_{\textrm{n}}^2C}{P_\textrm{AP} L(\textrm{dB})^{\textrm{(max)}}_\textrm{(local)}}\right)+\iota+E_{\textrm{s}}$, we have
\begin{flalign}\label{fla_dc}
L(\textrm{dB})^{\textrm{(max)}}_\textrm{(local)}\!=\! \frac{f_{\textrm{op}}(KAR_{\textrm{th}}\!+\!\xi R_{\textrm{th}}\!-\!\iota\!-\!E_{\textrm{s}}) \!+\!\sigma_\textrm{n}^2C\eta(f_{\textrm{op}}\!-\!KR_{\textrm{th}})}{\eta (f_{\textrm{op}}-KR_{\textrm{th}})P_{\textrm{AP}}}. \tag{D.2}
\end{flalign}}

\textcolor[rgb]{0.00,0.00,0.00}{Similarly, $E_\textrm{id}$ and $E_\textrm{u-f}$ are independent of $L(\textrm{dB})$, and the harvested energy $E_\textrm{eh}$ can be expressed as a function of $L(\textrm{dB})$, i.e.,
\begin{equation}\label{ls}
  E_\textrm{eh}(L(\textrm{dB}))=\eta P_{\textrm{AP}}L(\textrm{dB})\left(1-\frac{\sigma_{\textrm{n}}^2D}{P_\textrm{AP} L(\textrm{dB})}\right)E+\iota, \tag{D.3}
\end{equation}
where $D=2^{\frac{R_{\textrm{th}}T_{\textrm{b}}}{CE}}-1$ and $E=T_{\textrm{b}}- \frac{R_{\textrm{th}}T_{\textrm{b}}}{B\log\left(1 + \frac{\left|h_{\textrm{u-f}}\right|^2P^{\textrm{(max)}}_{\textrm{u-f}}}{\sigma_{\textrm{s}}^2}\right)} - \frac{KR_{\textrm{th}}T_{\textrm{b}}}{f_{\textrm{fogop}}} - \frac{\beta R_{\textrm{th}}T_{\textrm{b}}}{B\log\left( 1 + \frac{\left|h_{\textrm{f-u}}\right|^2P^{\textrm{(max)}}_{\textrm{f-u}}}{\sigma_{\textrm{f}}^2}\right) }$. Also, by setting $(\frac{P^{\textrm{(max)}}_{\textrm{u-f}}}{F}+\xi) R_{\textrm{th}}T_{\textrm{b}}=\eta P_{\textrm{AP}}L(\textrm{dB})^{\textrm{(max)}}_\textrm{(offload)}\left(1-\frac{\sigma_{\textrm{n}}^2D}{P_\textrm{AP} L(\textrm{dB})^{\textrm{(max)}}_\textrm{(offload)}}\right)E+\iota+E_{\textrm{s}}$, where $F$ = $B\log\left(1 + \frac{\left|h_{\textrm{u-f}}\right|^2P^{\textrm{(max)}}_{\textrm{u-f}}}{\sigma_{\textrm{s}}^2}\right)$, one can get
\begin{flalign}\label{fla_dc}
L(\textrm{dB})^{\textrm{(max)}}_\textrm{(offload)} = \frac{(P^{\textrm{(max)}}_{\textrm{u-f}}+\xi F)R_{\textrm{th}}T_{\textrm{b}}+\eta\sigma_\textrm{n}^2DEF-\iota-E_{\textrm{s}}}{\eta EFP_{\textrm{AP}}}. \tag{D.4}
\end{flalign}}

\textcolor[rgb]{0.00,0.00,0.00}{Finally, $L(\textrm{dB})^{\textrm{(max)}}$ can be obtained by $L(\textrm{dB})^{\textrm{(max)}} =\textrm{max}\{L(\textrm{dB})^{\textrm{(max)}}_\textrm{(local)},L(\textrm{dB})^{\textrm{(max)}}_\textrm{(offload)}\}$. Therefore, proposition 4 is proved.}
\end{appendices}

\end{document}